\documentclass[%
 reprint,
 amsmath,amssymb,
 aps,onecolumn,notitlepage
pre,
]{revtex4-1}

\usepackage{multirow}
\usepackage[final]{graphicx}
\usepackage{float}
\usepackage{dcolumn}
\usepackage{bm}
\usepackage{hyperref}
\usepackage{cancel}
\hypersetup{
    colorlinks=true,
    linkcolor=blue,
    filecolor=magenta,      
    urlcolor=blue,
}

\newcommand{\sign}{\text{sign}}
\usepackage{soul}
\usepackage{ulem} 
\usepackage[caption=false]{subfig}
\usepackage{natbib}
\begin{document}
\preprint{APS/123-QED}

\title{The orientation dynamics of a massive ellipsoid in simple shear flow}

\author{Giridar Vishwanathan}
 \altaffiliation[This work began when the author came to JNCASR as a summer research fellow, while being enrolled at the ]{Department of Mechanical Engineering, Indian Institute of Technology Madras, Chennai 600036 India}
\affiliation{Department of Mechanical Science and Engineering, \\University of Illinois Urbana-Champaign, Urbana, USA 
}

\author{Sangamesh Gudda}
 \email{Email address for correspondence: sangameshguddabk@jncasr.ac.in}
\affiliation{Engineering Mechanics Unit, Jawaharlal Nehru Centre for Advanced Scientific Research,\\ Jakkur,
Bengaluru 560 064, India\\}

\author{Ganesh Subramanian}
 \email{Email address for correspondence: sganesh@jncasr.ac.in}
\affiliation{Engineering Mechanics Unit, Jawaharlal Nehru Centre for Advanced Scientific Research,\\ Jakkur,
Bengaluru 560 064, India\\}
\date{\today}
\begin{abstract}

The orientation dynamics of a massive rigid ellipsoid in simple shear flow of a Newtonian fluid is investigated in detail. 
The term `massive' refers to dominant particle inertia, as characterized by $St \gg 1$, $St = \dot{\gamma}l^2\rho_p/\nu \rho_f$ being the Stokes number; here, $\dot{\gamma}$ is the shear rate, $\nu$ is the kinematic viscosity, $l$ is a characteristic ellipsoidal dimension (taken to be the longest semi axis), and $\rho_p$ and $\rho_f$ are the particle and fluid densities, respectively. Fluid inertial effects are neglected, so the particle Reynolds number ($Re$) is zero. The equations of motion of the ellipsoid in this limit reduce to those governing an Euler top, supplemented by a weak viscous torque. The dynamics consists of a fast conservative motion on time scales of $O(\dot{\gamma}^{-1})$ that, for an ellipsoid, involves a combination of spin, precession and nutation, and a slower component driven by the viscous torque; the latter modulates the angular momentum and rotational kinetic energy on asymptotically longer time scales of $O(St{\dot{\gamma}}^{-1})$. The spheroid is a limiting case where the fast motion comprises only a spin and a precession. The separation of time scales for large $St$ allows for use of the method of multiple scales for a spheroid, and the method of averaging for a triaxial ellipsoid, to derive an autonomous system of ODE's that govern motion on a four-dimensional slow manifold consisting of the three angular momenta and the rotational kinetic energy.

 For a triaxial ellipsoid, there are three fixed points on the slow manifold, corresponding to uniform rotation of the ellipsoid about each of its principal axes aligned with the ambient vorticity. Rotation about the shortest axis is stable, in agreement with previous numerical work, with the fixed point being a stable node. The fixed point corresponding to rotation about the longest axis is a saddle point, while intermediate-axis-aligned rotation corresponds to a singular fixed point that only plays a passive role in organizing the dynamics in its vicinity. For spheroids, the singular fixed point merges with one of the other two, leading to only two fixed points, one a stable node and the other a saddle; the former corresponds to rotation about the shorter axis. 
 
\end{abstract}

\maketitle
\section{Introduction}
\subsection{Motivation}

In this paper, we examine the orientation dynamics of ellipsoids in an ambient shearing flow, in the limit where particle inertia is dominant, fluid inertia still being negligible. This limit arises for gas-solid systems owing to the particle density ($\rho_p$) being much greater than the fluid density ($\rho_f$). The analysis here is a first step towards understanding the orientation dynamics of anisotropic particles in the aforementioned limit of dominant particle inertia, and accordingly, the effect of gravity via a sedimentation-induced torque\cite{cox1965,khayatcox1989,dabade2015} is neglected, and the ambient flow is taken to be a simple shear flow; the more involved dynamics in the one-parameter family of planar linear flows\,(the parameter, $\alpha\,\in\,[-1,1]$, with $\alpha=-1$, $0$ and $1$ corresponding to rigid body rotation, simple shear and planar extension, respectively) will be reported in a later communication. When generalized to unsteady, in particular, stochastic linear flows, the analysis would be of relevance to predicting the orientation distribution of sub-Kolmogorov ice crystals in cloud turbulence, a topic of considerable interest in recent times \cite{siewert2014,voth2017,anand2020,pumir2022}. Numerous efforts in the literature that examine the distribution of massive particles in both homogeneous isotropic turbulence and model stochastic flow fields\cite{bec2006,olla2008,rani2014,dhariwal2017,kasbaoui2015,mehlig2016}, deal exclusively with spherical particles, where the phenomena of interest such as preferential concentration and clustering in extension-dominant regions arise solely due to translational inertia\cite{jaehun2005,sundaramcollins1,sundaramcollins2}. Ours is the first analytical effort to examine inertia associated with orientation degrees of freedom, albeit in the simpler setting of a simple shear flow. 

A direct motivation for the analysis here is a series of numerical investigations examining the orientation dynamics of massive spheroids \cite{lundell2010,challabotla2015} and triaxial ellipsoids \cite{lundell2011} in a simple shear flow, described in more detail below. Indirect motivation arises the large body of work on the dynamics of inertial spheroids in turbulent channel flow. There is considerable evidence, even in the absence of gravity, of particle inertia playing a crucial role on the spheroid orientation dynamics in the vicinity of the wall\,(the viscous and buffer layers) where the mean shear becomes dominant\cite{marchioli2016,zhao2019mapping,MichelArcen2021}. The model problem solved here will help separate the roles of the mean shear, and superposed stochastic velocity gradient fluctuations, on the stability of a particular rotation mode\,(tumbling vs spinning vs kayaking, etc).

The relevant non-dimensional parameters are the Reynolds($Re=\frac{ l^2 \dot{\gamma} }{\nu}$) and Stokes($St=\frac{\rho_p l^2 \dot{\gamma}}{\rho_f \nu}$) numbers, and the limiting scenario examined here corresponds to $Re=0$ and $St \gg 1$, with the particle aspect ratios being arbitrary. Here, $\rho_f$  and $\nu$ are the fluid density and kinematic viscosity, respectively, $\rho_p$ is the particle density and $\dot{\gamma}$ is a characteristic shear rate. $l$ is taken to be the largest particle dimension which essentially determines the scale of the Stokesian disturbance velocity field. For an ellipsoid, the semi-axis lengths are $a,b$ and $c$\,($a>b>c$), and $l$ equals $a$; spheroids have semi-axes $(a,a,c)$, being characterized by a single aspect ratio $\kappa=a/c$; $\kappa < 1$ and $\kappa > 1$ for oblate and prolate spheroids, respectively. 

\subsection{Literature Review}\label{litrev}

The analysis of an inertialess torque-free ellipsoid($St=0$) in an ambient linear flow, in the creeping flow limit ($Re=0$), was first carried out by Jeffery \cite{jeffery1922} who derived its angular velocity as a function of the instantaneous orientation using ellipsoidal harmonics. For a spheroid in simple shear flow, the angular velocities simplify considerably owing to decoupling of the spin component, and allow for an analytical solution of the equations governing the two orientation variables. The solution shows that a spheroid rotates in a family of closed trajectories, now known as Jeffery orbits, and that are characterized by an orbit constant $C$ ranging from $0$ (the spinning or log-rolling mode) to $\infty$ (the tumbling mode). In planar linear flows, a spheroid continues to rotate in closed trajectories so long as $\alpha<\kappa^2\,(1/\kappa^2)$ for oblate\,(prolate) spheroids\cite{hinch1972,marath2018}.

Deviations from inertialess Jeffery orbit dynamics have been studied in several analytical efforts, often in an attempt to resolve the rheological degeneracy associated with a dilute Stokesian suspension of axisymmetric particles. Subramanian and Koch examined the effect of weak particle inertia\,($Re = 0$, $St \ll 1$), and the effect of both particle and fluid inertia on neutrally buoyant particles\,($Re = St \ll 1$) in the slender fiber \cite{subramanian2005}, and near-sphere \cite{subramanian2006} limits through use of the generalized reciprocal theorem; the additional influence of a sedimenting torque due to gravity was also probed in both cases. Dabade et al.\cite{dabade2016} investigated the effect of weak particle and fluid inertia on a neutrally buoyant spheroid of an arbitrary aspect ratio in simple shear flow. Inertia always caused a prolate spheroid to drift towards the tumbling mode. For an oblate spheroid, however, there exists a critical aspect ratio ($\approx 0.14$) below which the final state (tumbling or spinning) depends on its initial orientation. Above this aspect ratio, an oblate spheroid always tends to a spinning state about the ambient vorticity, in agreement with an independent investigation by Einarsson et al.\cite{einarsson2015}. For smaller aspect ratios, the partitioning between the spinning and tumbling modes, and thence, the elimination of rheological indeterminacy, depends crucially on stochastic reorientation mechanisms. Marath and Subramanian \cite{marath2018} extended the aforementioned study to planar linear flows, and examined the inertial orientation dynamics in the portion of the $\alpha-\kappa$ plane below the threshold curves $\alpha = \kappa^{\pm 2}$ mentioned above, which corresponds to a closed-trajectory topology in the Stokesian limit. Regions with initial-orientation-dependent behavior were identified, and the indeterminacy removed by considering a specific mechanism of stochastic orientation fluctuations - rotary Brownian motion. In the said efforts, the inertial orientation dynamics was also given a thermodynamic interpretation in $C-\kappa-Re\,Pe_r$ space\cite{dabade2016,dwivedi2017,marath2018}, $Pe_r$ here being the rotational Peclet number that characterizes the importance of Brownian motion reative to shear induced rotation. The tendency of inertia to (locally) stabilize both tumbling and spinning modes, over a certain range of aspect ratios, leads to a `tumbling-spinning' region of coexistence, which was found to be largest in extent for simple shear flow\cite{marath2018}. 

The inertial orientation dynamics in the aforementioned studies was for small $Re$ and $St$, in which case the separation between the Jeffery rotation and inertial drift time scales allowed for a one-dimensional description of the inertial orientation dynamics in terms of an orbital drift, defined as the rate of change of the orbit constant $C$ with time measured in units of the Jeffery period; $C$ being conserved in the Stokes limit. In this paper, we examine the opposite limit viz.\ $St\rightarrow \infty$. For $St = \infty$, corresponding to the absence of the suspending fluid, there are multiple conserved parameters analogous to the orbit constant above - the three components of the angular momentum vector ($\vec{L}$) and the rotational kinetic energy ($E$). For large but finite $St$, one expects the orientation dynamics to be describable in terms of a slow drift of these parameters.

While the orientation dynamics of inertialess triaxial ellipsoids was originally examined by Gierszewski and Chaffey\,(1978)\cite{chaffey1978}, the first detailed analysis was by  Hinch and Leal \cite{leal1979} via a study of the orientational Poincar\'e maps for a wide range of ellipsoid axis ratios. In contrast to spheroids, the dynamics was found to be quasi-periodic with trajectories in the Poincar\'e section, comprising the nutation and spin Euler angles ($\theta$, $\psi$)\,(see section \ref{gove} below for definitions) appearing as densely filled closed curves. The quasi-periodic dynamics was analyzed in terms of the stability of the principal-axis-aligned rotations. Later, Yarin et al.\cite{yarin1997} explored the Poincar\'e maps of slender ellipsoids with one axis much longer than the other two, and showed that the dynamics is chaotic in much of orientation space. Chaotic trajectories therefore appear to coexist with quasi-periodic ones for all triaxial ellipsoids, with the chaotic regions being almost imperceptibly small for the ellipsoids with order unity aspect ratios considered by Hinch and Leal (1979).

Several recent numerical efforts have examined the roles of finite fluid and particle inertia on both prolate and oblate spheroids in simple shear flow. The methods used include Lattice Boltzmann simulations (LBM) \cite{qi2003,huang2012,rosen2014,rosen2015}, and to a lesser extent, the distributed Lagrangian-multiplier fictitious-domain method \cite{yu2007}, and have led to a rich variety of dynamical behavior on the $Re\!-\!St$ plane. For instance, the sequence of transitions in the orientation dynamics, with increasing inertia, is tumbling, quasi-Jeffery orbit, log-rolling, inclined rolling, inclined kayaking and stationary alignment for a prolate spheroid; and spinning, inclined spinning and stationary alignment for an oblate spheroid. Despite this richness, certain features predicted by the weak-inertia analyses above persist. For instance, analogous to the degenerate scenario identified for small $Re$, by Einarsson et al. \cite{einarsson2015} and Dabade et al. \cite{dabade2016}, locally stable states co-exist at finite $Re$, with the particular state accessed depending on initial orientation. In fact, even the rather elaborate bifurcation sequences above have been found for small $Re$ and $St$, in the neighborhood of the threshold curves $\alpha = \kappa^{\pm 2}$, based on an analysis\cite{Pulkit2023} that goes beyond the orbital drift framework used in Dabade {\it et al.}\cite{dabade2016} and Marath {\it et al.}\cite{dwivedi2017,marath2018}. The effect of a fluid-inertia-induced slowdown on the time periods of both spheroid and ellipsoid rotation 
has also been examined\cite{mao2014,marath2017,rosen2017}.

The efforts most relevant to the present analysis are those geared towards understanding the role of particle inertia alone (over a wide range of $St$) on the orientation dynamics of prolate spheroids\cite{lundell2010}, oblate spheroids\cite{challabotla2015} and triaxial ellipsoids \cite{lundell2011} in simple shear flow. For prolate and oblate spheroids with small $St$, the fast dynamics was found to be similar to that of a Jeffery orbit, with a superimposed slow drift towards the flow-gradient plane (prolate) or vorticity axis (oblate), as predicted by the analytical efforts above. For large $St$, the dynamics is profoundly different. Prolate spheroids were found to rapidly precess about an inclined axis, which slowly tilts towards the ambient vorticity direction. Concomitantly, the angle of precession increases to $\pi/2$, as a result of which the spheroid approaches a steady aspect-ratio-independent rate of rotation in the flow-gradient plane\cite{lundell2010}. In contrast, oblate spheroids were found to transition from an initial state of rest to spinning about the symmetry axis, which in turn precesses about the ambient vorticity, approaching it for long times, with the rate of spin asymptoting to half the ambient vorticity
\cite{challabotla2015}. For both prolate and oblate spheroids starting from rest, at large $St$, there is an initial transient associated with a transition from Jeffery-orbit-like rotations to the aforementioned precessional dynamics; the duration of this transient becomes progressively shorter, in comparison to the precessional regime, with increasing $St$. The study on triaxial ellipsoids \cite{lundell2011} showed that, for intermediate $St$, minor deviations from axisymmetry lead to chaotic dynamics for slender ellipsoids. 
For sufficiently large $St$, however, the orientation dynamics is regularized, with all initial orientations appearing to converge to a final state corresponding to vorticity-aligned rotation about the shortest axis, in a manner similar to that for spheroids above\cite{challabotla2015,lundell2010}.

From the above review of literature, we find that while there have been many analytical efforts probing deviations from the Jeffery case for small $St$ and $Re$, studies dealing with the large $St$ limit are far fewer, and almost entirely numerical, despite the orientation dynamics being qualitatively different. The present work is thus the first analytical effort dealing with the large-$St$ orientation dynamics of a single triaxial ellipsoid in a steady shear flow. The organization of the article is as follows. 
In section \ref{gove}, the set of differential equations governing the dynamics of an ellipsoid and a spheroid are written down for arbitrary $St$, and their principal features discussed. Herein, we also validate our results\,(obtained from numerically integrating the said equations) against those known from earlier efforts, both analytical for small $St$ \cite{dabade2016} and numerical for large $St$ \cite{lundell2010}. As a prelude to the discussion on the dynamics for large but finite $St$, the limiting cases of $St=0$ (the Jeffery top) and $St=\infty$ (the Euler top) 
are discussed in section \ref{asympcase}. The analysis of a massive spheroid in simple shear flow is undertaken in section \ref{lsds}. The leading order variations of the angular momenta and the rotational kinetic energy on the slow time scale of $O(St\dot{\gamma}^{-1})$ are obtained via the method of multiple scales. Two fixed points on the slow manifold are identified, and analyzed for their stability. Rotation about the shorter axis aligned with vorticity is found to be stable, for both prolate and oblate spheroids, in agreement with earlier numerical investigations\cite{lundell2010,challabotla2015}. The $L_z-E$ phase plane is argued to be a convenient means of representing the results for different aspect ratios and $St$. In section \ref{lsde}, the method of averaging is used to derive the slow manifold equations for a triaxial ellipsoid in simple shear flow. The analysis of these equations reveals three fixed points, and the one corresponding to rotation with the shortest axis aligned with vorticity is again found to be stable. As for the spheroid, the ellipsoidal orientation dynamics for different axis ratio pairs is studied by means of $L_z-E$ phase portraits. 
Finally, section \ref{DAC} states the principal conclusions. Herein, we also provide a scaling arguments that, for a particle starting from rest, examine the relative lengths of the initial inertialess transient and the asymptotic large-$St$ regime. Appendix \ref{LOA} contains the evaluation of all averages necessary for derivation of the slow manifold equations, for both the spheroid and the ellipsoid, based on the for the classical Euler-top solution in section \ref{asympcase}. In Appendix \ref{FOD}, we evaluate the $O(1/St)$ oscillatory correction to the leading order secular dynamics of the spheroid, for the restricted case where the spheroid angular momentum is aligned with the ambient vorticity.



\begin{figure}
\includegraphics[scale=0.9]{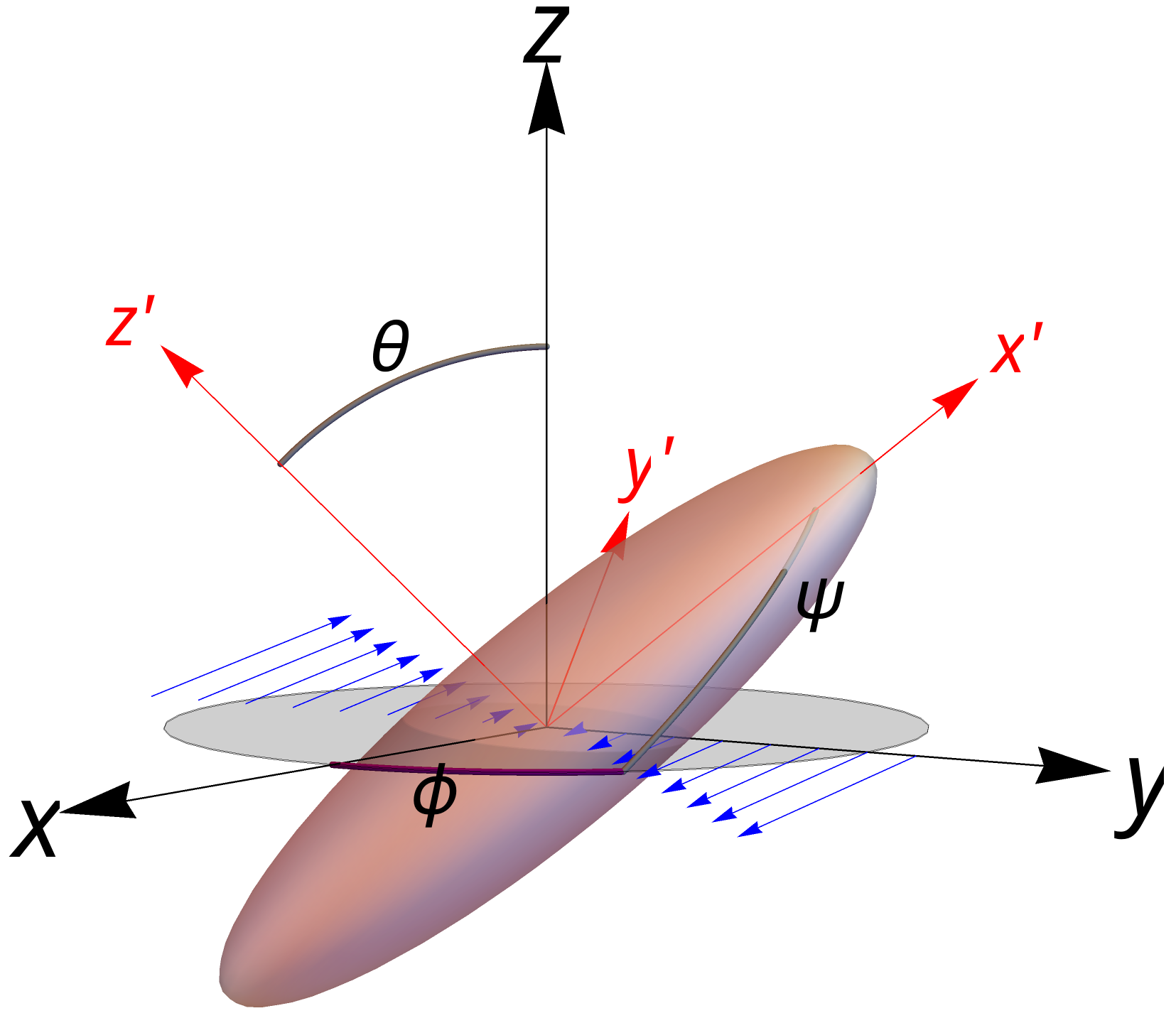}
\caption{A schematic view of the space-fixed and body-fixed coordinate systems for a triaxial ellipsoid in simple shear flow.}
\label{ellipsoid_in_shear_flow}
\end{figure}

\section{Governing Equations}\label{gove}

The equations governing the dynamics of an inertial ellipsoid in a simple shear flow can be obtained by equating the net torque on the ellipsoid to the rate of change of angular momentum; the flow, gradient and vorticity directions of the simple shear define the space-fixed coordinates $(x,y,z)$. In the general case, the torque is a combination of a viscous torque \cite{jeffery1922,kim2013} with additional contributions arising from fluid inertia \cite{dabade2016,marath2017,marath2018} and gravity \cite{dabade2015}. As already stated in the introduction, we neglect both these effects; fluid inertia is sub-dominant in the regime examined ($\rho_p/\rho_f \gg 1$), while gravity is neglected for simplicity. 
With these assumptions, the non-dimensional equation governing ellipsoid rotation may be written as:
\begin{equation}\label{ultima}
St \frac{d(\bar{\bar{I}}\cdot \vec{\Omega})}{dt}=-8\pi\bar{\bar{X}}\cdot(\vec{\Omega}-\vec{\Omega}_J).
\end{equation}

In (\ref{ultima}), the moment of inertia tensor ($\bar{\bar{I}}$) relates the angular velocity ($\vec{\Omega}$) to the angular momentum\,($\vec{L}= \bar{\bar{I}} \cdot \vec{\Omega})$, and the rotational resistance tensor ($\bar{\bar{X}}$) relates the viscous torque\,($\vec{T}_{viscous}$) to the deviation of the particle angular velocity from the Jeffery angular velocity ($\vec{\Omega}_J$), the latter being attained in the inertialess limit (defined and discussed in section \ref{jeff} below). Both $\bar{\bar{I}}$ and $\bar{\bar{X}}$ are diagonal in the body-fixed coordinate frame $(x',y',z')$, with axes aligned with the principal axes of the ellipsoid in order of increasing length\,(see figure \ref{ellipsoid_in_shear_flow}). Their components along the $x'$-axis are $I_{1}={(4 \pi/{15}) a b c (b^2+c^2)/c^5}$ and 
$X_{1}=(2/3)a b c (b^2+c^2)/(c^3(b^2\beta_2+c^2 \beta_3))$, respectively, with the other two components obtained via a cyclic permutation of the semi-axes labels and indices; here, 
$\beta_1=\int^{\infty}_0(a b c)/([a^2+t]\Delta)\text{d}t$, with $\beta_2$ and $\beta_3$ in the expression for $X_1$ being obtained by replacing $a$ with $b$ and $c$, respectively, and with all three $\beta_i$'s 
being expressible in terms of incomplete elliptic integrals\cite{kim2013}.

The space-fixed and body-fixed unit vector triads are related as $(\hat{x}',\hat{y}',\hat{z}') = t_{ij}(\hat{x},\hat{y},\hat{z})^T$, where the rotation matrix
\begin{align}\label{tiju}
	t_{ij}&=
	\begin{pmatrix}
		\cos\psi\cos\phi-\sin\phi\sin\psi\cos\theta&\sin\phi\cos\psi+\sin\psi\cos\phi\cos\theta&\sin\theta\sin\psi\\
		-\sin\psi\cos\phi-\sin\phi\cos\psi\cos\theta&-\sin\phi\sin\psi+\cos\psi\cos\phi\cos\theta&\sin\theta\cos\psi\\
		\sin\theta\sin\phi&-\sin\theta\cos\phi&\cos\theta\\
	\end{pmatrix}.
\end{align}
Here, the ellipsoid orientation is specified by the Euler angles $(\theta,\phi,\psi)$ in accordance with the usual convention where $\theta$ is the angle between $z'$ and $z$, $\phi$ is the angle between the line of nodes and the $x$ axis, and $\psi$ is the angle between the $x'$ axis and the line of nodes\cite{goldstein2002}. The components of the angular velocity vector in body-fixed coordinates, $\vec{\Omega}=\omega_1 \hat{x'}+\omega_2 \hat{y'}+\omega_3 \hat{z'}$, are related to the rates of change of the Euler angles \cite{landau1976}. With $\dot{\theta}$ being the rate of nutation, $\dot{\phi}$ the rate of precession and $\dot{\psi}$ the rate of spin, one has:
\begin{align}
\omega_1 &=(\dot{\phi}\sin{\theta}\sin{\psi}+\dot{\theta}\cos{\psi}),\label{omg1def}\\
\omega_2 &= (\dot{\phi}\sin{\theta}\cos{\psi}-\dot{\theta}\sin{\psi}),\\
\omega_3 &=(\dot{\phi}\cos{\theta}+\dot{\psi}).\label{omg3def}
\end{align}

Using (\ref{omg1def}-\ref{omg3def}) in (\ref{ultima}), one obtains the following system of second-order differential equations that govern the orientation dynamics of a triaxial ellipsoid: 
\begin{align}
&(St I_1)[\ddot{{\phi}}\sin\theta\sin\psi +\dot{\phi}\dot{\theta}\cos\theta\sin\psi+\dot{\phi}\dot{\psi}\sin\theta\cos\psi+{\ddot{\theta}}\cos\psi-\dot{\psi}\dot{\theta}\sin\psi]\label{fth}\\\nonumber&+St(I_3-I_2)[-\dot{\phi}\dot{\theta}\cos\theta\sin\psi+\dot{\phi}\dot{\psi}\sin\theta\cos\psi-\dot{\psi}\dot{\theta}\sin\psi+\dot{\phi}^2\cos\theta\sin\theta\cos\psi]\\\nonumber&=(-8\pi X_1)[\dot{\phi}\sin\theta\sin\psi+\dot{\theta}\cos\psi-\omega_{1,J}],\nonumber\\
&(St I_2)[\ddot{{\phi}}\sin\theta\cos\psi +\dot{\phi}\dot{\theta}\cos\theta\cos\psi-\dot{\phi}\dot{\psi}\sin\theta\sin\psi-{\ddot{\theta}}\sin\psi-\dot{\psi}\dot{\theta}\cos\psi] \label{fph}\\&+St(I_1-I_3)[\dot{\phi}\dot{\theta}\cos\theta\cos\psi+\dot{\phi}\dot{\psi}\sin\theta\sin\psi+\dot{\psi}\dot{\theta}\cos\psi+\dot{\phi}^2\cos\theta\sin\theta\sin\psi]\nonumber\\&=(-8\pi X_2)[\dot{\phi}\sin\theta\cos\psi-\dot{\theta}\sin\psi-\omega_{2,J}],\nonumber\\
&(St I_3)[\ddot{{\psi}}+{\ddot{\phi}}\cos\theta-\dot{\phi}\dot{\theta}\sin\theta],\label{fps}+St(I_2-I_1)[(\dot{\phi}^2\sin^2\theta-\dot{\theta}^2)\cos\psi\sin\psi+\dot{\phi}\dot{\theta}\sin\theta\cos2\psi] \\  &=(-8\pi X_3)[\dot{\psi}+\dot{\phi}\cos\theta-\omega_{3,J}],
\nonumber\end{align}
where the $\omega_{i,J}$'s, the Jeffery angular velocity components along the individual ellipsoid axes, are given in section \ref{asympcase}.

For an oblate spheroid, obtained in the limit $a =b$, there is a degeneracy involved in the choice of a principal axis in the plane perpendicular to $z'$\,(the symmetry axis). We take the $x'$ axis to be along the line of nodes for purposes of convenience, the direction of the $y'$ axis being fixed as a result, and this being equivalent to setting $\psi = 0$ in (\ref{tiju}). 
Note that, although $\psi$ is a degenerate degree of freedom for a spheroid, the rate of spin $\dot{\psi}$ is still relevant via the gyroscopic restoring torque. The resulting simpler expressions for the angular velocity components are:
\begin{align}
\omega_1 &=\dot{\theta},\label{omg1def_spheroid}\\
\omega_2 &= \dot{\phi}\sin{\theta},\label{omg2def_spheroid}\\
\omega_3 &=\dot{\phi}\cos{\theta}+\dot{\psi},\label{omg3def_spheroid}
\end{align}
while the simpler set of equations governing oblate spheroid rotation, obtained using $\psi= 0$ in (\ref{fth}-\ref{fps}), is:
\begin{align}
\frac{4\pi St}{15}[(\frac{1}{\kappa^2}-\kappa^2)\dot{\phi}^2\sin\theta \cos\theta +(\frac{1}{\kappa^2}+1)\ddot{\theta}+2 \dot{\psi}\dot{\phi}\sin{\theta}]&=-8\pi Y_c(\dot{\theta}-\omega_{1,J}),\label{fsptho}
\\\frac{4\pi St}{15}[2\dot{\phi}\dot{\theta}\cos\theta+(\frac{1}{\kappa^2}+1)\ddot{\phi}\sin{\theta}-2\dot{\psi}\dot{\theta}]&=-8 \pi Y_c(\dot{\phi}\sin{\theta}-\omega_{2,J}),\label{fsppho}
\\\frac{8\pi St}{15}\frac{2}{\kappa^2}(\ddot{\phi}\cos{\theta}+\ddot{\psi}-\dot{\theta}\dot{\phi}\sin{\theta})&=-8 \pi X_c (\dot{\psi}+\dot{\phi}\cos{\theta}-\omega_{3,J}). \label{fpsso}
\end{align}
The corresponding equations for a prolate spheroid may be obtained from (\ref{fsptho}-\ref{fpsso}) by replacing $St$ by $St/\kappa^2$. 
The resistance coefficients corresponding to axial\,($X_c$) and equatorial\,($Y_c$) rotations, that appear in (\ref{fsptho}-\ref{fpsso}), may now be written in terms of elementary functions. Those for an oblate spheroid are
$X_c=X_3=(2/3)(1-\kappa^2)^{3/2}/\kappa^3[\cos^{-1}{\kappa}-\kappa\sqrt{1-\kappa^2}]$ and $Y_c=X_1=X_2=(2/3)(1-\kappa^2)^{3/2}(1+\kappa^2)/\kappa^3[\kappa\sqrt{1-\kappa^2}-(-1+2\kappa^2)\cos^{-1}{\kappa}]$; those for a prolate spheroid are $X_c=X_3=(2/3)(1-\kappa^{-2})^{3/2}/[\kappa\sqrt{\kappa^2-1}-\cosh^{-1}{\kappa}]$ and $Y_c=X_1=X_2=(2/3)(1-\kappa^{-2})^{3/2}(\kappa^2+1)/[-\kappa\sqrt{\kappa^2-1}+(2\kappa^2-1)\cosh^{-1}{\kappa}]$\cite{kim2013}.

\subsection{Validation of Numerical Method}

\begin{figure}
\subfloat[\label{delTvslogSt} $\Delta T$ vs $St$]{\includegraphics[scale=0.5,keepaspectratio=true,trim={0 0 0 1cm},clip]{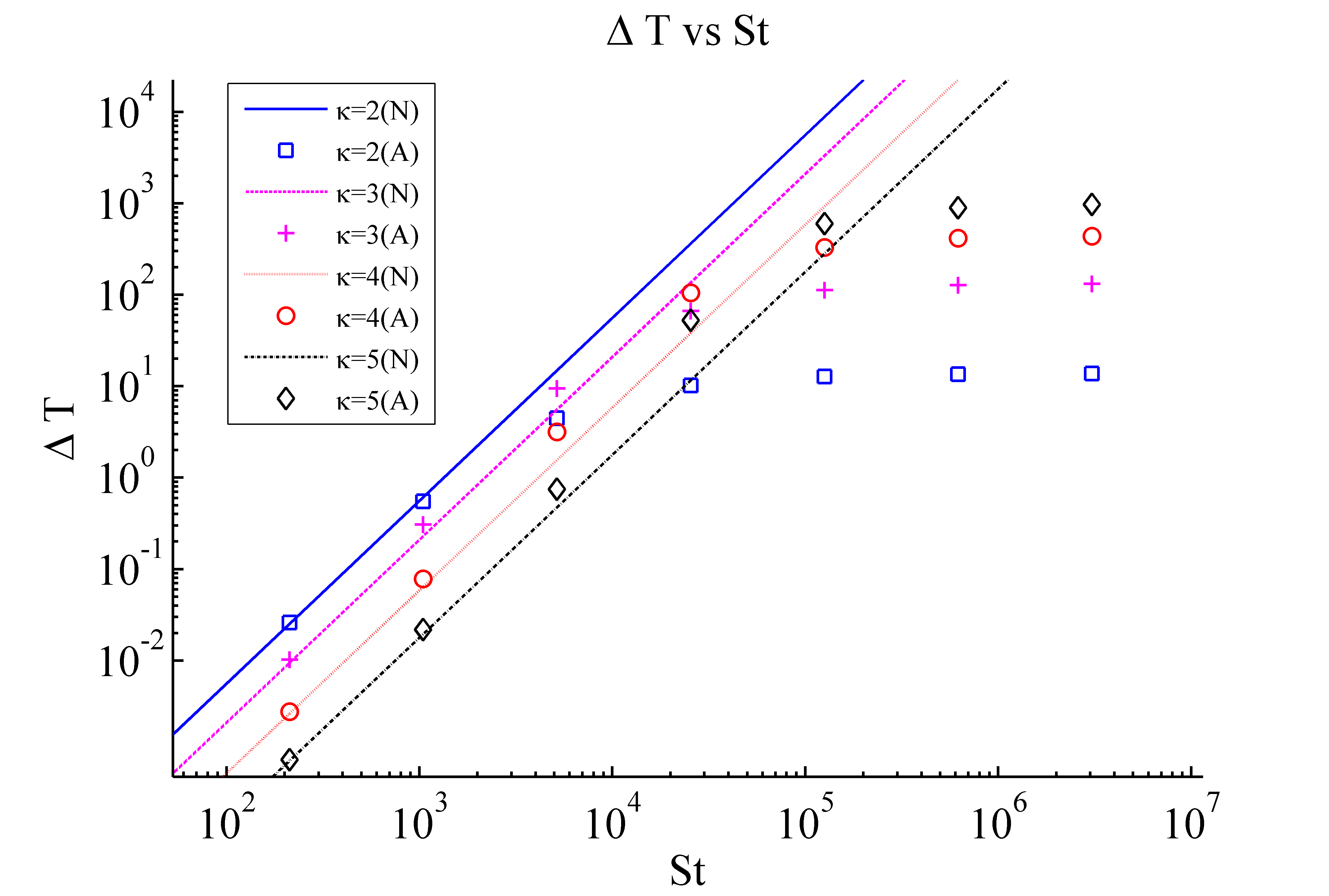}}
\subfloat[\label{TvslogSt} $T$ vs $St$] {\includegraphics[scale=0.5,trim={0 0 0 1cm},clip]{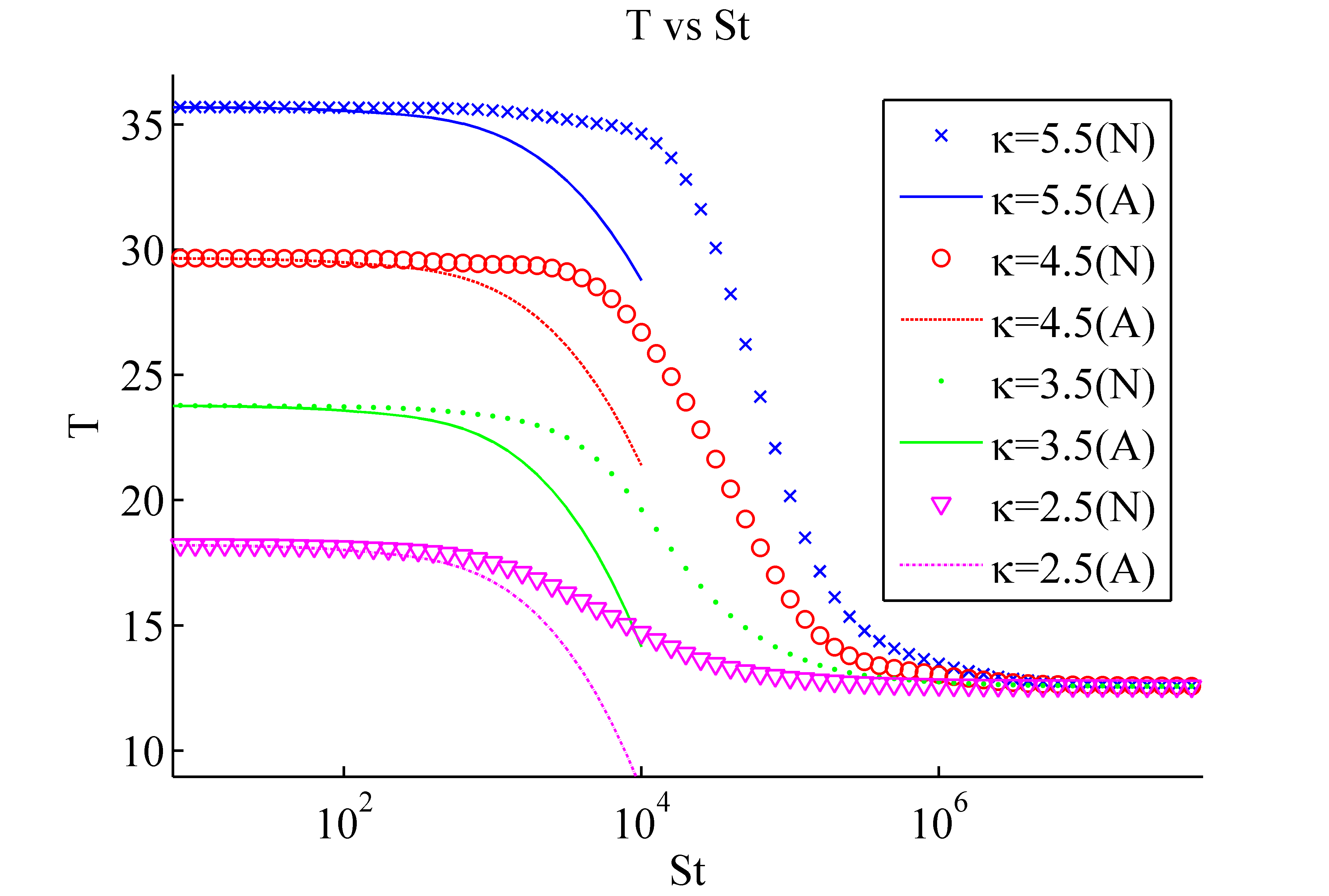}
}\\
\caption{Figure \ref{delTvslogSt} shows how the deviation of the period of rotation from the Jeffery time period varies with $St$, for different spheroidal (prolate) aspect ratios, and its comparison with the analytical prediction in (\ref{vivtp})\cite{dabade2016} . In Figure \ref{TvslogSt} we plot the transition from the $\kappa$-dependent Jeffery time period ($St \ll 1$) to the $\kappa$-independent Euler limit ($St \gg 1$) \cite{lundell2011}.}
\end{figure}

We numerically integrate the system of equations for a prolate spheroid, as has also been done previously\cite{lundell2011,lundell2010,challabotla2015}, and validate our integration routine by comparison both with earlier theoretical\,(for small $St$; see \cite{dabade2016}) and numerical\,(for large $St$; see \cite{lundell2010}) results. Figure \ref{delTvslogSt} compares the numerically obtained increase in the time period $\Delta T$\,(relative to the Jeffery value), of a tumbling prolate spheroid, to the O($St^2$) increase predicted by Dabade et al. \cite{dabade2016}, given by:
\begin{equation}\label{vivtp}
\Delta T=-St^2 \frac{\pi[(1-2\kappa^2)\cosh^{-1}{\kappa}+\kappa \sqrt{\kappa^2-1}]^2}{100\kappa^2(\kappa^4-1)(\kappa+1)^2}.
\end{equation}

The numerical results in Figure \ref{delTvslogSt} deviate from the analytical asymptotes for $St$ of order unity. This deviation is postponed to larger $St$ for increasingly slender prolate spheroids since, as indicated above, the measure of particle inertia for such spheroids is $St/\kappa^2$ rather than $St$. 
Figure \ref{TvslogSt} shows the transition of the rotation time period, again for a tumbling prolate spheroid, from the Jeffery value ($2\pi(\kappa+\kappa^{-1})$; see section \ref{asympcase} below) to an aspect-ratio-independent limit ($4\pi$
) with increasing $St$, in agreement with Lundell et.al \cite{lundell2010}.

\section{The limiting cases: the Jeffery and the Euler tops}\label{asympcase}

For $St=0$ or $\infty$, equations (\ref{fth}-\ref{fps}) for an ellipsoid, or equations (\ref{fsptho}-\ref{fpsso}) for a spheroid, reduce to classical solutions well known in literature. These are discussed in what follows.

\subsection*{An inertialess ellipsoid in simple shear flow\,($ St = Re=0$): the Jeffery top}\label{jeff}

In the limit $St \rightarrow 0$, (\ref{ultima}) reduces to a viscous-torque-free constraint at each instant in time, which gives $\vec{\Omega}=\vec{\Omega}_J$, with the body-fixed components of $\vec{\Omega}_J$ 
 given by\cite{jeffery1922}:
\begin{align}
\omega_{1,J}=\frac{b^2(t_{22}t_{31})-c^2(t_{32}t_{21})}{(b^2+c^2)}\label{jw1},\\
\omega_{2,J}=\frac{c^2(t_{32}t_{11})-a^2(t_{12}t_{31})}{(c^2+a^2)}\label{jw2},\\
\omega_{3,J}=\frac{a^2(t_{12}t_{21})-b^2(t_{22}t_{11})}{(a^2+b^2)}\label{jw3}.
\end{align}
The $t_{ij}$'s in (\ref{jw1}-\ref{jw3}), as functions of the Euler angles, have been defined in section \ref{gove}. The orientation dynamics may then be obtained by expressing the $\omega_{i,J}$'s in terms of the rates of change of the Euler angles using (\ref{omg1def}-\ref{omg3def}), and integrating the resulting system of  ODEs. 

Ellipsoid trajectories in simple shear flow were characterized numerically by Hinch and Leal \cite{leal1979}, for a wide range of aspect ratios, via $\theta-\psi$ Poincar\'e sections ($0\leq \theta \leq \pi/2, 0\leq \psi \leq \pi/2$) corresponding to $\phi=n \pi$ - the choice of a $\theta-\psi$ plane derives from the secularly increasing nature of $\phi$; on account of symmetry relations, the aforementioned intervals in the  $\theta-\psi$ plane are subsets of the actual intervals of variation\,($0\leq \theta \leq \pi, 0\leq \psi \leq 2\pi$). Invariant sets on this plane, corresponding to principal-axis-aligned rotations, were identified: the edge $\theta=0$ and the fixed points $(\theta,\psi)\equiv(\pi/2,0)$ and $(\theta,\psi)\equiv(\pi/2,\pi/2)$ correspond, respectively, to rotation of the ellipsoid with its longest, intermediate and shortest axis aligned with the ambient vorticity. In these configurations, the ellipsoid rotation is independent of the length of the vorticity-aligned axis\cite{prabhu_2022}, and therefore, identical to a tumbling spheroid with an aspect ratio equal to the ratio of the ellipsoid axes in the flow-gradient plane. 
For the order unity aspect ratios studied by the above authors, trajectories of non-aligned ellipsoids appeared on the Poincar\'e sections as densely filled closed curves, corresponding to quasiperiodic behavior with two incommensurate frequencies. When the frequencies are well separated, as is true when the ellipsoid cross-section is nearly circular, the larger frequency corresponds to rotation along nearly closed Jeffery-like orbits and the smaller one to a slower drifting motion between a pair of limiting orbits. For certain subsets of trajectories, the latter lie on either side of the flow-gradient plane, implying that a triaxial ellipsoid crosses the flow-gradient plane during the course of its rotation; a behavior prohibited by symmetry for spheroids. 
\begin{figure}
\subfloat[\label{slender_prolate} Slender prolate ellipsoid $(b/a=1/10,c/a=1/20)$]{\includegraphics[scale=0.45]{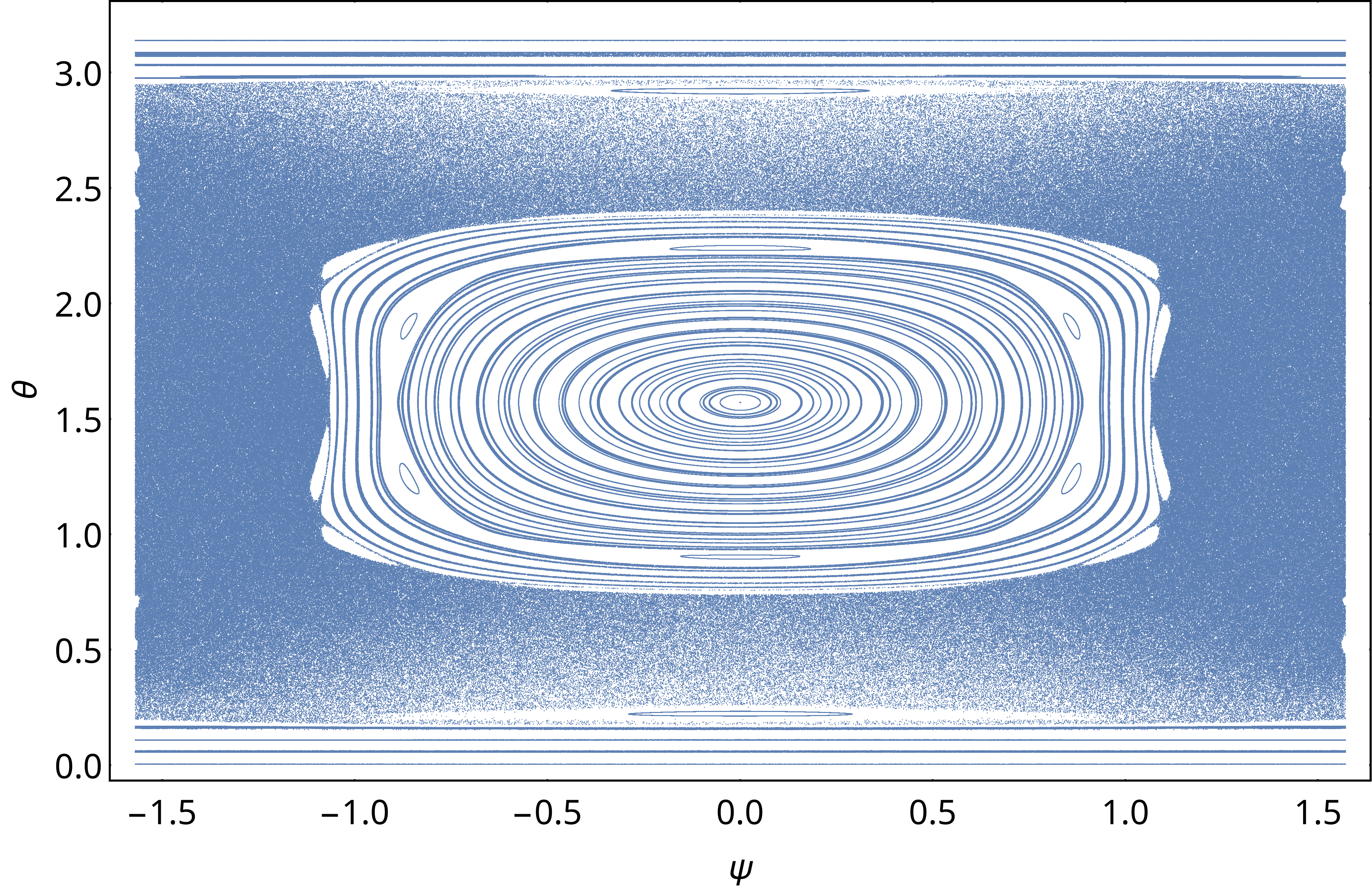}}
\subfloat[\label{fat_prolate} Fat prolate ellipsoid $(b/a=1.2/3,c/a=1/3)$]{\includegraphics[scale=0.45]{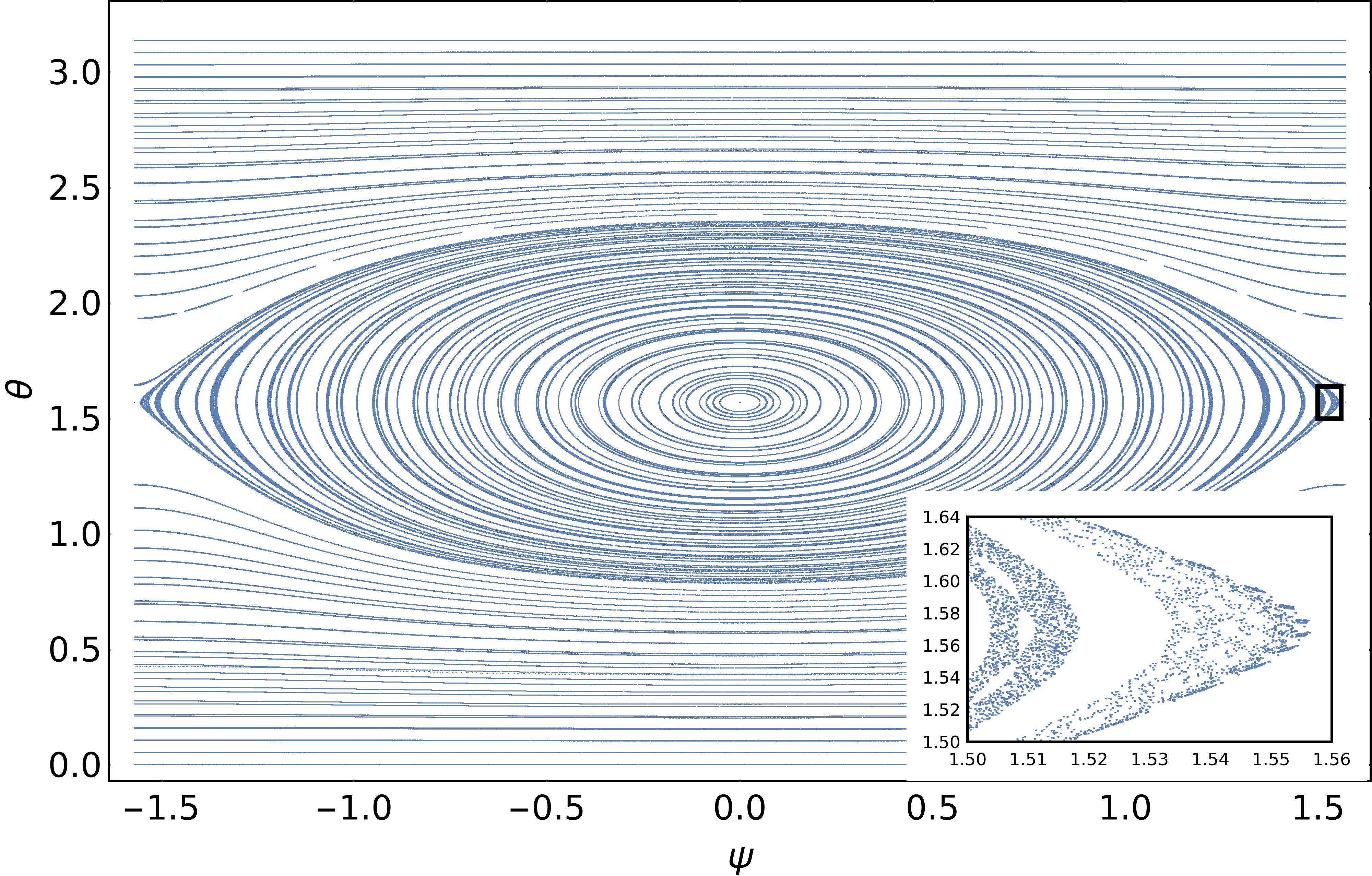}}

\subfloat[\label{flat_oblate} Thin oblate ellipsoid $(b/a=1/2,c/a=1/20)$]{\includegraphics[scale=0.45]{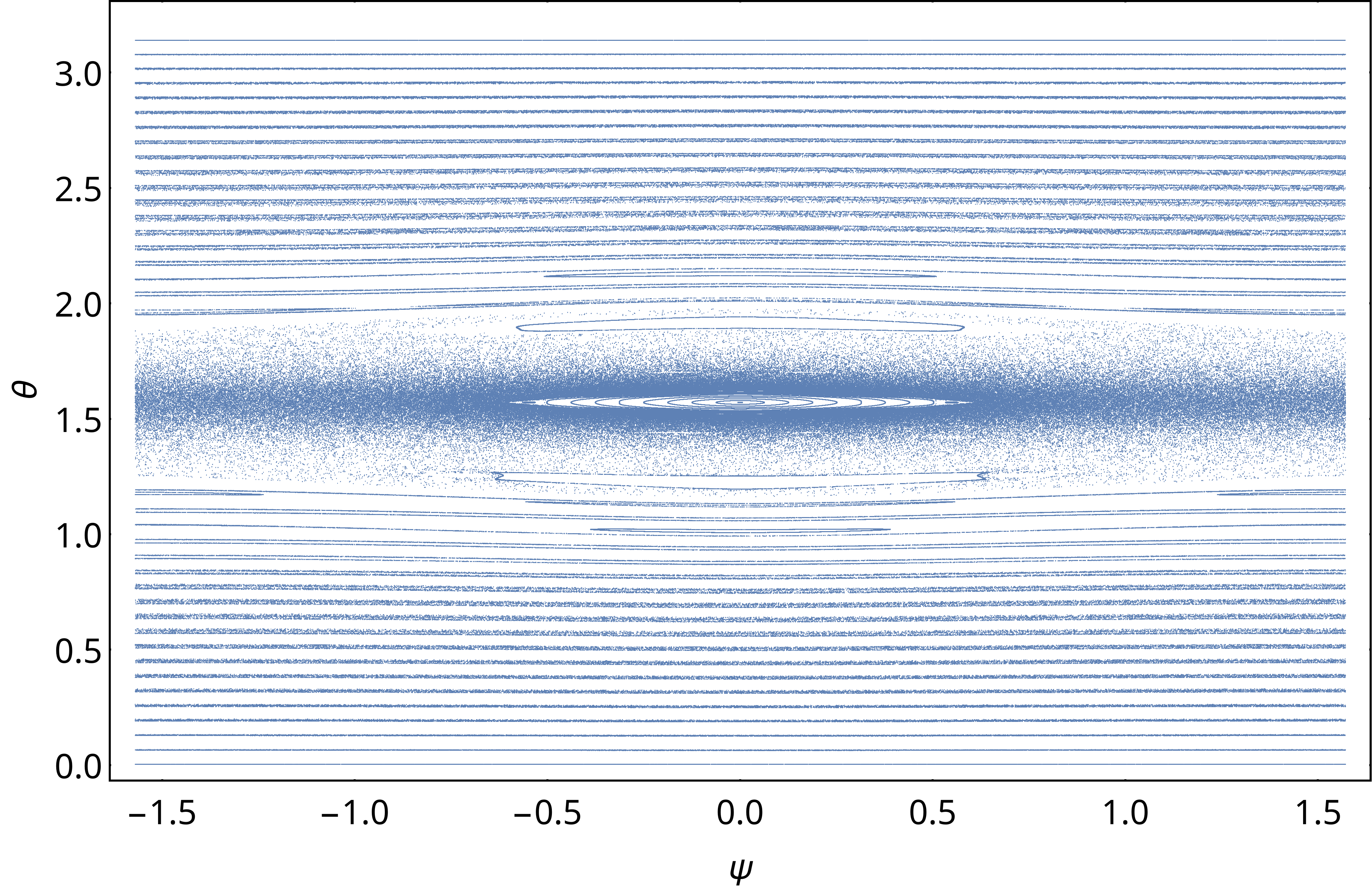}}
\subfloat[\label{fat_oblate} Fat oblate ellipsoid $(b/a=2/2.1,c/a=1/2.1)$]{\includegraphics[scale=0.45]{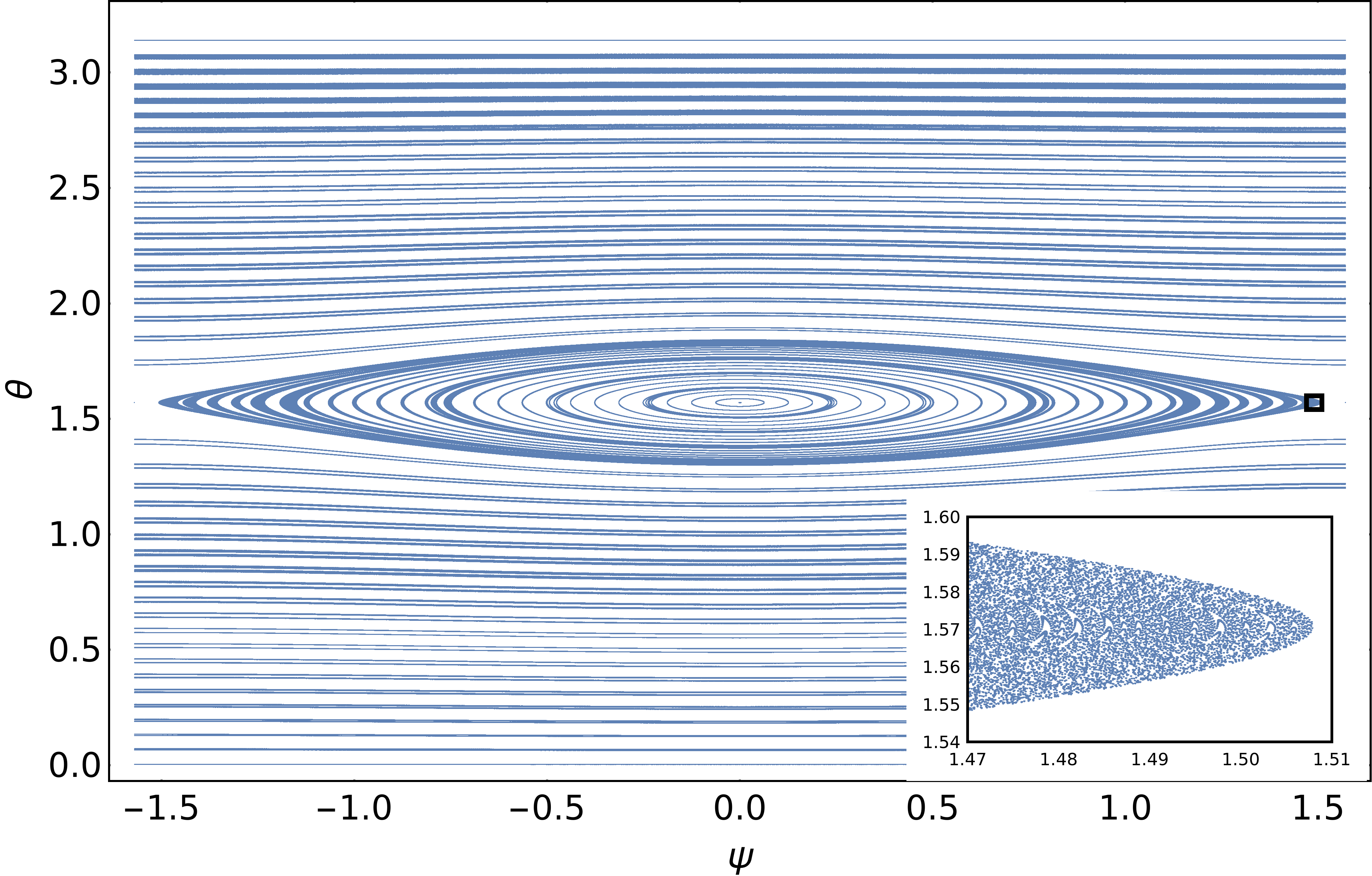}}
\caption{The $\theta-\psi$ Poincare sections for inertialess ellipsoids with different aspect ratios. Chaotic trajectories are dominant for both slender and flat ellipsoid; for ellipsoids with order unity aspect ratios, the magnified views highlight chaotic dynamics in the vicinity of the hyperbolic points.}
\label{poincare_sections}
\end{figure}

A later detailed examination of slender\,(large $c/a$ and moderate $b/a\neq 1$) triaxial ellipsoids by Yarin et al. \cite{yarin1997} revealed that the dynamics is not entirely quasiperiodic, and that a large subset of the trajectories exhibits chaotic dynamics. The chaos may be understood as arising from the absence of a separation between the (long) Jeffery period of such ellipsoids and the time scale of crossings of the flow-gradient plane (in course of the aforementioned drifting motion) - such crossings occur in an unpredictable manner during the long period of near-alignment with the flow-vorticity plane. In fact, chaotic and quasiperiodic trajectories coexist for any ellipsoid, the proportion of these two classes of trajectories being a function of the axis ratios. This is illustrated via the Poincare sections in Figs.\ref{poincare_sections}a-d which, for purposes of clarity, are shown over the larger interval $\theta \in [0,\pi]$. From these figures, chaotic trajectories for ellipsoids with order unity aspect ratios are seen to be restricted to thin stochastic layers along separatrices that connect the saddle points at $(\theta,\psi) \equiv (\frac{\pi}{2},\pm \frac{\pi}{2}$), both of these corresponding to the shortest-axis-aligned rotation\,(see magnified views in Figs.\ref{poincare_sections}b and d).
As the ellipsoid becomes increasingly slender\,($b/a, c/a \rightarrow 0$ in Fig.\ref{poincare_sections}a) or flat\,($c/a, c/a \rightarrow 0$ in Fig.\ref{poincare_sections}c), the stochastic layers become thicker, with a concomitant shrinking of the region corresponding to quasiperiodic dynamics.

The dynamics is much simpler for spheroids owing to a de-coupling of the spin component. Equations (\ref{jw1}-\ref{jw3}) reduce to the following simpler forms:
\begin{align}
\omega_{1,J}=&-\frac{(\kappa^2-1)\sin2\theta \sin2\phi}{4(\kappa^2+1)}\label{spheroid_wj1},\\
\omega_{2,J}=&-\frac{(\sin^2{\phi}+\kappa^2\cos^2{\phi})}{\kappa^2+1}\sin\theta\label{spheroid_wj2},\\
\omega_{3,J}=&-\frac{\cos\theta}{2}\label{spheroid_wj3},
\end{align}
where (\ref{spheroid_wj3}) implies that the spheroid merely spins at a rate commensurate with the ambient vorticity projected along its axis. Solving the remaining two equations lead to the following rates of change of the Euler angles:
\begin{align}\label{jwphi}
\dot{\phi} &= \frac{(\sin^2{\phi}+\kappa^2\cos^2{\phi})}{\kappa^2+1},
\\\label{jwtheta}
\dot{\theta}&= \frac{(\kappa^2-1)\sin{2\phi}\sin{2\theta}}{4(\kappa^2+1)}.
\\
\intertext{Equations (\ref{jwphi}) and (\ref{jwtheta}) may be integrated in closed form to obtain the so-called Jeffery orbits. The generic orbit is a spherical ellipse traversed by the spheroid orientation vector, with a time period of $2\pi(\kappa+\kappa^{-1})$, and is defined by:}
\label{phitraj}
\tan{\phi} &= -\kappa\tan{\frac{t}{\kappa+1/\kappa}},
\\\label{thetraj}
\tan{\theta}&= \frac{C\kappa}{(\sin^2{\phi}+\kappa^2 \cos^2{\phi})^2}.
\end{align}

Jeffery orbits are thus parameterized by an orbit constant $C$ that ranges from zero to infinity. The limiting orbits $C=0$ and $C=\infty$ correspond to the spinning\,(or log-rolling) and tumbling modes, respectively. When $C=0$, the spheroid spins about its symmetry axis aligned with the ambient vorticity, and at a rate equalling half the vorticity. For $C=\infty$, it tumbles with its axis in the flow-gradient plane at a rate that is a function of its orientation. The angular velocity of a tumbling prolate\,(oblate) spheroid is minimum for the flow\,(gradient)-aligned orientation; as implied by the discussion of the near-aligned phase in the ellipsoid dynamics above, the angular velocity minimum decreases to zero in the limit of slender prolate\,($\kappa \rightarrow \infty$) and flat oblate\,($\kappa \rightarrow 0$) spheroids.

\subsection*{Torque-Free rotation of an ellipsoid ($St=\infty$): The Euler top}\label{symtop}
In the limit $St\rightarrow \infty$, the viscous torque term in (\ref{ultima}) becomes negligible, and ellipsoid rotation is governed by the Euler equations of classical dynamics, given by \cite{landau1976}:

\begin{equation}\label{eulg}
I_i\dot{\omega}_i+\epsilon_{ijk}I_k\omega_j\omega_k=0
\end{equation} 

  
The angular momentum, $\vec{L} = I_1\omega_1\hat{x}'+I_2\omega_2\hat{y}'+I_3\omega_3\hat{z}'$, and rotational kinetic energy, $E=I_i\omega_i^2/2$, are now constants of motion. The general solution of (\ref{eulg}) may be given in terms of the Jacobi elliptic functions \cite{gradshteyn2007,landau1976} as:    
\begin{align}\label{eq:omg1}
\omega_1&=\sqrt{\frac{2EI_3-L^2}{I_1(I_3-I_1)}}cn(\tau,k),\\
\label{eq:omg2}
\omega_2&=\sqrt{\frac{2EI_3-L^2}{I_2(I_3-I_2)}}sn(\tau,k),\\
\label{eq:omg3}
\omega_3&=\sqrt{\frac{L^2-2EI_1}{I_3(I_3-I_1)}}dn(\tau,k),
\end{align}
where $\tau=t\sqrt{\frac{(I_3-I_2)(L^2-2EI_1)}{I_1I_2I_3}}$ and $k=[\frac{(I_2-I_1)(2EI_3-L^2)}{(I_3-I_2)(L^2-2EI_1)}]^{\frac{1}{2}}$\,($0\leq k^2 \leq 1$) are the argument and modulus of the elliptic functions, respectively.
All three angular velocity components are periodic, with $\omega_1$ and $\omega_2$ having a zero mean with period $T=4K(k)$, and $\omega_3$ having a non-zero mean with period $T/2$; here, $K(k)=\int^{\pi/2}_0(1-k^2\sin^2{\chi})^{-1/2}\text{d}\chi$ is the complete elliptic integral of the first kind \cite{gradshteyn2007}. The ellipsoid trajectory may be obtained in an angular-momentum-aligned coordinate system with its $z$-axis along $\vec{L}$, and with the transformation matrix, that relates this coordinate system to the space-fixed one, being given by: 
\begin{align}\label{fodij}
	\alpha_{ij}=
	\begin{pmatrix}
		\frac{L_{y}}{L_{xy}}&\frac{L_{x} L_{z}}{L L_{xy}}&\frac{L_{x}}{L}\\
		-\frac{L_{x}}{L_{xy}}&\frac{L_{y} L_{z}}{L L_{xy}}&\frac{L_{y}}{L}\\
		0&-\frac{L_{xy}}{L}&\frac{L_{z}}{L}\\
	\end{pmatrix}.
\end{align}
 Note that $\vec{L}$ is the only physical vector in the angular-momentum-aligned coordinate system. The arbitrariness involved in the choice of the other two axes does not, however, affect the dynamics for large but finite $St$, analyzed in sections \ref{lsds} and \ref{lsde}. The angular-momentum-aligned and body-fixed unit vector triads may be related as $(\hat{x}',\hat{y}',\hat{z}') = t_{ij,LO}(\hat{x}_{LO},\hat{y}_{LO},\hat{z}_{LO})^T$, where $t_{ij,LO}$ is given by an expression analogous to (\ref{tiju}), but in terms of a new set of Euler angles ($\theta_{LO},\phi_{LO},\psi_{LO}$); one has the relation $t_{ij}^T=\alpha_{ij}t_{ij,LO}^T$ between the transformation matrices.


Resolving $\vec{L}$ along the principal axes, one obtains: 
\begin{align}\label{omg1d}
L\sin{\theta_{LO}}\sin{\psi_{LO}}&=I_1\omega_1, \\
L\sin{\theta_{LO}}\cos{\psi_{LO}}&=I_2\omega_2, \\ \label{omg3d}
L\cos{\theta_{LO}}&=I_3\omega_3. 
\end{align}

Equations (\ref{omg1d}-\ref{omg3d}) may now be used along with the analogs of (\ref{omg1def}-\ref{omg3def}), with ($\theta,\phi,\psi$) replaced by ($\theta_{LO},\phi_{LO},\psi_{LO}$), to obtain the spin and nutation angles as the following functions of time:
\begin{align}\label{eq:nut}
\cos\theta_{LO}&=\sqrt{\frac{I_3(L^2-2EI_1)}{L^2(I_3-I_1)}}dn(\tau,k),\\
\label{eq:rota}
\tan\psi_{LO}&=\sqrt{\frac{I_1(I_3-I_2)}{I_2(I_3-I_1)}}\frac{cn{(\tau,k)}}{sn{(\tau,k)}},
\end{align}
which show that $\theta_{LO}$ and $\psi_{LO}$ (mod $2\pi$) are periodic functions with the same period $T/2$. The precession angle may be written down as the sum of a periodic ($\phi_{LO,1}$) and a secular ($\phi_{LO,2}$) component:
\begin{equation}
\phi_{LO}=\phi_{LO,1}+\phi_{LO,2},\label{eq:prec}
\end{equation}
where
\begin{align}\label{eq:prec1}
\exp(2i\phi_{LO,1})&=\frac{\vartheta_{4}(2t/T-i\zeta,q)}{\vartheta_{4}(2t/T+i\zeta,q)},\\\label{eq:prec2}
\phi_{LO,2}&=\frac{Lt}{I_1}-\frac{2it\vartheta_{4}'(i\zeta,q)}{T\vartheta_{4}(i\zeta,q)}.
\end{align}
In (\ref{eq:prec1}-\ref{eq:prec2}), $\vartheta_4$ is the Jacobi Theta function of the fourth kind \cite{maf1953}, $\vartheta_4'$ its derivative, $q=\exp{(-\pi K(\sqrt{1-k^2})/K(k))}$ the elliptic nome and $\zeta$ is defined by the relation 
$sn(2i\zeta K(k),k)=i\sqrt{\frac{I_3(L^2-2EI_1)}{I_1(2EI_3-L^2)}}$. Note that $a>b>c$ implies $I_3>I_2>I_1$, and as a result $L$\,(the angular momentum magnitude) and $E$ must satisfy $2EI_1 \leq L^2 \leq 2EI_3$. The expressions (\ref{eq:omg1}-\ref{eq:prec2}) above are only valid when $2EI_2 \leq L^2 \leq 2EI_3$, a constraint necessary for the modulus to lie in the unit interval. The expressions for the complementary case, $2EI_1 \leq L^2 < 2EI_2$, are obtained by interchanging the suffixes $1$ and $3$.

\begin{figure}
\subfloat[\label{euler_top_nutation} Trajectory topology of an Euler top; $(a,b,c)\equiv(1,2,2.3)$ 
$(\theta_0,\phi_0,\psi_0,\dot{\theta_0},\dot{\phi_0},\dot{\psi_0}) \equiv(1.05,0,0,2.5,15,4)$ ]
{\includegraphics[scale=0.25]{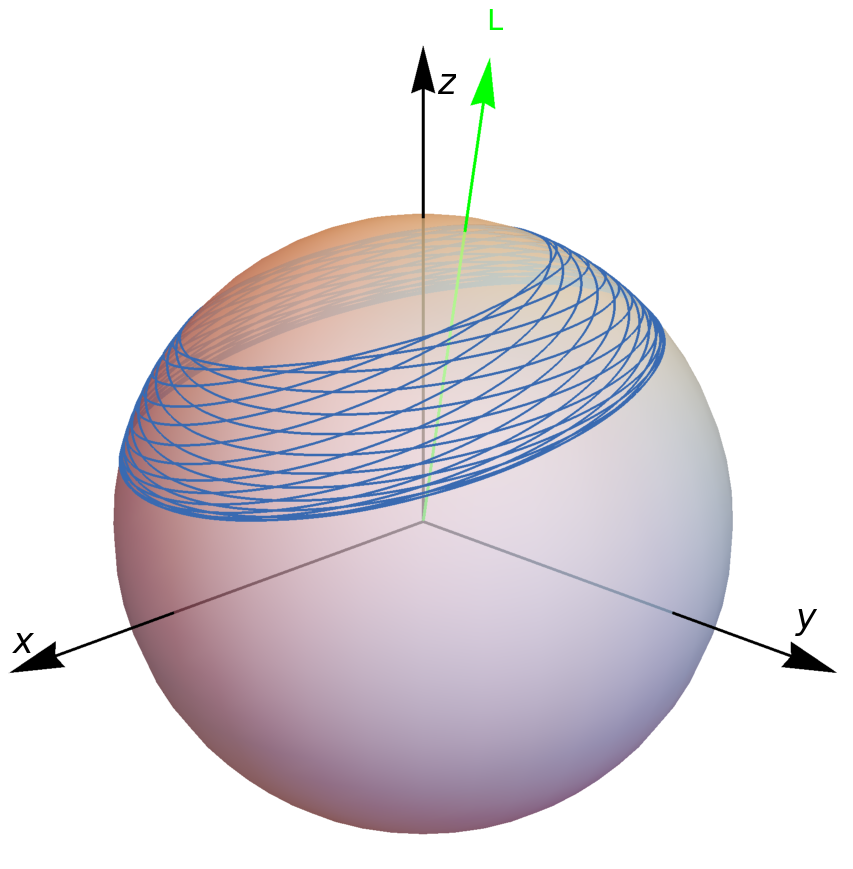}}
 \hspace{0.07\linewidth}
\subfloat[\label{symmetric_top_nutation} Trajectory topology of a symmetric Euler top; $(\kappa=2.3)$ $(\theta_0,\phi_0,\psi_0,\dot{\theta_0},\dot{\phi_0},\dot{\psi_0}) \equiv(1.05,0,0,2.5,15,4)$]{\includegraphics[scale=0.25]{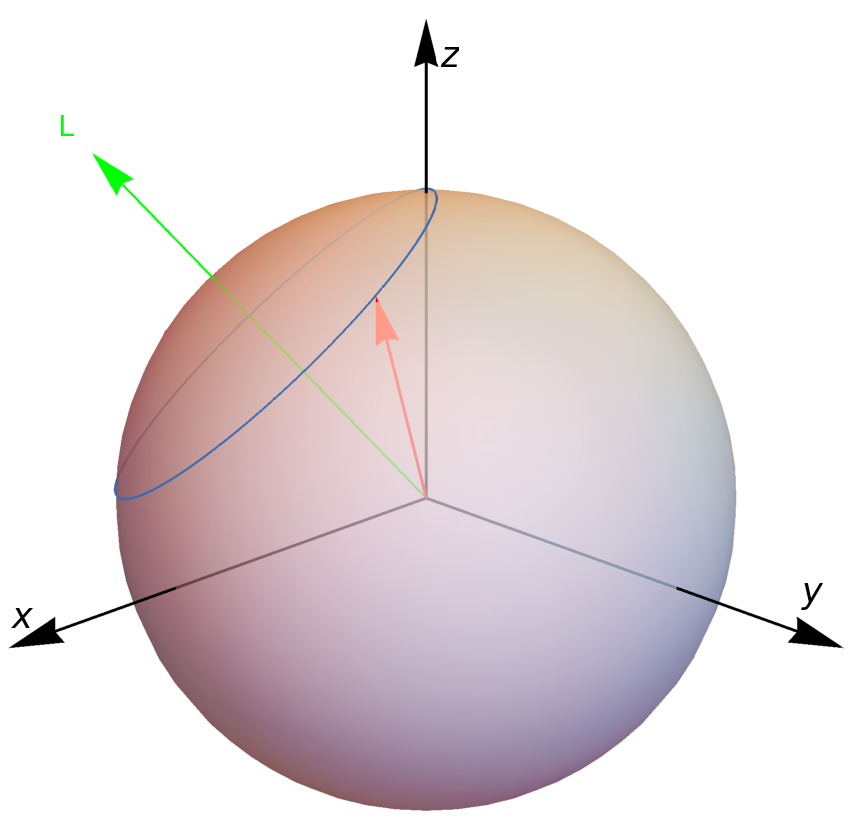}}
\caption{Unit-sphere trajectories for the general and symmetric Euler tops.}
\end{figure}

Despite the analytical complexity of the general orientation dynamics, simple special configurations exist. These are the principal-axis-aligned rotations, each of which corresponds to $L^2/2E$ equalling the relevant principal moment of inertia. It is well known that rotations about the shortest ($L^2/2E=I_3$) and longest ($L^2/2E=I_1$) axes are neutrally stable, while the intermediate-axis-aligned rotation($L^2/2E=I_2$) is a saddle point, leading to the so-called tennis racket instability\cite{ashbaugh_twisting_1991}. Qualitative features of the solution trajectories in $(I_1\omega_1, I_2\omega_2, I_3\omega_3)$ space may, in fact, be inferred without the detailed expressions above\cite{landau1976,goldstein2002}. The ellipsoid trajectory, the polhode, is the curve of intersection of the kinetic energy ellipsoid (${(I_i\omega_i)^2}/{2EI_i}=1$) and the angular momentum sphere ($(I_i\omega_i)^2/L^2=1$). Stability of the principal rotations may then be understood by perturbing the angular momentum magnitude\,(sphere radius) while keeping the energy constant. For the longest and shortest axes, the perturbed polhode is a closed curve around the original fixed point\,(the principal rotation), implying neutral stability. For the intermediate axis, there is an initial exponential-in-time departure from the fixed point, in accordance with its saddle-like character. The physical trajectory of any ellipsoid axis can be understood as a combination of precession, spin and nutation. Both the nutation angle\,(\ref{eq:nut}) and the tangent of the spin angle\,(\ref{eq:rota}) are periodic with period $T/2$, this synchronization arising from the existence of one more constant of motion\,($4$ as opposed to $3$) than that necessary for integrability, and ensuring that trajectories in the $\theta_{LO}\!-\!\psi_{LO}\,(\text{mod}\,2\pi)$ plane are discrete sequences of points rather than densely filled curves. 
The precession angle consists of a periodic component with period $T$, and a linear-in-time component that increases by $2\pi$ over $T'\,(T'=2\pi t/\phi_{LO,2})$. 
$T$ and $T'$ are incommensurate in general, and the motion quasiperiodic, so an ellipsoid axis never returns to its original position. Figure \ref{euler_top_nutation} shows the generic unit-sphere trajectory of an Euler top where the aforesaid incommensurability leads to the ellipsoid orientation vector densely filling a circular band for sufficiently long times.

For $I_1=I_2$ as is the case for an oblate spheroid, the modulus $k$ equals zero, and the elliptic functions in (\ref{eq:omg1}) and (\ref{eq:omg2}) reduce to the corresponding trigonometric ones, with $\omega_3$\,(the spheroid spin) in (\ref{eq:omg3}) reducing to a constant. For a prolate spheroid with $I_2=I_3$, the modified modulus ($I_1$ and $I_3$ switched in the expression for $k$ above) is again zero, and an analogous simplification occurs. The spheroid corresponds to the so-called symmetric Euler top which precesses at a constant rate about the space-fixed vector $\vec{L}$. The rate of spin is also a constant, as mentioned above, and there is no nutation. The trajectory of the spheroid axis, unlike the ellipsoid, is thus periodic, and along a circle in the plane normal to $\vec{L}$, as shown in figure \ref{symmetric_top_nutation}. Note, however, that the body-fixed orthonormal triad does not return to its original configuration in general, since the rates of precession and spin continue to be incommensurate\,(although this is irrelevant on account of axial symmetry). Denoting the axial and equatorial moments of inertia as $I_a$ and $I_e $($I_3=I_2=I_e$, $I_1=I_a$ for a prolate spheroid; $I_1=I_2=I_e$, $I_3=I_a$ for an oblate spheroid), the angular velocities in (\ref{eq:omg1}-\ref{eq:omg3}) reduce to $\omega_1 =\sqrt{\frac{2EI_a-L^2}{I_e(I_a-I_e)}} \cos\tau$ and $\omega_2 = \sqrt{\frac{2EI_a-L^2}{I_e(I_a-I_e)}}\sin\tau$, and $\omega_3 = \eta$, which lead to the following expressions for the Euler angles in the angular-momentum-aligned coordinate system:   
\begin{align}\label{spherloth}
\cos{\theta_{LO}}&=\frac{I_a\eta}{L},\\\label{spherlops}
\dot{\psi}_{LO}&=-\frac{(I_a-I_e)\eta}{I_e},\\\label{spherloph}
\phi_{LO}&=\frac{Lt}{I_e},\\\label{EetaL}
E&=\frac{L^2-I_a(I_a-I_e)\eta^2}{2I_e},
\end{align}
where $\eta$ determines the angle of precession for a fixed angular momentum; note that $ -L/I_a\leq\eta\leq L/I_a$ from (\ref{spherloth}). As already mentioned earlier, $\psi_{LO}$ itself is of no significance, and only $\dot{\psi}_{LO}$ enters the orientation dynamics. The sign of $\dot{\psi}_{LO}$ in (\ref{spherlops}) can be understood from the balance between centrifugal forces and gyroscopic forces\cite{subramanian2006}. For a prolate spheroid, centrifugal forces act to move $\vec{\Omega}$ away from $\vec{L}$. Therefore, $\vec{\Omega}$ must lie between $\vec{L}$ and the spheroid axis, so the gyroscopic force opposes the centrifugal force, in turn implying that the spin must be positive; that is, a precessing prolate spheroid spins faster than the hypothetical rigidly rotating one. Conversely, in oblate spheroids, $\vec{L}$ must lie in between the spheroid axis and $\vec{\Omega}$, leading to a negative (relative) spin. 

The only relevant time period for a spheroid is that associated with precession ($T'=2\pi I_e/L$). The torque-free trajectories are termed Euler orbits, and the surface swept out by the spheroid axis is referred to as the Euler cone, characterized by its angle ($\theta_{LO}$) and orientation\,(specified by the unit vector $\hat{L}$). Equation (\ref{EetaL}) implies that it is sufficient to know any two of $\eta$, $E$ and $L$ in order to specify a particular orbit. The stability of the principal rotations may again be analyzed based on geometrical considerations, the constant-energy ellipsoid now replaced by a spheroid. Perturbations from axial rotation leads to polhodes that encircle the corresponding fixed point, implying neutral stability. 

\section{Large-$St$ Dynamics in simple shear flow: Spheroids}\label{lsds}

\begin{figure}
\includegraphics[scale=0.15]{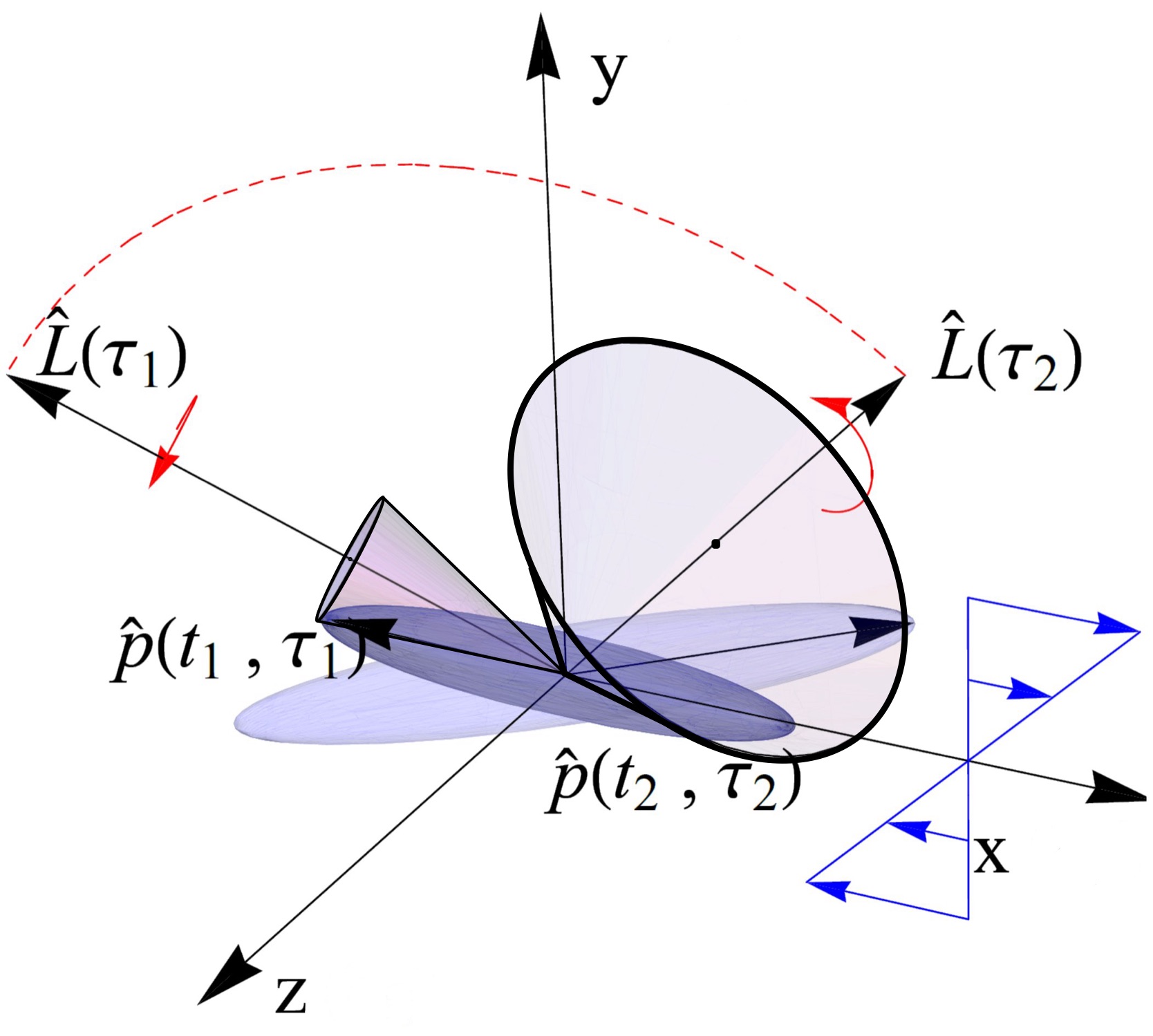}
\caption{A schematic illustrating spheroid orientation dynamics, for large but finite $St$, in an ambient simple shear flow. The evolution of the Euler cone swept by the rapidly precessing spheroid orientation vector ($\hat{p}$), from time $\tau_1$ to $\tau_2$, is shown.} 
\label{eulconedrift}
\end{figure}

For large but finite $St$, the orientation dynamics on time scales of $O(\dot{\gamma}^{-1})$ is similar to the Euler top in section \ref{symtop}. However, $\vec{L}$ and $E$ are no longer constants of motion, but instead are slow variables that evolve on an asymptotically longer time scale of $O(St \dot{\gamma}^{-1})$ on account of the weak viscous torque. This slow variation amounts to the Euler cone tilting and changing its angle, on the said time scale, as illustrated in figure \ref{eulconedrift}. The large-$St$ scenario thus lends itself to a multiple time scales analysis\cite{kevorkian1996,subramanian_multiple_2004}. Strictly speaking, the fast time scale is $O(\Omega^{-1})$, $\Omega$ being the instantaneous angular velocity, and the inverse shear rate estimate above is due to $\Omega_J$ in (\ref{ultima}) being $O(\dot{\gamma}^{-1})$. However, if $\Omega\gg\Omega_J$, $\Omega$ decreases exponentially fast until $\Omega \sim \Omega_J$ (discussed further in section \ref{numress} below). In the opposite limit, $\Omega\ll\Omega_J$, which is the case for a spheroid starting from rest, we show in section \ref{DAC} that the spin-up time to an angular velocity of $O(\Omega_J )$ is $O(St^{\frac{1}{2}}\dot{\gamma}^{-1})$, that is still asymptotically small compared to the aforementioned slow time scale of $O(St \dot{\gamma}^{-1})$. Hence, barring an initial transient, it is reasonable to use $O(\dot{\gamma}^{-1})$ as a representative scale for the fast dynamics.

The spheroid serves as a good starting point for the large-$St$ asymptotics, owing to the simplicity of the infinite-$St$ dynamics (see (\ref{spherloth}-\ref{EetaL}) in section \ref{symtop}). The simplicity arises on one hand due to the Euler-top solution being expressible in terms of trigonometric functions, and on the other hand, due to there only being a single characteristic time scale (that associated with precession). This allows for use of the method of multiple scales \cite{kevorkian1996,subramanian_multiple_2004} to analyze the leading order evolution on the slow time scale. The Euler-top solution for a triaxial ellipsoid involves two incommensurate time scales\,(those associated with spin/nutation and precession), and does not lend itself to the usual multiple scales formalism. The slow dynamics in this case is analyzed in section \ref{lsde} using the method of averaging. The torque on a spheroid in expanded form, in terms of the axial\,($X_c$) and equatorial\,($Y_c$) resistance coefficients, is given by: 
\begin{equation}\label{vistors}
\vec{T}_{viscous}=-\frac{8\pi}{St}[Y_c(\omega_1-\omega_{1,J})\hat{x}'+Y_c(\omega_2-\omega_{2,J})\hat{y}'+X_c(\omega_3-\omega_{3,J})\hat{z}'],
\end{equation}
where $X_c$ and $Y_c$ have been defined for both prolate and oblate spheroids in section \ref{gove}. The equations governing the rates of change of the slow variables $L_x,L_y,L_z$ and $\eta$ (taken to be the fourth degree of freedom, in preference to $E$), are obtained by resolving (\ref{vistors}) along the space-fixed directions, and the spheroid axis\,($\hat{z}'$), respectively, and using these components in (\ref{ultima}), to obtain:

\begin{align}\label{levosx}
\dot{L_x}&=-\frac{8\pi}{St}[t_{11} Y_c(\omega_1-\omega_{1,J})+t_{21} Y_c(\omega_2-\omega_{2,J})+t_{31} X_c(\omega_3-\omega_{3,J})],
\\\label{levosy}
\dot{L_y}&=-\frac{8\pi}{St}[t_{12} Y_c(\omega_1-\omega_{1,J})+t_{22} Y_c(\omega_2-\omega_{2,J})+t_{32} X_c(\omega_3-\omega_{3,J})],
\\\label{levosz}
\dot{L_z}&=-\frac{8\pi}{St}[t_{13} Y_c(\omega_1-\omega_{1,J})+t_{23} Y_c(\omega_2-\omega_{2,J})+t_{33} X_c(\omega_3-\omega_{3,J})],
\\\label{nos}
\dot{\eta}&=-\frac{8\pi X_c}{St I_a}[\omega_3-\omega_{3,J}].
\end{align}

Note that the rate of change of $E$ may be derived from (\ref{levosx}-\ref{nos}) as:
\begin{equation}\label{evose}
\dot{E}=-\frac{8\pi}{St}[Y_c(\omega_1-\omega_{1,J})\omega_1+Y_c(\omega_2-\omega_{2,J})\omega_2+X_c(\omega_3-\omega_{3,J})\omega_3].
\end{equation}

The $\omega_i$'s in (\ref{levosx}-\ref{nos}) may be related to the space-fixed components of $\vec{L}$ using the relations $\omega_i = t_{ij}L_j/I_i$, and the $\omega_{i,J}$'s can be cast in terms of the Euler angles using (\ref{spheroid_wj1}-\ref{spheroid_wj3}). Implementing these substitutions, and writing down the explicit expressions for the $t_{ij}$'s, one obtains:
\begin{align}\label{lxs2}
\dot{L_x}&=-\frac{8\pi}{St}[\frac{Y_c}{I_e}L_x+(X_c-\frac{Y_c I_a}{I_e})\eta\sin{\theta}\sin{\phi}+\frac{1}{2}(X_c-Y_c\frac{I_a}{I_e})\cos\theta\sin\theta\sin\phi],
\\\label{lys2}
\dot{L_y}&=-\frac{8\pi}{St}[\frac{Y_c}{I_e}L_y-(X_c-\frac{Y_c I_a}{I_e})\eta\sin{\theta}\cos{\phi}-\frac{1}{2}(X_c-Y_c(1-\frac{I_a}{2I_e}))\cos\theta\sin\theta\cos\phi],
\\\label{lzs2}
\dot{L_z}&=-\frac{8\pi}{St}[\frac{Y_c}{I_e}L_z+(X_c-\frac{Y_c I_a}{I_e})\eta\cos{\theta}+\frac{X_c}{2}\cos^2{\theta}+\frac{Y_c I_a}{2I_e}(\sin^2{\theta}\sin^2{\phi}+\kappa^2\sin^2{\theta}\cos^2{\phi})],
\\\label{ns2}
\dot{\eta}&=-\frac{8\pi X_c}{St I_a}[\eta+\frac{1}{2}\cos{\theta}].
\end{align}
The rates of change of the slow variables in (\ref{lxs2}-\ref{ns2}) is $O(St^{-1})$, implying that an $O(1)$ change occurs only over time scales of $O(St\dot{\gamma}^{-1})$, in agreement with the scaling arguments above. However, the system (\ref{lxs2}-\ref{ns2}) is not closed owing to the dependence on ($\theta$,$\phi$). For large $St$, this dependence may be eliminated by exploiting the fact that the short-time dynamics corresponds to the symmetric Euler top\,(with slow variables regarded as constants). To formally derive the system of equations involving the slow variables alone, we  proceed using the method of multiple scales\cite{kevorkian1996}. Within this framework, the slow variables are expanded in the form: $L_x=L_x^{(0)}(t_2)+ \frac{1}{St}L_x^{(1)}(t_1,t_2)$, $L_y=L_y^{(0)}(t_2)+ \frac{1}{St} L_y^{(1)}(t_1,t_2)$, $L_z=L_z^{(0)}(t_2)+ \frac{1}{St} L_z^{(1)}(t_1,t_2)$, $\eta=\eta^{(0)}(t_2)+ \frac{1}{St} \eta^{(1)}(t_1,t_2)$. Here, $t_1=t$ and $t_2=\frac{t}{St}$ denote the fast and slow time variables, with the leading terms assumed to evolve solely on the slow time scale, and the higher order corrections assumed to be periodic in $t_1$; the periodicity of the fast-time dynamics arises from the Euler-top solution. Substitution in (\ref{lxs2} - \ref{ns2}) leads to:
\begin{align}\label{lxms}
\frac{\partial{{L_x^{(1)}}}}{\partial t_1}+\frac{d{{L_x^{(0)}}}}{d t_2}&=-8\pi[\frac{Y_c}{I_e}{L_x^{(0)}}+(X_c-\frac{Y_c I_a}{I_e}){\eta^{(0)}}\sin{\theta}\sin{\phi}+\frac{1}{2}(X_c-Y_c\frac{I_a}{I_e})\cos\theta\sin\theta\sin\phi],
\\\label{lyms}
\frac{\partial{{L_y^{(1)}}}}{\partial t_1}+\frac{d{{L_y^{(0)}}}}{d t_2}&=-8\pi[\frac{Y_c}{I_e}{L_y^{(0)}}-(X_c-\frac{Y_c I_a}{I_e}){\eta^{(0)}}\sin{\theta}\cos{\phi}-\frac{1}{2}(X_c-Y_c(1-\frac{I_a}{2I_e}))\cos\theta\sin\theta\cos\phi],
\\\label{lzms}
\frac{\partial{{L_z^{(1)}}}}{\partial t_1}+\frac{d{{L_z^{(0)}}}}{d t_2}&=-8\pi[\frac{Y_c}{I_e}{L_z^{(0)}}+(X_c-\frac{Y_c I_a}{I_e}){\eta^{(0)}}\cos{\theta}+\frac{X_c}{2}\cos^2{\theta}+\frac{Y_c I_a}{2I_e}(\sin^2{\theta}\sin^2{\phi}+\kappa^2\sin^2{\theta}\cos^2{\phi})],
\\\label{nms}
\frac{\partial{{\eta^{(1)}}}}{\partial t_1}+\frac{d{{\eta^{(0)}}}}{dt_2}&=-\frac{8\pi X_c}{I_a}[{\eta^{(0)}}+\frac{1}{2}\cos\theta].
\end{align}
The Euler angles in (\ref{lxms}-\ref{nms}) are given by (\ref{spherloth}-\ref{EetaL}) with $t$ replaced by $t_1$, and this $t_1$-dependence may in turn be eliminated by averaging  (\ref{lxms}-\ref{nms}) over a single precessional period, with $t_2$ fixed. Defining the symmetric Euler-top averages as $\langle f(t_1,t_2)\rangle=\frac{1}{T'}\int^{T'}_0 {f(t_1,t_2)\text{d}t_1}$, one obtains:

\begin{align}\label{lxmsavg}
\frac{d{{L_x^{(0)}}}}{d t_2}&=-8\pi[\frac{Y_c}{I_e}{L_x^{(0)}}+(X_c-\frac{Y_c I_a}{I_e}){\eta^{(0)}} \langle\sin{\theta}\sin{\phi}\rangle +\frac{1}{2}(X_c-Y_c\frac{I_a}{I_e}) \langle\cos\theta\sin\theta\sin\phi\rangle ],
\\\label{lymsavg}
\frac{d{{L_y^{(0)}}}}{d t_2}&=-8\pi[\frac{Y_c}{I_e}{L_y^{(0)}}-(X_c-\frac{Y_c I_a}{I_e}){\eta^{(0)}} \langle\sin{\theta}\cos{\phi}\rangle -\frac{1}{2}(X_c-Y_c(1-\frac{I_a}{2I_e})) \langle\cos\theta\sin\theta\cos\phi\rangle ],
\\\label{lzmsavg}
\frac{d{{L_z^{(0)}}}}{dt_2}&=-8\pi[\frac{Y_c}{I_e}{L_z^{(0)}}+(X_c-\frac{Y_c I_a}{I_e}){\eta^{(0)}} \langle\cos{\theta}\rangle +\frac{X_c}{2}\langle\cos{\theta}^2\rangle +\frac{Y_c I_a}{2I_e}( \langle(\sin{\theta}\sin{\phi})^2 \rangle+\kappa^2 \langle (\sin{\theta}\cos{\phi})^2\rangle )],
\\\label{nmsavg}
\frac{d{{\eta^{(0)}}}}{d t_2}&=-\frac{8\pi X_c}{I_a}[{\eta^{(0)}}+\frac{1}{2} \langle\cos\theta\rangle].
\end{align}
The fast-time averages of the functions of Euler angles in (\ref{lxmsavg}-\ref{nmsavg}) can be determined by first expressing these angles, in terms of the leading order slow variables, using 
\begin{equation}\label{alphatranss}
\begin{pmatrix} 
&\sin{\theta}\sin{\phi}\\
-&\sin{\theta}\cos{\phi}\\
&\cos\theta\\
\end{pmatrix}
=(\alpha_{ij})^{(0)}\cdot\begin{pmatrix}
&\bar{h}\sin{(\frac{L^{(0)}}{I_e}t_1+\phi_0)}\\
-&\bar{h}\cos{(\frac{L^{(0)}}{I_e}t_1+\phi_0)}\\
&h\\
\end{pmatrix},
\end{equation}
where $h=\cos{\theta_{LO}}^{(0)}=I_a \eta^{(0)}/L^{(0)}$, $\bar{h}=\sqrt{1-h^2}$. The matrix transformation that relates the angular-momentum-aligned and space-fixed coordinate systems, $\alpha_{ij}$ has been defined earlier\,(the superscript $0$ implies the $L$'s in its elements have been replaced by $L^{(0)}$'s); details with regard to the evaluation of these averages are given in Appendix \ref{LOA}.
Using (\ref{alphatranss}) to compute the averages, the system of equations governing evolution on the four-dimensional slow manifold is given by: 
\begin{align}\label{lxsmain}
\frac{d{L_x^{(0)}}}{dt_2}=&-8\pi {L_x^{(0)}}[(\frac{X_c h^2}{I_a}+\frac{Y_c \bar{h}^2}{I_e})+(\frac{X_c I_e-Y_c I_a}{2I_e})(\frac{{L_z^{(0)}}}{2{L^{(0)}}^2})(2h^2-\bar{h}^2)],
\\\label{lysmain}
\frac{d{L_y^{(0)}}}{dt_2}=&-8\pi {L_y^{(0)}}[(\frac{X_c h^2}{I_a}+\frac{Y_c \bar{h}^2}{I_e})+(\frac{X_c I_e+Y_c I_a-2Y_c I_e}{2I_e})(\frac{{L_z^{(0)}}}{2{L^{(0)}}^2})(2h^2-\bar{h}^2)],
\\\label{lzsmain}
\frac{d{L_z^{(0)}}}{dt_2}=&-8\pi[(\frac{X_c h^2}{I_a}+\frac{Y_c \bar{h}^2}{I_e}){L_z^{(0)}}+\frac{{L_x^{(0)}}^2}{{L^{(0)}}^2}(\frac{X_c\bar{h}^2}{4}+\frac{Y_c\bar{h}^2\kappa^2}{2(\kappa^2+1)}+\\&\frac{Y_ch^2}{(\kappa^2+1)})+\frac{{L_y^{(0)}}^2}{{L^{(0)}}^2}(\frac{X_c\bar{h}^2}{4}+\frac{Y_c\bar{h}^2}{2(\kappa^2+1)}+\frac{Y_ch^2\kappa^2}{(\kappa^2+1)})+\frac{{L_z^{(0)}}^2}{{L^{(0)}}^2}(\frac{X_ch^2}{2}+\frac{Y_c\bar{h}^2}{2})], \nonumber
\\\label{nsmain}
\frac{d{\eta^{(0)}}}{dt_2}=&-\frac{8\pi X_c {\eta^{(0)}}}{I_a}[1+\frac{{L_z^{(0)}} I_a }{2 {L^{(0)}}^2}].
\end{align}
The above system is 
valid for $St\gg1$ when $\Omega$ is $O(\dot{\gamma}^{-1})$, as argued previously, and 
will now be used to examine the orientation dynamics for large but finite $St$.   



Equating the rates of change in (\ref{lxsmain}-\ref{nsmain}) to zero, one finds the following fixed points:
\begin{align}
({L_x^{(0)}},{L_y^{(0)}},{L_z^{(0)}},{\eta^{(0)}})&\equiv(0,0,0,-\frac{I_a}{2},\pm\frac{1}{2}), \label{spinmode_sph}\\
({L_x^{(0)}},{L_y^{(0)}},{L_z^{(0)}},{\eta^{(0)}})&\equiv(0,0,0,-\frac{I_e}{2},0). \label{tumbmode_sph}
\end{align}
Both the fixed points in (\ref{spinmode_sph}) denote spheroid rotation with its axis aligned with the ambient vorticity, and accordingly, correspond to the spinning mode. The one in (\ref{tumbmode_sph}) denotes vorticity-aligned rotation with the spheroid axis in the flow-gradient plane, and is termed the tumbling mode. Both rotations occur with a constant angular velocity of $-1/2$, consistent with the numerical results presented earlier (see figure \ref{TvslogSt}\cite{lundell2010}); although, this constancy is in contrast to the variable angular velocity that characterizes inertialess tumbling, with this variation increasing for extreme aspect ratios. The rotational kinetic energies\,($E^{(0)}$) are $I_a/8$ and $I_e/8$ for the spinning and tumbling modes, respectively. We now examine the stability of these fixed points.



On linearizing (\ref{lxsmain}-\ref{nsmain}) about the tumbling-mode fixed point, one finds uncoupled equations for each of the slow variables, which immediately leads to the following eigenvalues: 
$$(\lambda_1,\lambda_2,\lambda_3,\lambda_4)\equiv\bigg(-\frac{8\pi}{I_e}(\frac{Y_c \kappa^2}{\kappa^2+1}+\frac{X_c}{2}),-\frac{8\pi}{I_e}(\frac{Y_c }{\kappa^2+1}+\frac{X_c}{2}),-\frac{8\pi Y_c}{I_e},-\frac{8\pi X_c}{I_a}(1-\frac{I_a}{I_e})\bigg).$$
$\lambda_1,\lambda_2$ and $\lambda_3$ are always negative irrespective of $\kappa$, since both $Y_c$ and $X_c$ are positive for spheroids, this arising from the positive definiteness of the viscous energy dissipation \cite{kim2013}. In contrast, $\lambda_4$ is positive for an oblate spheroid, but negative for a prolate one, implying that the tumbling mode is a stable node (saddle point) for prolate (oblate) spheroids. 
The linearization for the spinning mode leads to independent equations for $L_x^{(0)}$ and $L_y^{(0)}$, but a coupled 2D system for $L_z^{(0}$ and $\eta^{(0)}$, and the resulting eigenvalues are given by: 
$$(\lambda_1,\lambda_2,\lambda_3,\lambda_4)\equiv \bigg(-\frac{16\pi Y_c}{I_a(\kappa^2+1)},-\frac{16\pi Y_c \kappa^2}{I_a(\kappa^2+1)},-\frac{8\pi X_c}{I_a},-\frac{16\pi Y_c}{I_a}(\frac{I_a}{I_e}-1)\bigg).$$
Analogous to the tumbling case above, $\lambda_1,\lambda_2$ and $\lambda_3$ are always negative. $\lambda_4$ is now negative for an oblate spheroid but positive for a prolate one, implying that the spinning mode is a stable node (saddle point) for oblate (prolate) spheroids. In summary, vorticity-aligned-rotation about the shorter axis is a stable node, and that about the longer axis is a saddle point, for any massive spheroid in a simple shear flow. 

\begin{figure}
\subfloat[\label{LxvstvarStS1} $L_{x}/I_e$ vs $t_2$ for different $St$; ($\kappa=2$) $(\dot{\theta}_0, \dot{\phi}_0,\dot{\psi}_0,\theta_0,\phi_0) \equiv(1,1,1,\pi/3,\pi/6)$ ]{\includegraphics[scale=0.45]{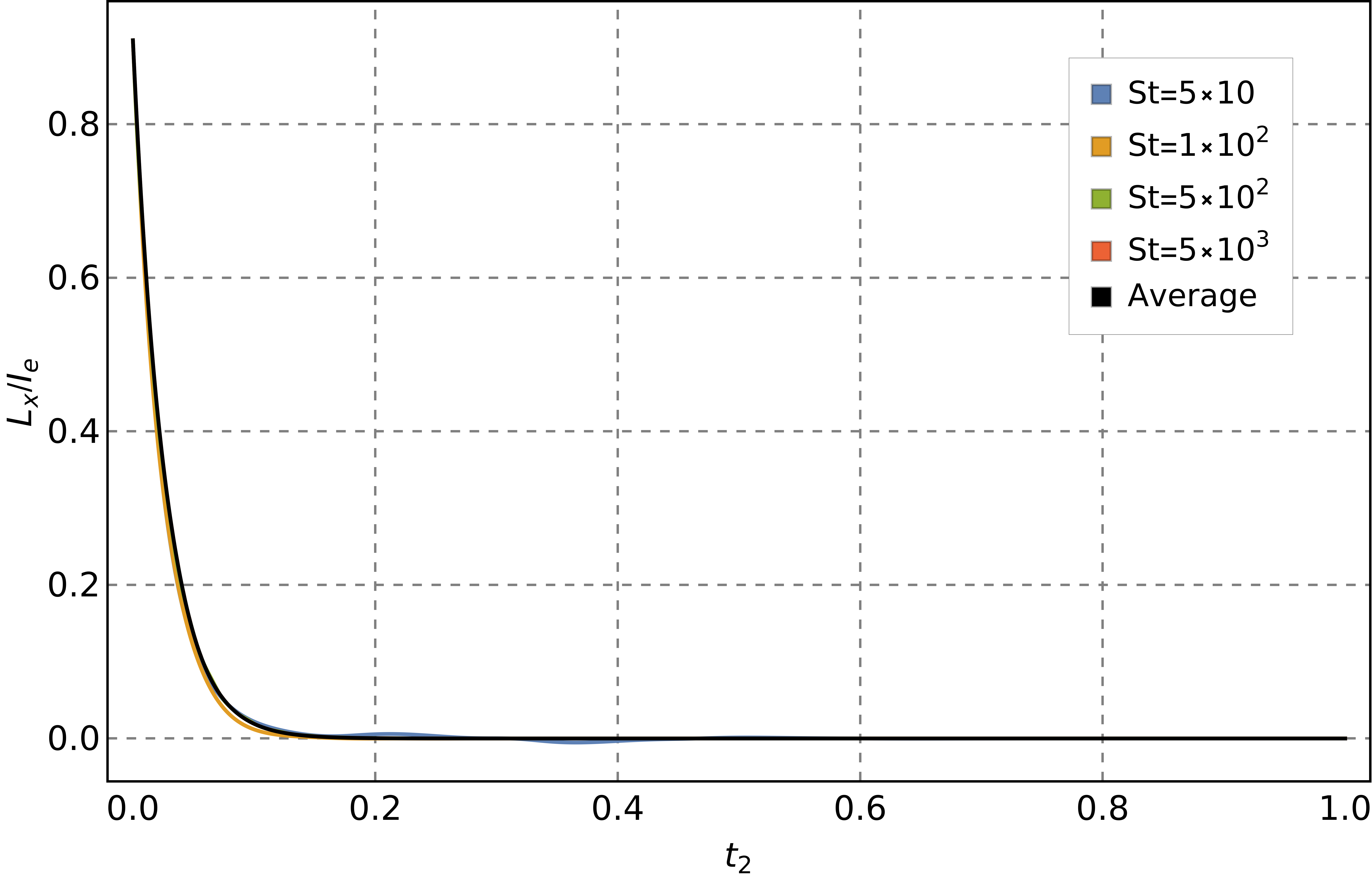}}
\subfloat[\label{LyvstvarStS1} $L_{y}/I_e$ vs $t_2$ for different $St$; ($\kappa=2$) $(\dot{\theta}_0, \dot{\phi}_0,\dot{\psi}_0,\theta_0,\phi_0) \equiv(1,1,1,\pi/3,\pi/6)$]{\includegraphics[scale=0.45]{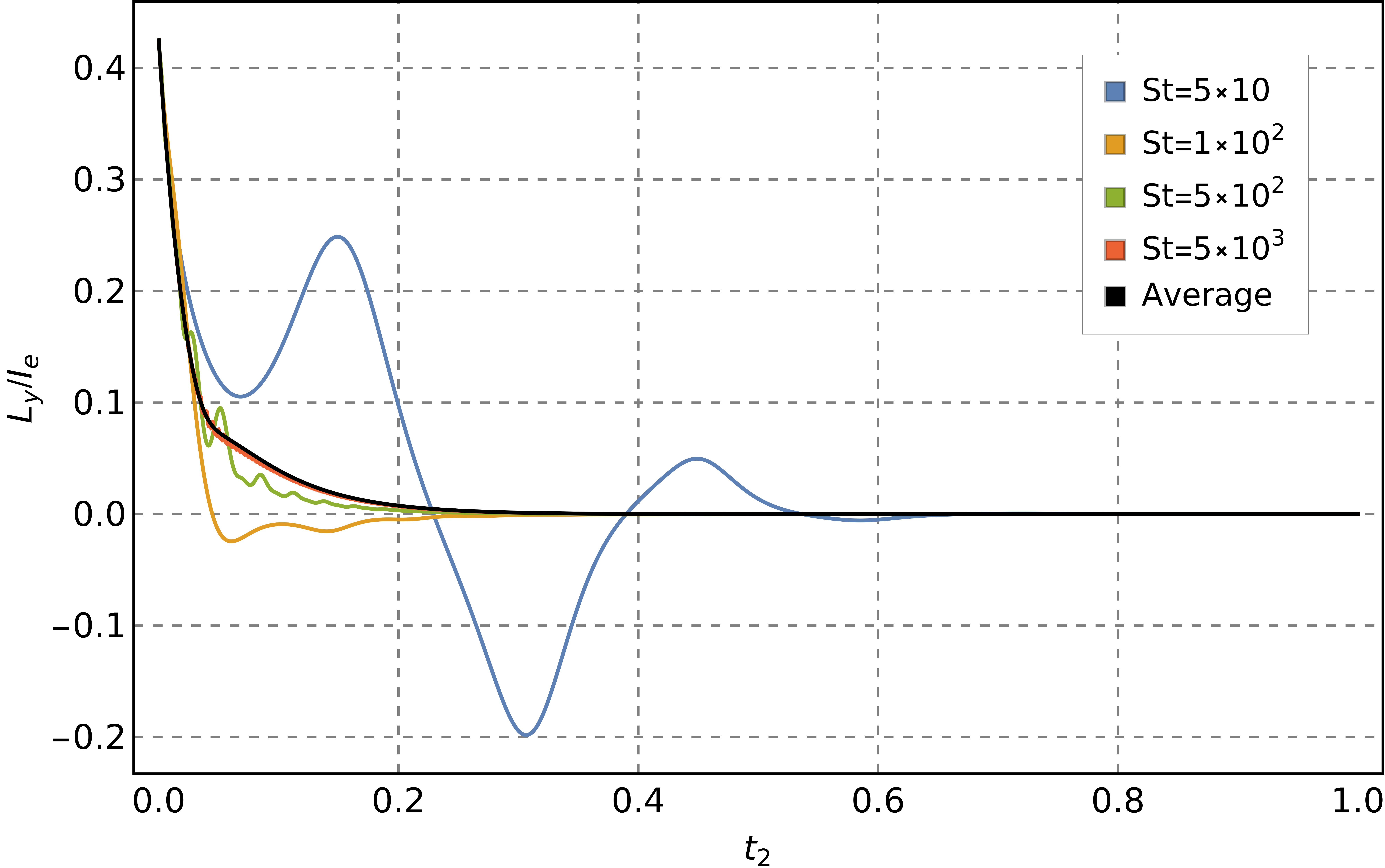}}\\
\subfloat[\label{LzvstvarStS1} $L_z/I_e$ vs $t_2$ for different $St$ ; ($\kappa=2$) $(\dot{\theta}_0, \dot{\phi}_0,\dot{\psi}_0,\theta_0,\phi_0) \equiv(1,1,1,\pi/3,\pi/6)$ ]{\includegraphics[scale=0.45]{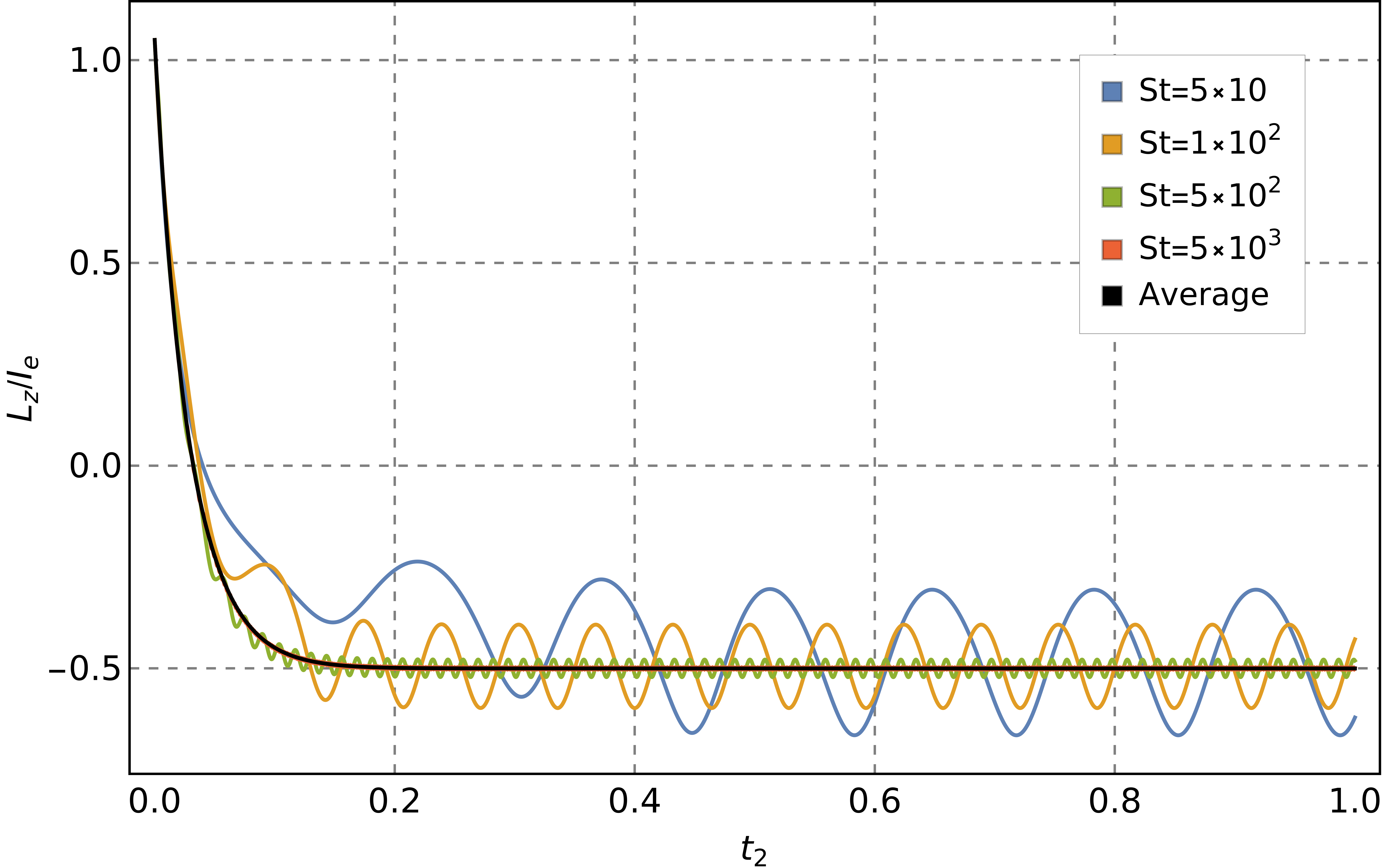}}
\subfloat[\label{EvstvarStS1} $E/I_e$ vs $t_2$ for different $St$; ($\kappa=2$) $(\dot{\theta}_0, \dot{\phi}_0,\dot{\psi}_0,\theta_0,\phi_0) \equiv(1,1,1,\pi/3,\pi/6)$]{\includegraphics[scale=0.45]{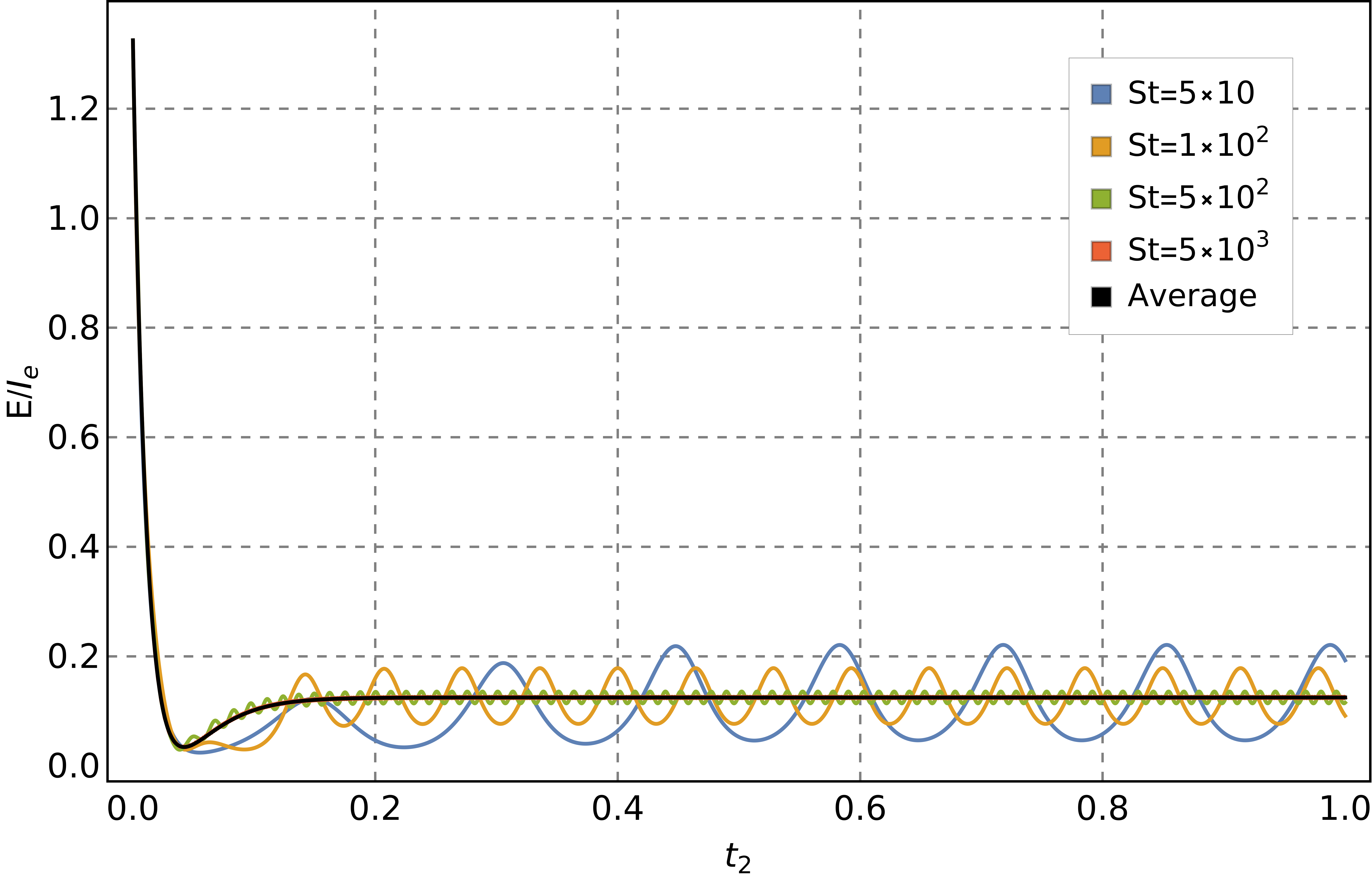}}
\caption{Time evolution of the normalized slow variables compared with full-solution trajectories, for a prolate spheroid with $\kappa = 2$, for two different sets of initial conditions. For long times, all trajectories converge to $(L_x/I_e,L_y/I_e,L_z/I_e,E/I_e)\equiv(0,0,-1/2,1/8)$, corresponding to the (stable) tumbling mode.}
\label{elevS}
\end{figure}

\begin{figure}
\subfloat[\label{LxvstvarStS2} $L_{x}/I_e$ vs $t_2$ for different $St$; ($\kappa=2$) $(\dot{\theta}_0, \dot{\phi}_0,\dot{\psi}_0,\theta_0,\phi_0) \equiv(0,-0.4,0.6,3.14,0)$ ]{\includegraphics[scale=0.45]{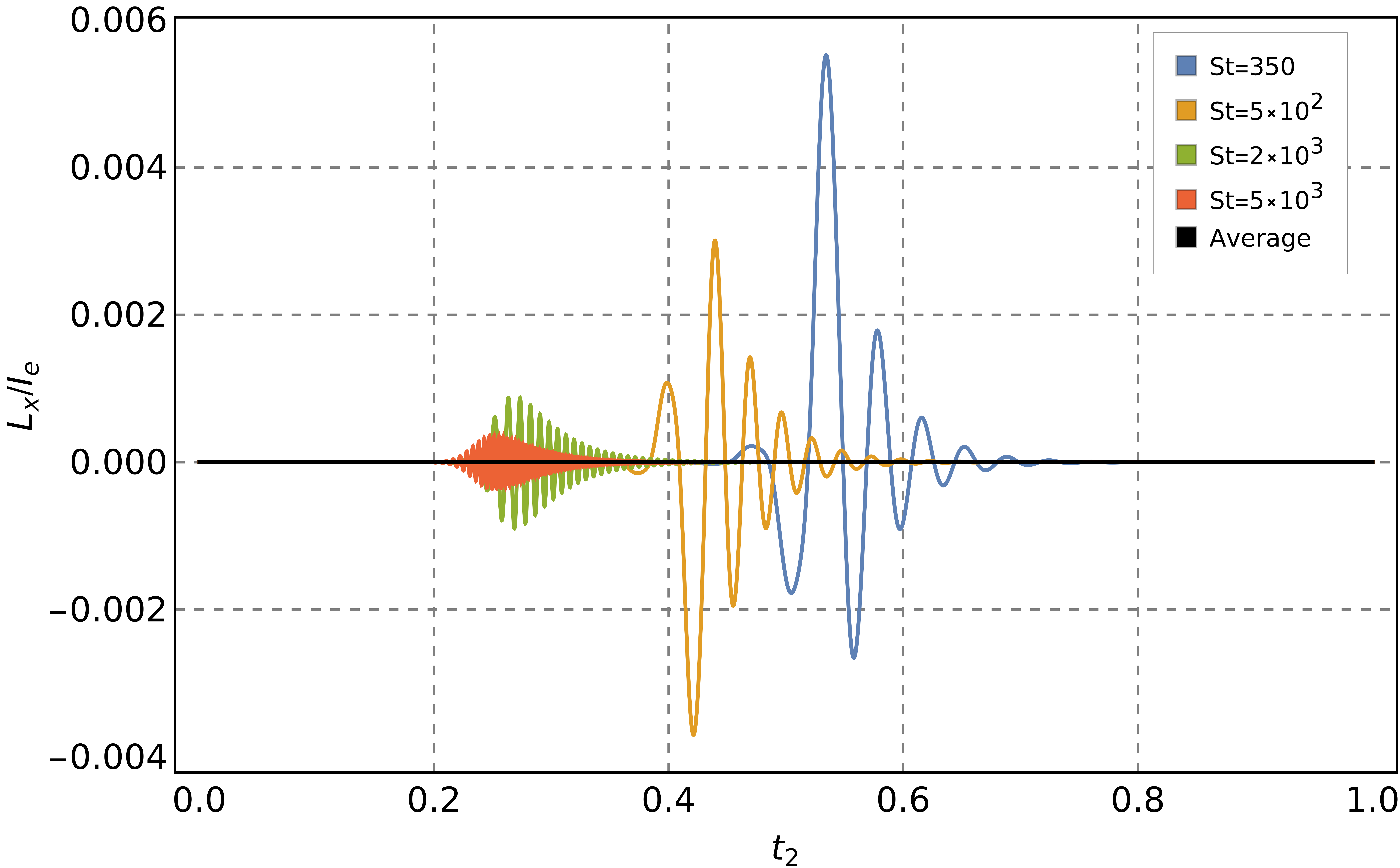}}
\subfloat[\label{LyvstvarStS2} $L_{y}/I_e$ vs $t_2$ for different $St$; ($\kappa=2$) $(\dot{\theta}_0, \dot{\phi}_0,\dot{\psi}_0,\theta_0,\phi_0) \equiv(0,-0.4,0.6,3.14,0)$ ]{\includegraphics[scale=0.45]{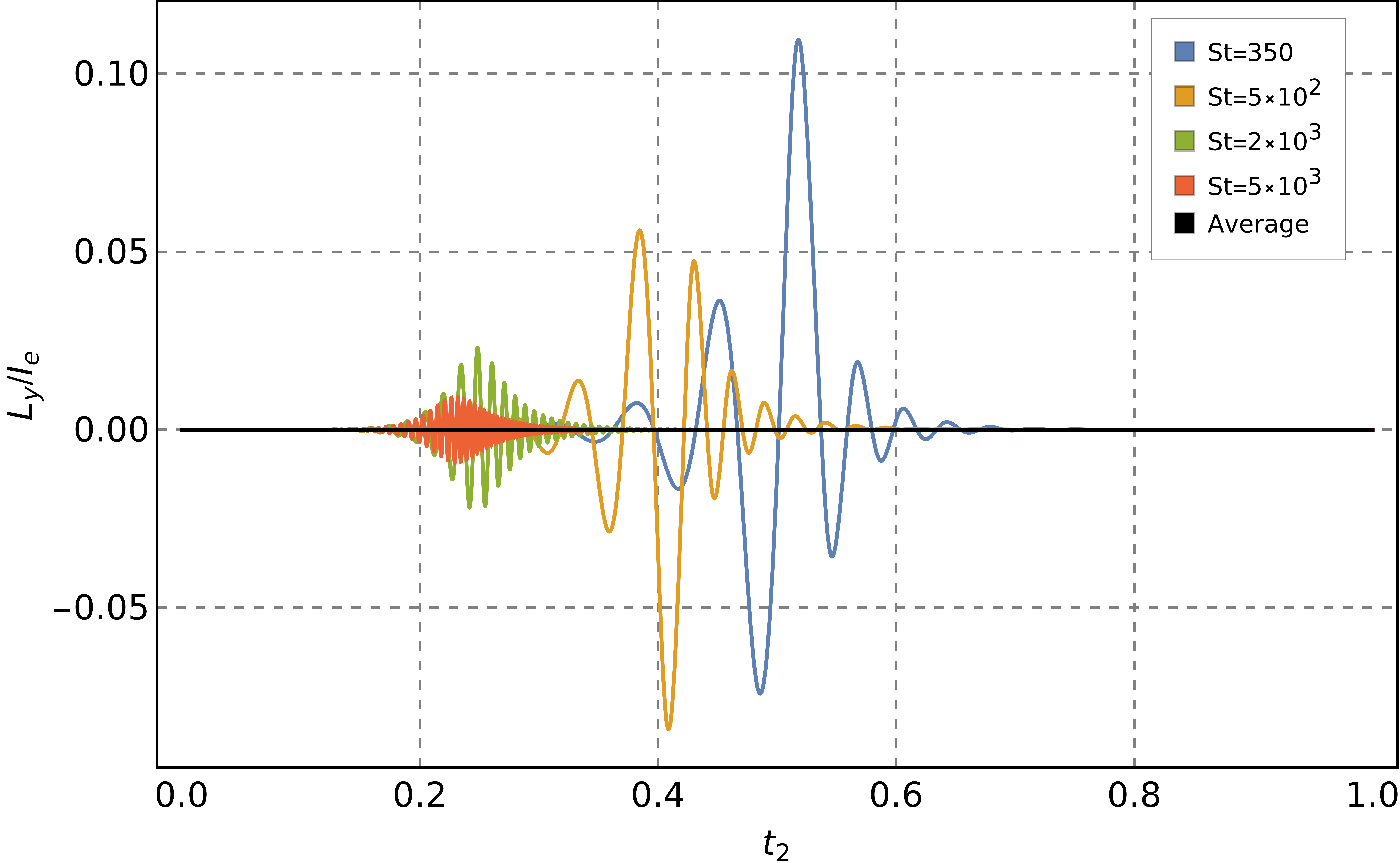}}\\
\subfloat[\label{LzvstvarStS2} $L_z/I_e$ vs $t_2$ for different $St$ ; ($\kappa=2$)  $(\dot{\theta}_0, \dot{\phi}_0,\dot{\psi}_0,\theta_0,\phi_0) \equiv(0,-0.4,0.6,3.14,0)$ ]{\includegraphics[scale=0.45]{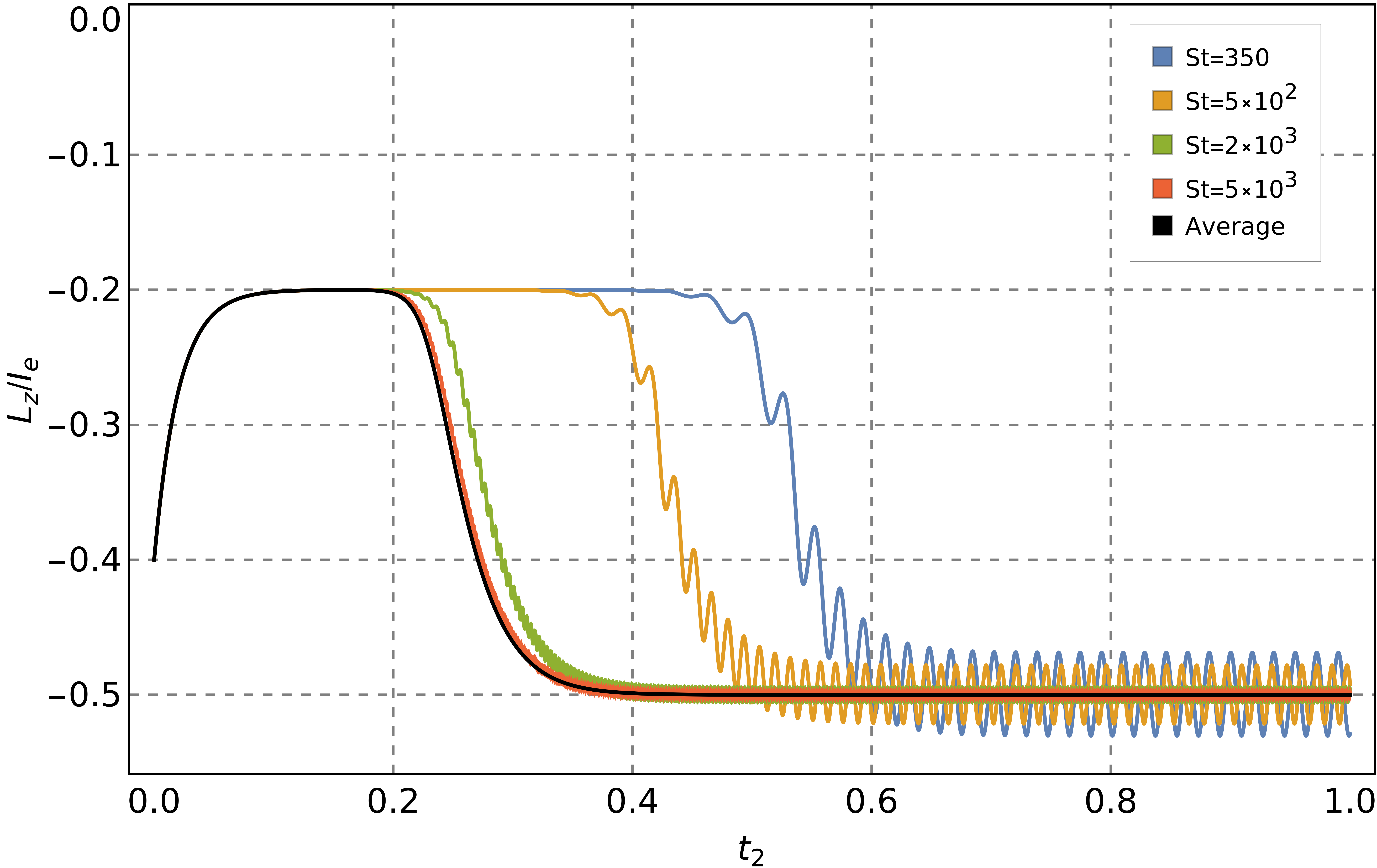}}
\subfloat[\label{EvstvarStS2} $E/I_e$ vs $t_2$ for different $St$; ($\kappa=2$)  $(\dot{\theta}_0, \dot{\phi}_0,\dot{\psi}_0,\theta_0,\phi_0) \equiv(0,-0.4,0.6,3.14,0)$ ]{\includegraphics[scale=0.45]{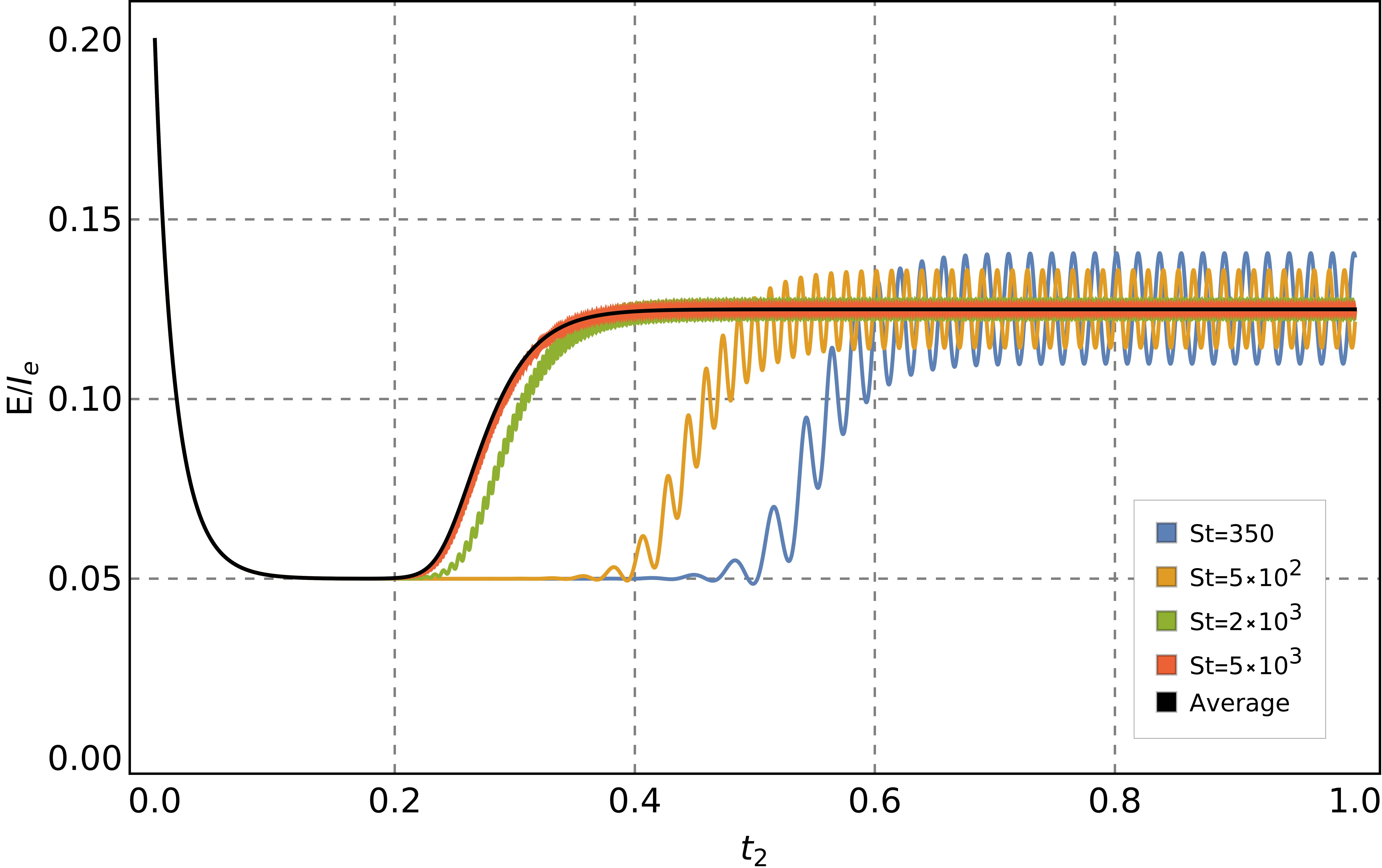}}
\caption{Time evolution of the normalized slow variables compared with full-solution trajectories, for a prolate spheroid with $\kappa = 2$, for an initial condition with an $O(\dot{\gamma})$ angular velocity. For long times, all trajectories converge to $(L_x/I_e,L_y/I_e,L_z/I_e,E/I_e)\equiv(0,0,-1/2,1/8)$, corresponding to the (stable) tumbling mode, although the $L_z$ and $E$ plots exhibit intermediate plateaus, corresponding to a long time spent in the vicinity of the spinning mode. }
\label{elevS2}
\end{figure}

The analysis above only yields the local structure of the slow manifold in the neighborhood of the two fixed points. In order to assess the global topology of slow-manifold trajectories, we now analyze the time dependent solutions of (\ref{lxsmain}-\ref{nsmain}). To begin with, we compare the slow-manifold solutions, as a function of time, to those obtained from a numerical integration of the full system for different $St$; note that obtaining the former is much easier on account of the elimination of the fast time scale that renders the full system increasingly stiff for large $St$. The time evolution for two sets of initial conditions\,(which include specifying the initial angular velocities for the full system) is shown in Figures \ref{LxvstvarStS1}-\ref{EvstvarStS1} and \ref{LxvstvarStS2}-\ref{EvstvarStS2}, for a prolate spheroid with $\kappa=2$. For sufficiently long times, in all cases, $L_x$ and $L_y$ decrease to zero, while $L_z$ and $E$ converge to $-I_e/2$ and $I_e/8$, respectively, in agreement with the fixed-point analysis above. It is worth noting that the second initial condition has deliberately been chosen such that the trajectory spends a long but finite time in the neighborhood of the the saddle point\,(spinning mode), before finally asymptoting to the stable node; this manifests as an extended intermediate plateau in Figures \ref{LzvstvarStS2} and \ref{EvstvarStS2}. As will be seen below, the above long-time behavior is true for almost all initial conditions, except for the set of measure zero that lie on the (3D)\,stable manifold of the spinning mode. It is also evident from figures \ref{LxvstvarStS1}-\ref{EvstvarStS2} that the deviation of the full solution trajectories, from the slow-manifold ones, has an oscillatory character. 
From the structure of the multiple scales expansions given above, this oscillatory variation is expected to be a fast-time-scale contribution. Accordingly, the time period and amplitude of the oscillations are seen to decrease with increasing $St$. In general, the slow-manifold solution gives an accurate prediction for $St \gtrsim 100$, athough the full solution follows the slow trajectory on average even for $St = 50$. It is shown in Appendix \ref{FOD} (using a WKB formalism) that accounting for the fast $O(St^{-1})$ oscillations leads to better agreement with the exact dynamics down to smaller $St$. 

\subsection{The $L_z-E$ Phase Plane: Spheroid}\label{numress}

The functional complexity of the right-hand-side of (\ref{lxsmain}-\ref{nsmain}) precludes a direct analytical approach. The nature of solutions  may, however, be better understood in the large angular momentum limit, characterized by $L^{(0)} \gg I_e\,(I_a)$ for a prolate (an oblate) spheroid. The slow manifold equations simplify to:
\begin{align}\label{lxshed}
\frac{d{L_x^{(0)}}}{dt_2}&=-8\pi {L_x^{(0)}}[(\frac{X_c h^2}{I_a}+\frac{Y_c \bar{h}^2}{I_e})],\\\label{lyshed}
\frac{d{L_y^{(0)}}}{dt_2}&=-8\pi {L_y^{(0)}}[(\frac{X_c h^2}{I_a}+\frac{Y_c \bar{h}^2}{I_e})],\\\label{nshed}
\frac{d{\eta^{(0)}}}{dt_2}&=-\frac{8\pi X_c {\eta^{(0)}}}{I_a},\\\label{lzshed}
\frac{d{L_z^{(0)}}}{dt_2}&=-8\pi L_z^{(0)}[(\frac{X_c h^2}{I_a}+\frac{Y_c \bar{h}^2}{I_e})],
\end{align}
and it is readily shown that ${L^{(0)}}^2$ and $E^{(0)}$ decay as a sum of two exponentials with characteristic times given by $I_a/(16\pi X_c)$ and $I_e/(16\pi Y_c)$. Since the limit considered implies $\Omega \gg \Omega_J$, with $\Omega_J\sim O(1)$ and $L^{(0)}\sim I_e \Omega$\,(or $I_a\Omega)$, the exponential decay may also be understood directly from (\ref{ultima}) - neglecting $\Omega_J$ immediately leads to each angular velocity component decaying exponentially with time. Further, after some algebra, (\ref{lxshed}-\ref{nshed}) yields: $$\frac{d{L_x^{(0)}}}{L_x^{(0)}}=\frac{d{L_y^{(0)}}}{L_y^{(0)}}=\frac{d{L_z^{(0)}}}{L_z^{(0)}}=\frac{d{L^{(0)^2}}}{2L^{(0)^2}},$$
which implies that, in the large angular momentum limit, the Euler cone orientation remains virtually unchanged even as $L^{(0)}$ and $E^{(0)}$ decay to sufficiently low magnitudes, such that $L^{(0)} \sim I_e$. Interestingly, the limit $L^{(0)}_{z}\ll L^{(0)}$ also leads to (\ref{lxshed}-\ref{nshed}), 
implying that ${L^{(0)}}^2$ and $E^{(0)}$ again decay with an unchanged Euler-cone orientation, although confined to the flow-gradient plane. 
The discussion below is therefore limited to the cases with ${L^{(0)}}\sim I_e\sim L_z^{(0)}$. 

While the dynamics on the slow manifold may be understood based on plots showing the time evolution of the angular momenta and energy, such as those in figures \ref{LxvstvarStS1}-\ref{EvstvarStS2}, these are restricted to specific initial conditions, and making broad observations becomes cumbersome even for a fixed $\kappa$. Since the system of slow-manifold equations is autonomous, a more efficient method is to study the phase space representation in terms of the dependent variables alone, thereby covering the entire range of initial conditions. 
We address this by first considering the $L_z-E$ subspace, since trajectories starting with $(L_x^{(0)},L_y^{(0)}) \equiv (0,0)$ are restricted to this plane for all time. Furthermore, all the fixed points identified above lie in this plane, with the pair of degenerate spinning-mode points now being coincident. The $L_z-E$ plane is preferred over the $L_z-\eta$ plane because it generalizes to the case of a triaxial ellipsoid considered in the next section. Note that we refer to the phase space as the $L_z-E$ subspace (as opposed to $L_z^{(0)}-E^{(0)}$) since projections of the full-solution trajectories will also be shown alongside the slow-manifold ones later. For this reason, the superscript will be also avoided for the remainder of this section, and distinctions between the slow manifold and full solution will be made explicitly. 
\begin{figure}
{\includegraphics[scale=0.8]{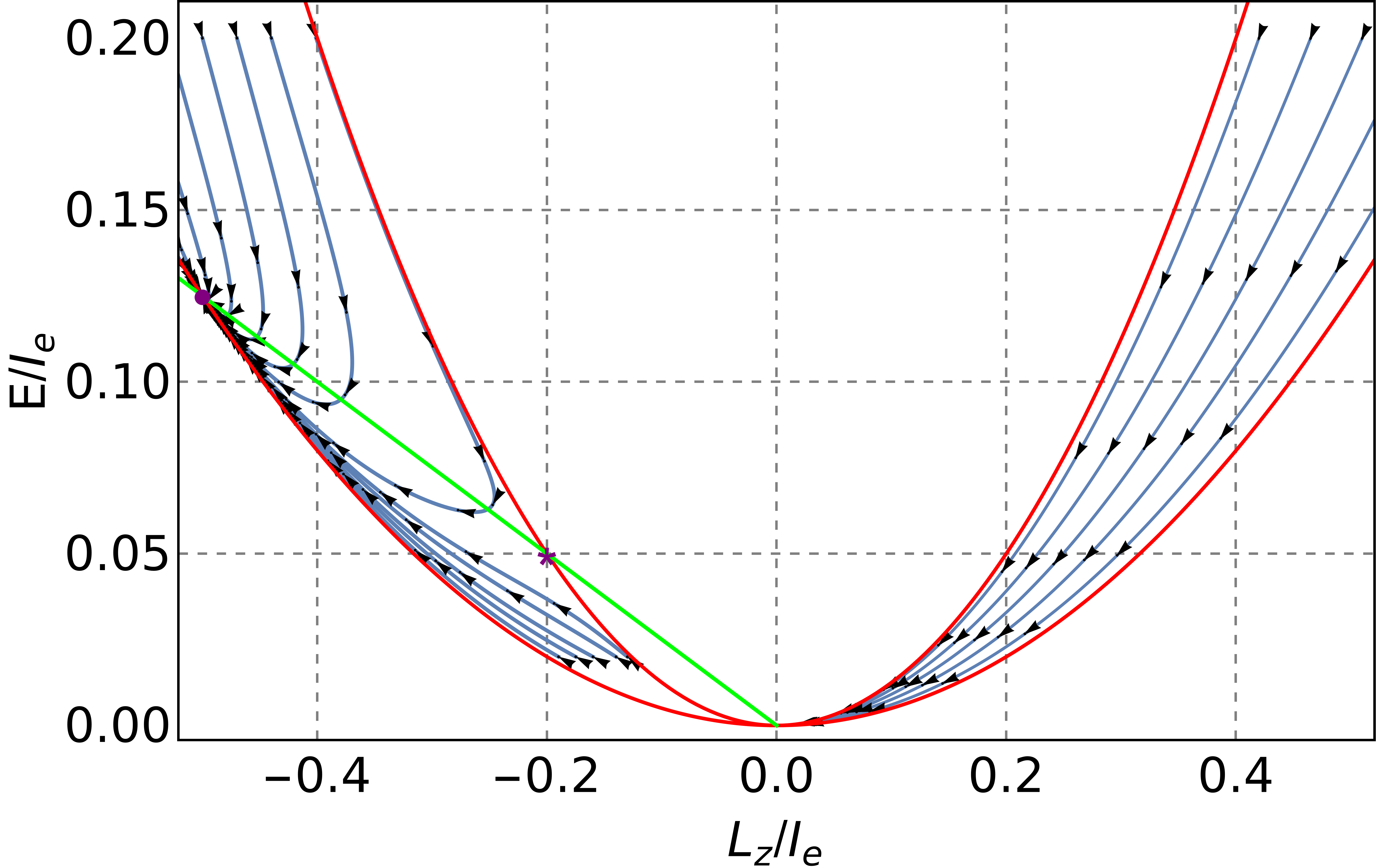}}\\
\caption{The $L_z-E$ phase plane for $(L_{x0},L_{y0})\equiv(0,0)$, for a prolate spheroid ($\kappa=2$). The red parabolae, defined by $L_z^2=2EI_a$ and $L_z^2=2EI_e$, demarcate the physically accessible domain. Almost all slow-manifold trajectories for $L_z < 0$ converge to $(L_z/I_e,E/I_e)\equiv(-1/2,1/8)$\,(purple dot); the only exception being the stable manifold of the saddle point, $(L_z/I_e,E/I_e)\equiv(-1/2,1/8)$\,(purple asterisk), which coincides with the left-arm of the upper parabola. All slow-manifold trajectories for $L_z > 0$ converge to the pinch point $(L_z/I_e,E/I_e)\equiv (0,0)$; the green line, $(L_z/I_e)+4(E/I_e)=0$, intersects the bounding parabolae at the pair of fixed points, while passing through the pinch point.}
\label{LZEfullphase}
\end{figure}

Figure \ref{LZEfullphase} shows trajectories on the $L_z-E$ plane, obtained from solving the slow-manifold equations with $(L_x,L_y) \equiv (0,0)$; the $L_z$ and $E$ axes are normalized by $I_e$ for the prolate spheroid chosen.
The pair of red parabolae, demarcating the physically accessible domain, arise from the inequalities, $(L_z/I_e)^2<2(E/I_e)$ and $(L_z/I_e)^2>2(I_a/I_e)(E/I_e)$, and are projections of the (hyper)\,paraboloids in the full four-dimensional manifold. The stable node\,(purple `o')  corresponds to $(-1/2,1/8)$, and the saddle point\,(purple `*') to $(I_a/I_e)(-1/2,1/8)$; the corresponding inequalities and fixed-point coordinates for an oblate spheroid may be obtained by interchanging $I_e$ and $I_a$ above. The two parabolae touch at the pinch point $(L_z,E) \equiv (0,0)$, which divides the physical domain into positive and negative-$L_z$ halves. While the trajectory topology in the negative-$L_z$ half is non-trivial due to the two fixed points, almost all trajectories nevertheless converge to the stable node for long times. The only exception is the set of zero measure corresponding to initial conditions on the upper parabola, which serves as the stable manifold of the saddle point. In contrast, the positive-$L_z$ half consists entirely of trajectories that exhibit a monotonic decrease of both $L_z$ and $E$ along roughly parabolic paths, eventually terminating at the origin. The two halves of the $L_z-E$ plane are thus disconnected, this being an artifact of the assumption inherent in the derivation of the slow-manifold equations - that of the time scale associated with an Euler-orbit ($O(I_e/L\dot{\gamma}^{-1})$) being far smaller than that associated with the tilting of the Euler cone ($O(St\dot{\gamma}^{-1})$). This is no longer true when $L\ll I_e$ which, for $L_x=L_y=0$, occurs when $L_z\ll I_e$; taken together with the bounding parabolae, this implies $E \ll I_e$, so that the slow-manifold description breaks down in the neighborhood of the pinch point. 

\begin{figure}
\subfloat[\label{LZEfullphasefast} $L_z-E$ phase plot; $(\kappa=2)$ $(\phi_0,\dot{\theta}_0,\dot{\phi}_0,\dot{\psi}_0)\equiv(0,0, L_{z0}/I_1,L_{z0}\cos{\theta_0}(1/I_e-1/I_a))$ ]
{\includegraphics[scale=0.45]{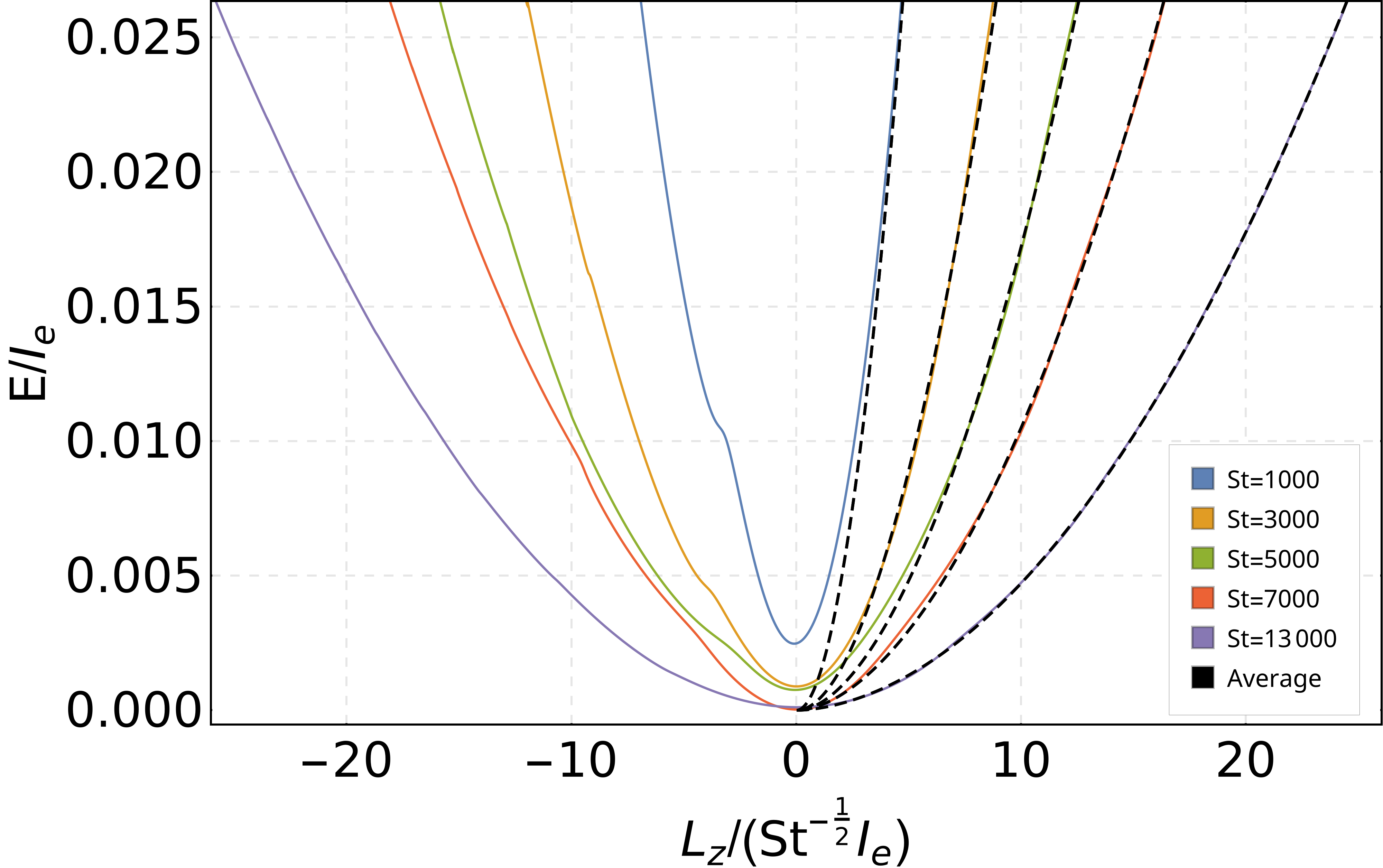}}
\subfloat[\label{LZEfullphasepinch} Dynamics near the pinch point; $(\kappa=2)$ $(\phi_0,\dot{\theta}_0,\dot{\phi}_0,\dot{\psi}_0)\equiv(0,0,L_{z0}/I_1,L_{z0}\cos{\theta_0}(1/I_e-1/I_a))$]{\includegraphics[scale=0.45]{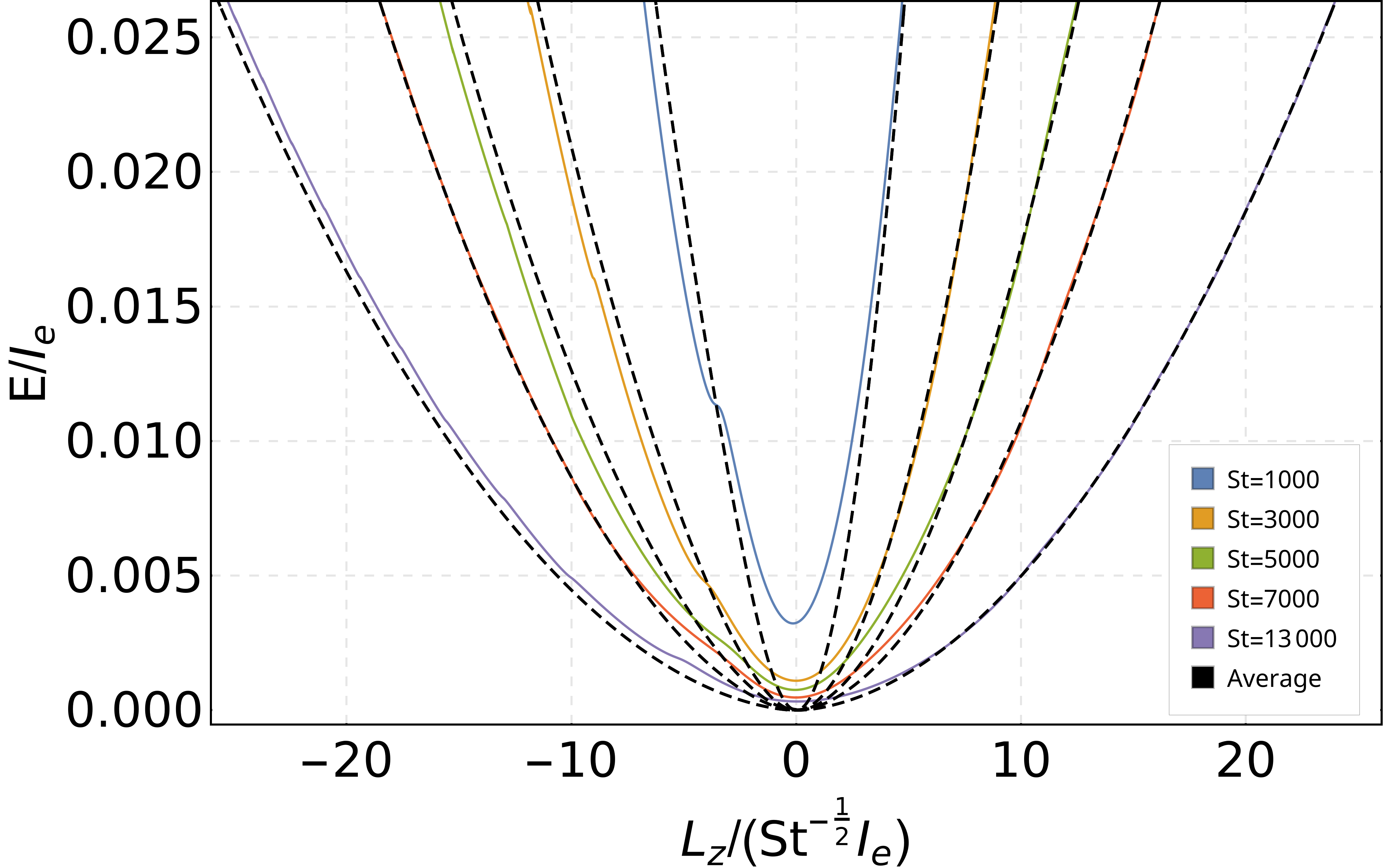}}
\caption{Comparison between slow manifold and full solution for trajectories with $(L_{x0},L_{y0})\equiv(0,0)$ , as well as for trajectories with small $L_{x0}$ and $L_{y0}$. Figure (\ref{LZEfullphasepinch}) shows the complete $L_z-E$ plot illustrating the good agreement across all four sets of trajectories for most of the plane. Figure (\ref{LZEfullphasefast}) contrasts the behavior near the pinch point. Slow manifold trajectories with $(L_{x0},L_{y0})\equiv(0,0)$ and $L_{z0}>0$ cannot cross the pinch point $(L_z,E)\equiv(0,0)$, and deviate from the full solution close to it. For trajectories with small but non-zero $L_{x0}$ and $L_{y0}$, good agreement is observed throughout.}
\end{figure}

Physically, the pinch point corresponds to the instantaneous state of rest, as the spheroid transitions from a rotation opposed to the shear, to one that is in the same sense. Figure \ref{LZEfullphasefast} illustrates the above breakdown by showing the comparison between trajectories obtained from an integration of the full system of equations, and those obtained from integrating (\ref{lxsmain}-\ref{nsmain}) with initial conditions confined to the $L_z-E$ plane. While the slow-manifold trajectories grind to a halt in the vicinity of the pinch point, the full-solution trajectories deviate from the slow-manifold ones, and cross over from the positive to the negative $L_z$-half, eventually converging to the stable node. For large $St$, this crossover occurs in an $O(St^{-\frac{1}{2}})$ neighborhood of the pinch point, the reason being discussed in section \ref{DAC} in the context of a spheroid starting from rest. Figure \ref{LZEfullphasepinch} shows the projected slow-manifold and full-solution trajectories, again in the neighborhood of the pinch point, but for small but finite initial values for $L_x$ and $L_y$. The slow-manifold trajectories do crossover from one $L_z$-half to another, this being possible due to $L_x$ and $L_y$ being non-zero. Further, the slow-manifold and full solution trajectories now agree well, over their entire length, for sufficiently large $St$. 

\begin{figure}
	\subfloat[$L_z-E$ phase plot; ($\kappa=2$) $(L_{x0},L_{y0})\equiv(0,0)$; 
	 $(\phi_0,\dot{\theta}_0,\dot{\phi}_0,\dot{\psi}_0)\equiv(0,0,L_{z0}/I_1,L_{z0}\cos{\theta_0}(1/I_e-1/I_a))$.  ]{\includegraphics[scale=0.54]{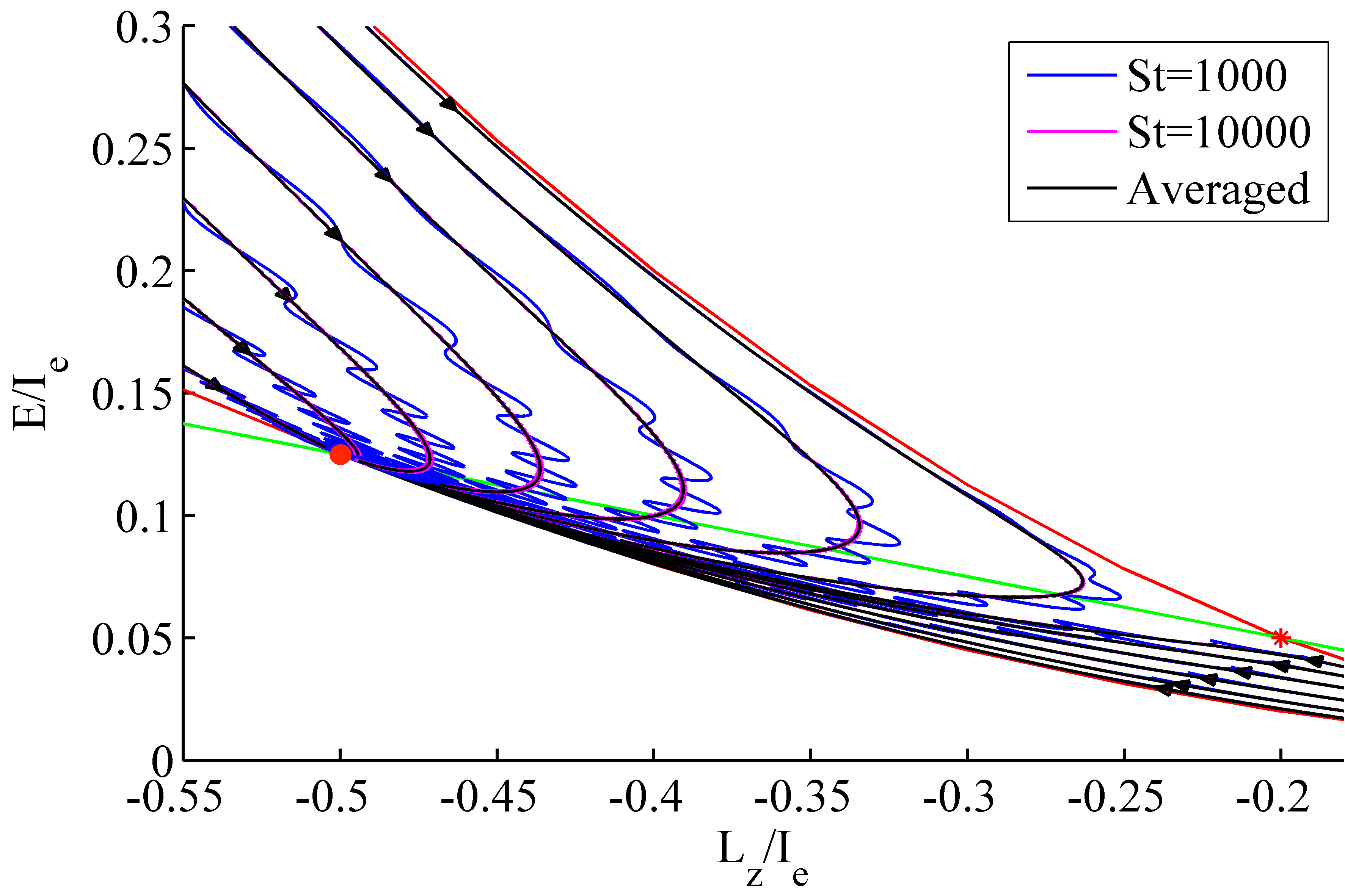}}
	\subfloat[$L_z-E$ phase plot; ($\kappa=3.33$) $(L_{x0},L_{y0})\equiv(0,0)$; 
	$(\phi_0,\dot{\theta}_0,\dot{\phi}_0,\dot{\psi}_0)\equiv(0,0,L_{z0}/I_1,L_{z0}\cos{\theta_0}(1/I_e-1/I_a))$.]{\includegraphics[scale=0.52]{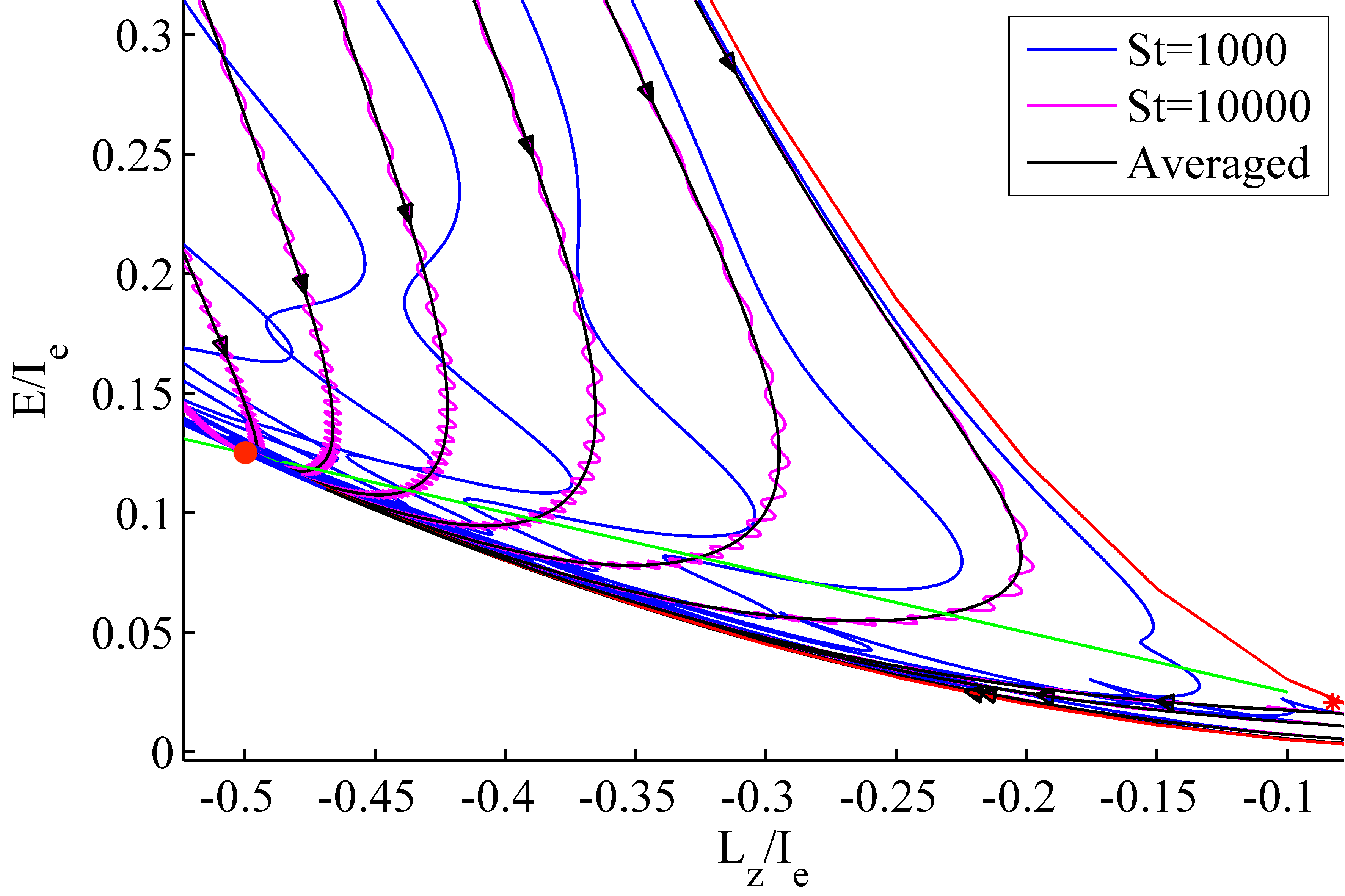}}\\
	\subfloat[$L_z-E$ phase plot; ($\kappa=0.5$) $(L_{x0},L_{y0})\equiv(0,0)$; 
	$(\phi_0,\dot{\theta}_0,\dot{\phi}_0,\dot{\psi}_0)\equiv(0,0,L_{z0}/I_1,L_{z0}\cos{\theta_0}(1/I_e-1/I_a))$.]{\includegraphics[scale=0.55]{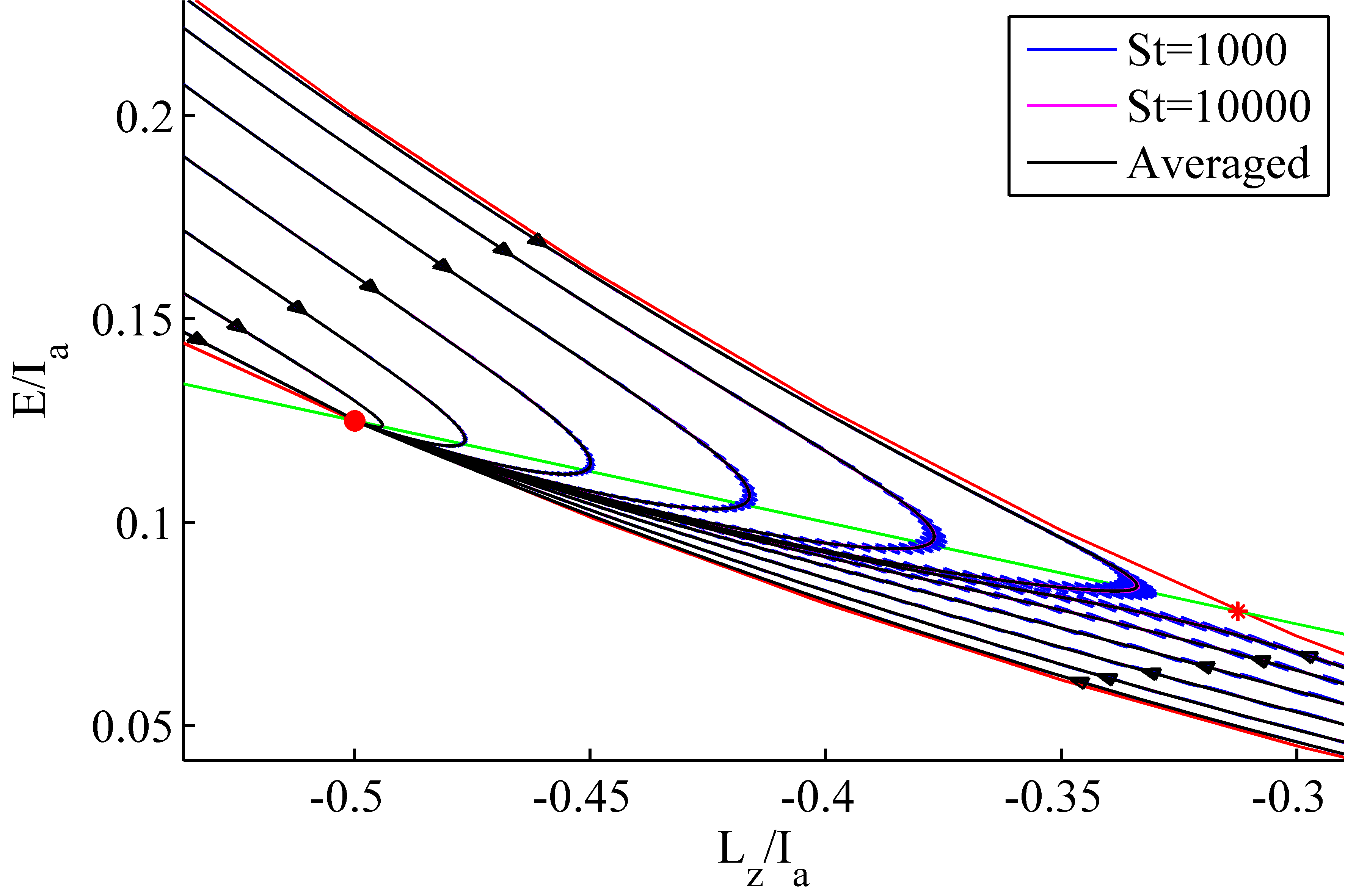}}
	\subfloat[$L_z-E$ phase plot; ($\kappa=0.3$) $(L_{x0},L_{y0})\equiv(0,0)$; 
	$(\phi_0,\dot{\theta}_0,\dot{\phi}_0,\dot{\psi}_0)\equiv(0,0,L_{z0}/I_1,L_{z0}\cos{\theta_0}(1/I_e-1/I_a))$.]{\includegraphics[scale=0.55]{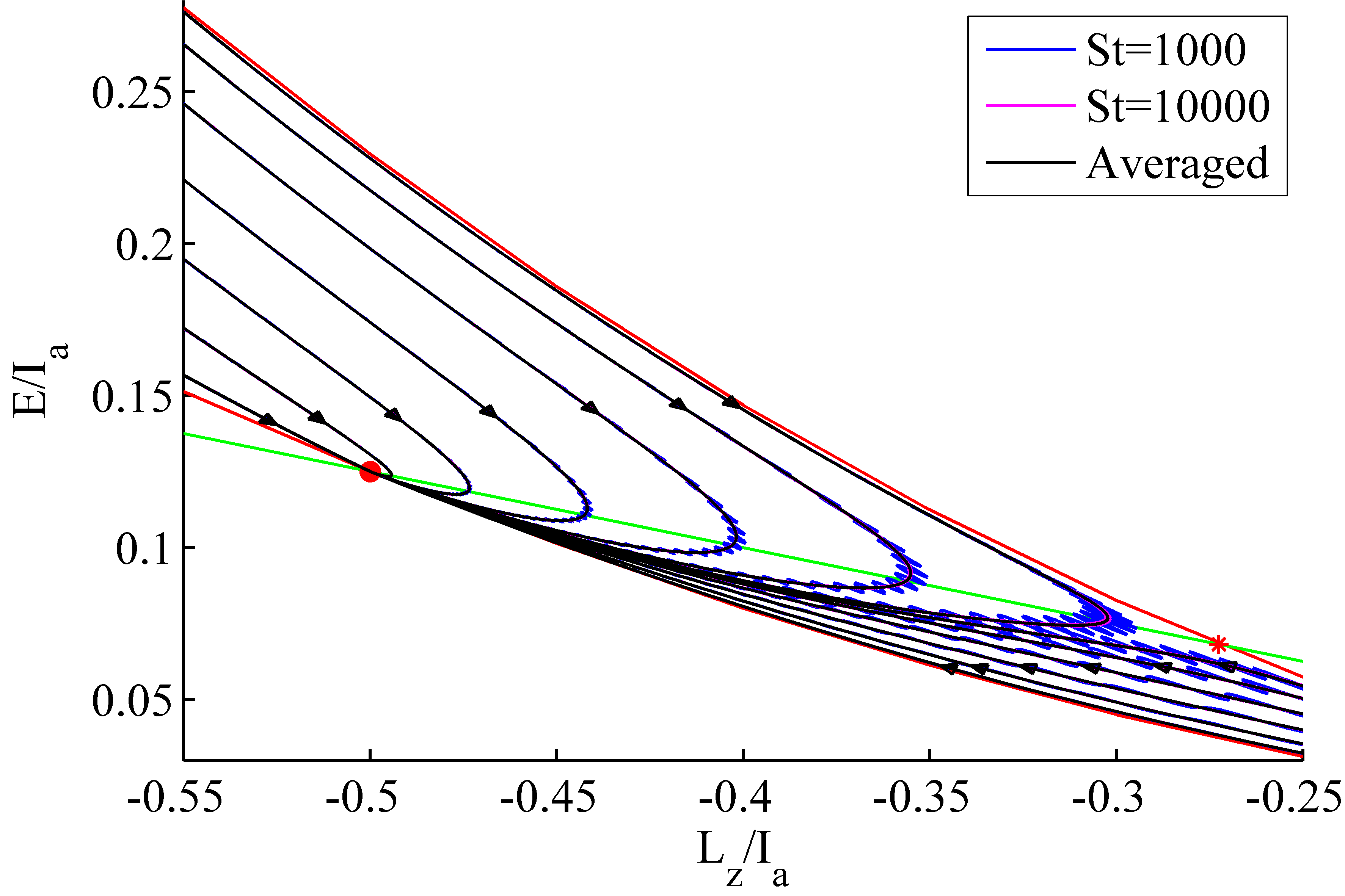}}
	\caption{$L_z-E$ phase plots for spheroids having different aspect ratios showing the comparison between slow-manifold (Averaged) with full solution, for $St = 1000, 10000$. The red * denotes the saddle point and red o denotes the stable node; the green line which passes through the fixed points, is $L_z+4E=0$. Phase plots for prolate and oblate spheroids of all aspect ratios are qualitatively similar. For a fixed $St$, deviation from the slow manifold approximation decreases as $\kappa$ approaches unity.}
	\label{splze}
\end{figure}

\begin{figure}
	\subfloat[$L_z$ vs $t/St^*$; $St^*=2.071$,\newline $(L_{x0},L_{y0},L_{z0},E_{0})\equiv(0,0,-1,0.6)$]{\includegraphics[scale=0.55]{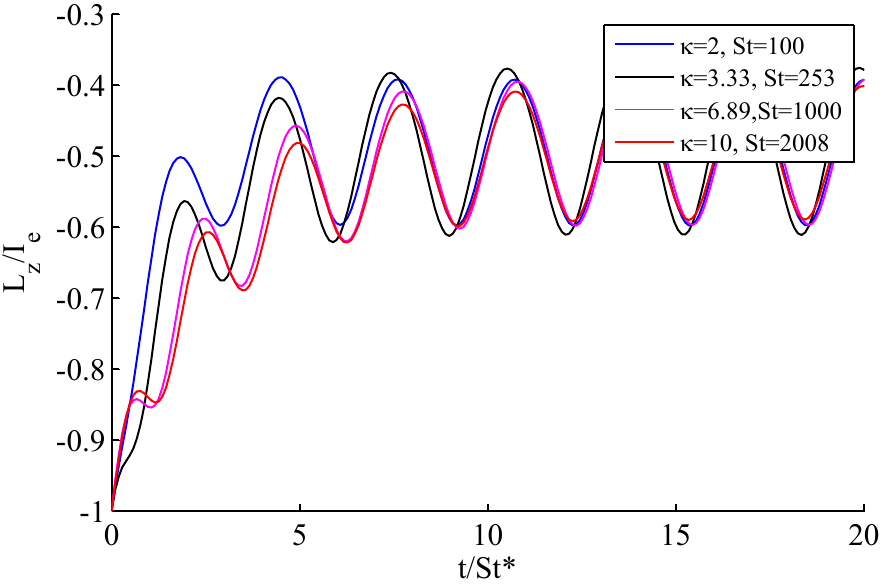}}
	\subfloat[$E$ vs $t/St^*$; $St^*=2.071$, \newline$(L_{x0},L_{y0},L_{z0},E_{0})\equiv(0,0,-1,0.6)$]{\includegraphics[scale=0.55]{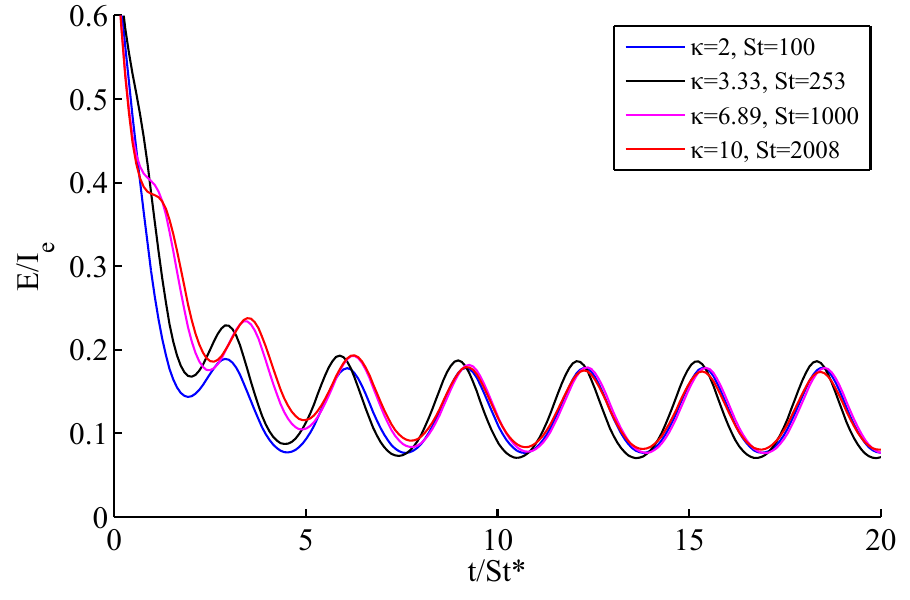}}
	\caption{The evolution of the normalized angular momentum and energy for spheroids with identical $St^*$, but varying $\kappa$ and $St$, is shown. The long-time evolution is seen to be a function of $St^*$ alone. }
	\label{steff}
\end{figure}

Figure \ref{splze} shows a comparison of slow-manifold and (projected)\,full-solution trajectories, on the $L_z-E$ phase plane, for different $St$ and $\kappa$. 
Fewer trajectories are shown compared to Figure \ref{LZEfullphase}, with the $St$ chosen large enough\,($\gtrsim 500$), so as to avoid the overlap of neighboring full-solution trajectories. 
While the slow-manifold approximation works well in all cases, one nevertheless notes, for a given large $St$, that increasing shape anisotropy leads to a greater deviation from the slow-manifold predictions. This is seen for $(\kappa,St)\equiv(3.33,1000)$ where the full-solution trajectory oscillations have a greater amplitude and a lower frequency. Both these effects may be attributed to the nominal Stokes number not accounting for the $\kappa$-dependence of the moment of inertia and resistance tensors. This may be addressed by defining an effective Stokes number $St^*=I_a St/(8\pi X_c)$ \cite{challabotla2015}, as is seen from Figure \ref{steff} which compares the normalized temporal evolution\,(full solution trajectories) of $L_z$ and $E$ for spheroids with different $\kappa$ and $St$, such that $St^*$ remains the same. Although the plots in Figure \ref{steff}a and b do not overlap completely for small times, the approach to equilibrium for similar initial conditions is similar despite $St$ changing by an order of magnitude; further, the comparison for long times remains good regardless of $\kappa$. It is worth noting that $St^* \approx St\ln \kappa/\kappa^2$ for $\kappa \rightarrow \infty$, while $St^* \approx St\kappa$ for $\kappa \rightarrow 0$, and therefore, for a given eccentricity, $St^*$ for oblate spheroids is greater than that for the complementary prolate ones. The asymmetry in $\kappa$-scaling between the slender-fiber and flat-disk limits also explains the contrasting levels of agreement for $\kappa = 3.33$ and $\kappa = 0.3$, in figure \ref{splze}. Note that weakening particle inertial effects, with increasing $\kappa$, were already seen in Figure \ref{delTvslogSt} in the context of the time period comparison.

\begin{figure}
	\subfloat[\label{lzen0} $L_z-E$ projection for varying $L_{x0}/L_{y0}$;\newline ($\kappa=2$) $(L_0/I_e,E_0/I_e)\equiv(2/3,2/9)$]{\includegraphics[scale=0.55]{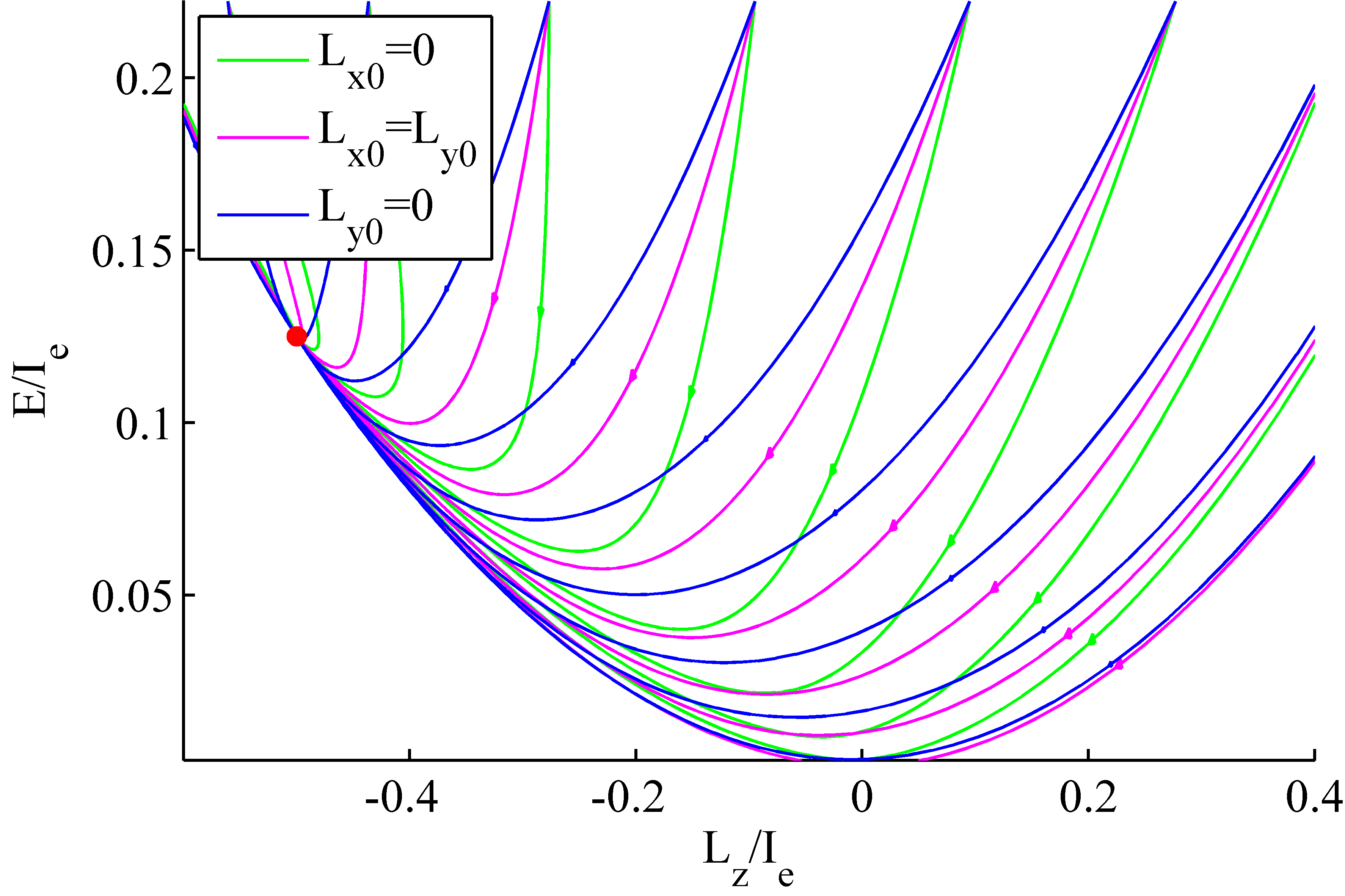}}		\subfloat[\label{lzlxyn0} $L_z-(L_x^2+L_y^2)^{1/2}$ projection for varying $L_{x0}/L_{y0}$;\newline ($\kappa=2$) $(L_0/I_e,E_0/I_e)\equiv(2/3,2/9)$]{\includegraphics[scale=0.55]{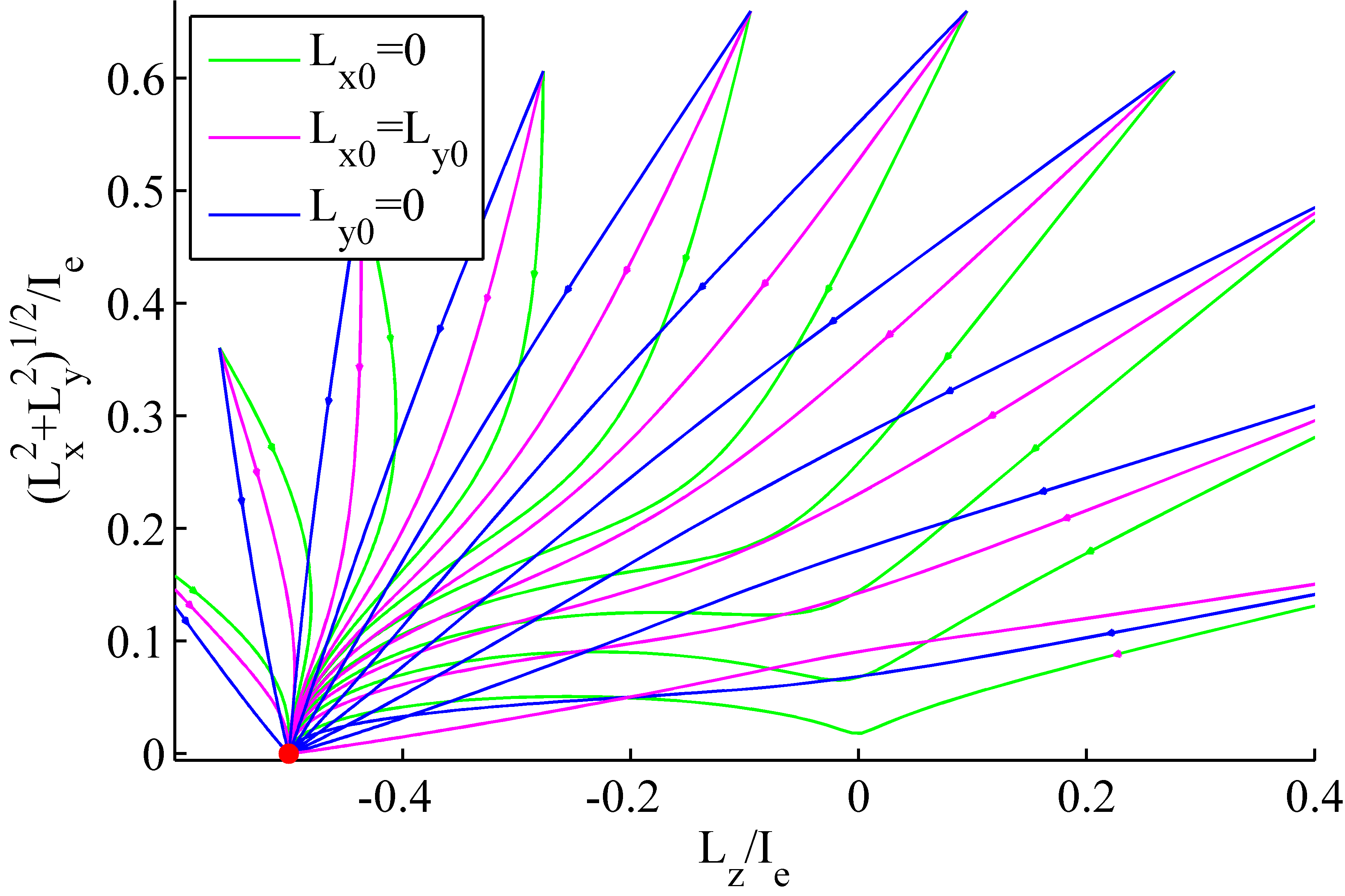}}\\
	\subfloat[\label{lzen0_5}$L_z-E$ projection for varying $L_{x0}/L_{y0}$;\newline ($\kappa=2$) $(L_0/I_e,E_0/I_e)\equiv(2/3,0.305)$]{\includegraphics[scale=0.55]{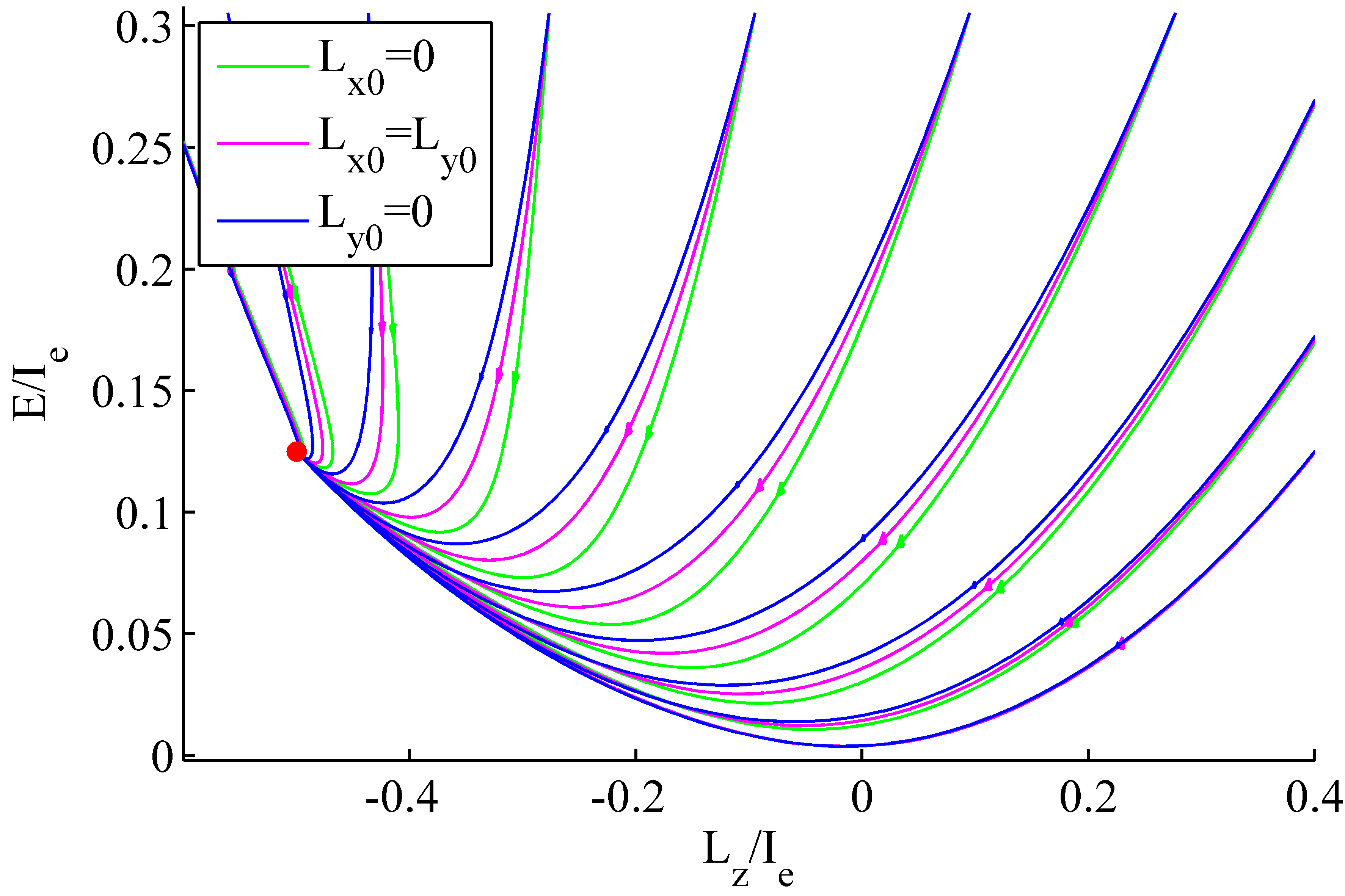}}
	\subfloat[\label{lzlxyn0_5}$L_z-(L_x^2+L_y^2)^{1/2}$ projection for varying $L_{x0}/L_{y0}$;\newline ($\kappa=2$) $(L_0/I_e,E_0/I_e)\equiv(2/3,0.305)$]{\includegraphics[scale=0.55]{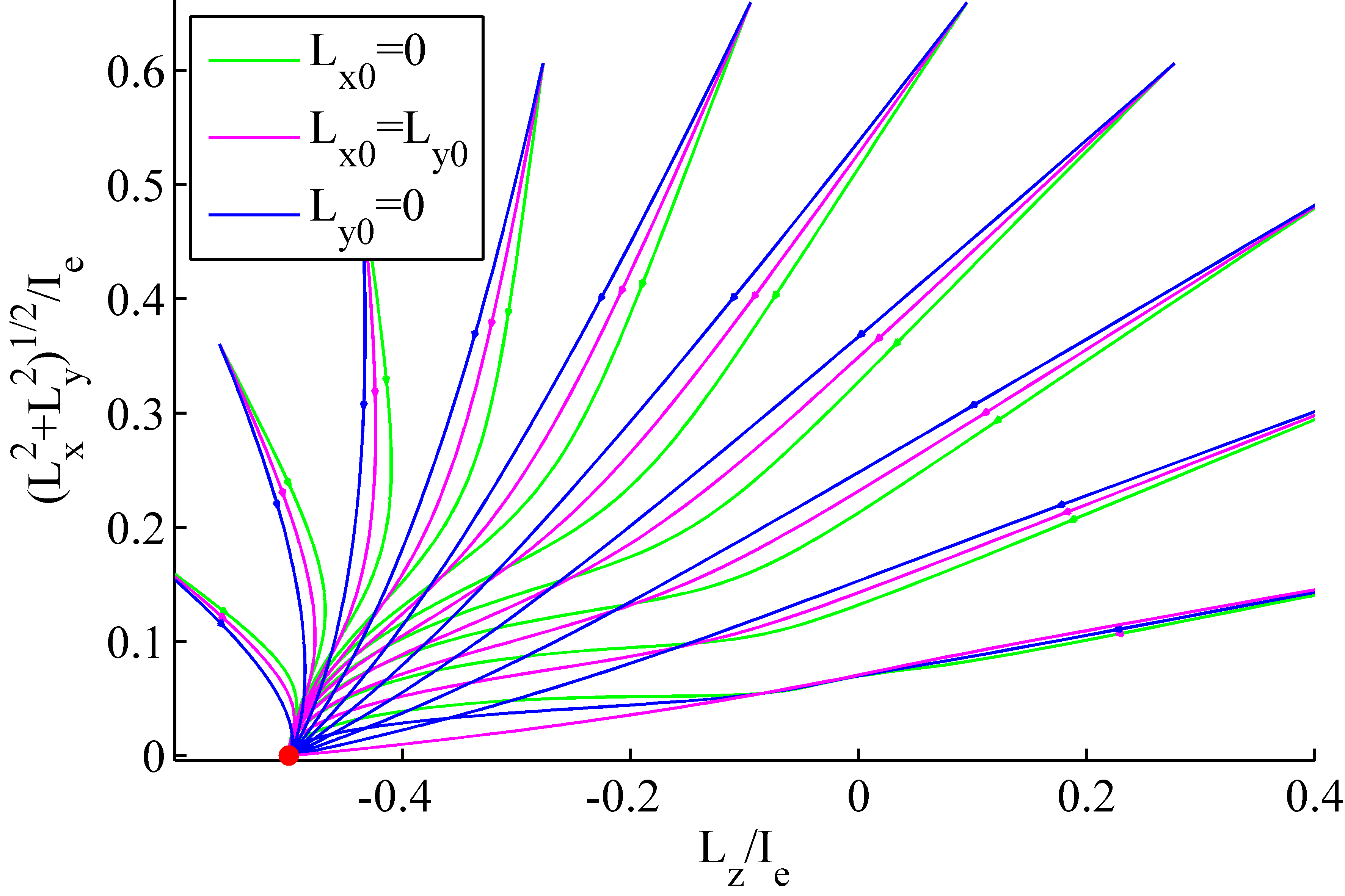}}\\ 
	\subfloat[\label{lzen0_87b} $L_z-E$ projection for varying $L_{x0}/L_{y0}$;\newline ($\kappa=2$) $(L_0/I_e,E_0/I_e)\equiv(2/3,0.472)$]{\includegraphics[scale=0.55]{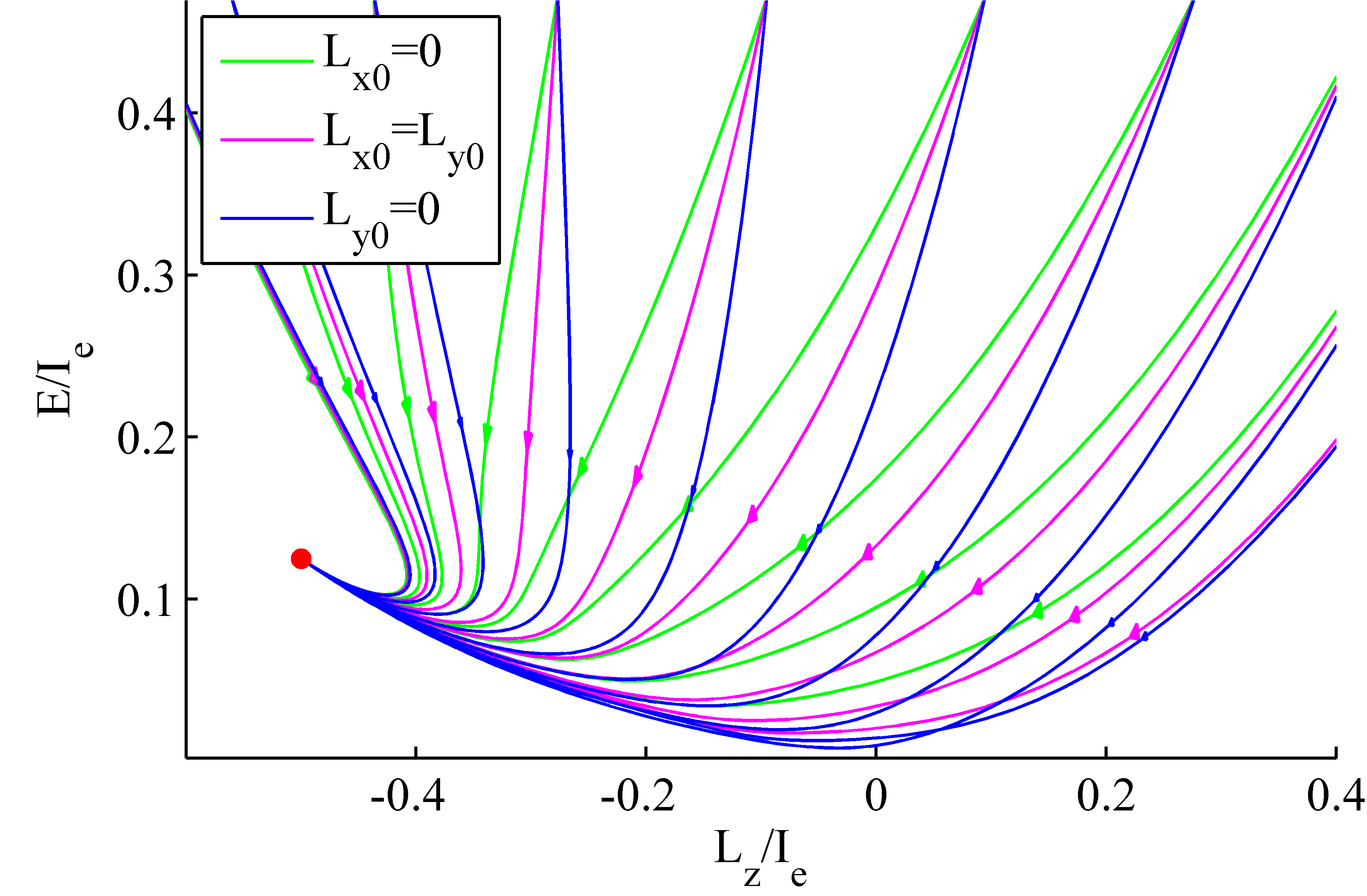}}
	\subfloat[\label{lzlxyn0_87} $L_z-(L_x^2+L_y^2)^{1/2}$ projection for varying $L_{x0}/L_{y0}$;\newline ($\kappa=2$) $(L_0/I_e,E_0/I_e)\equiv(2/3,0.472)$]{\includegraphics[scale=0.55]{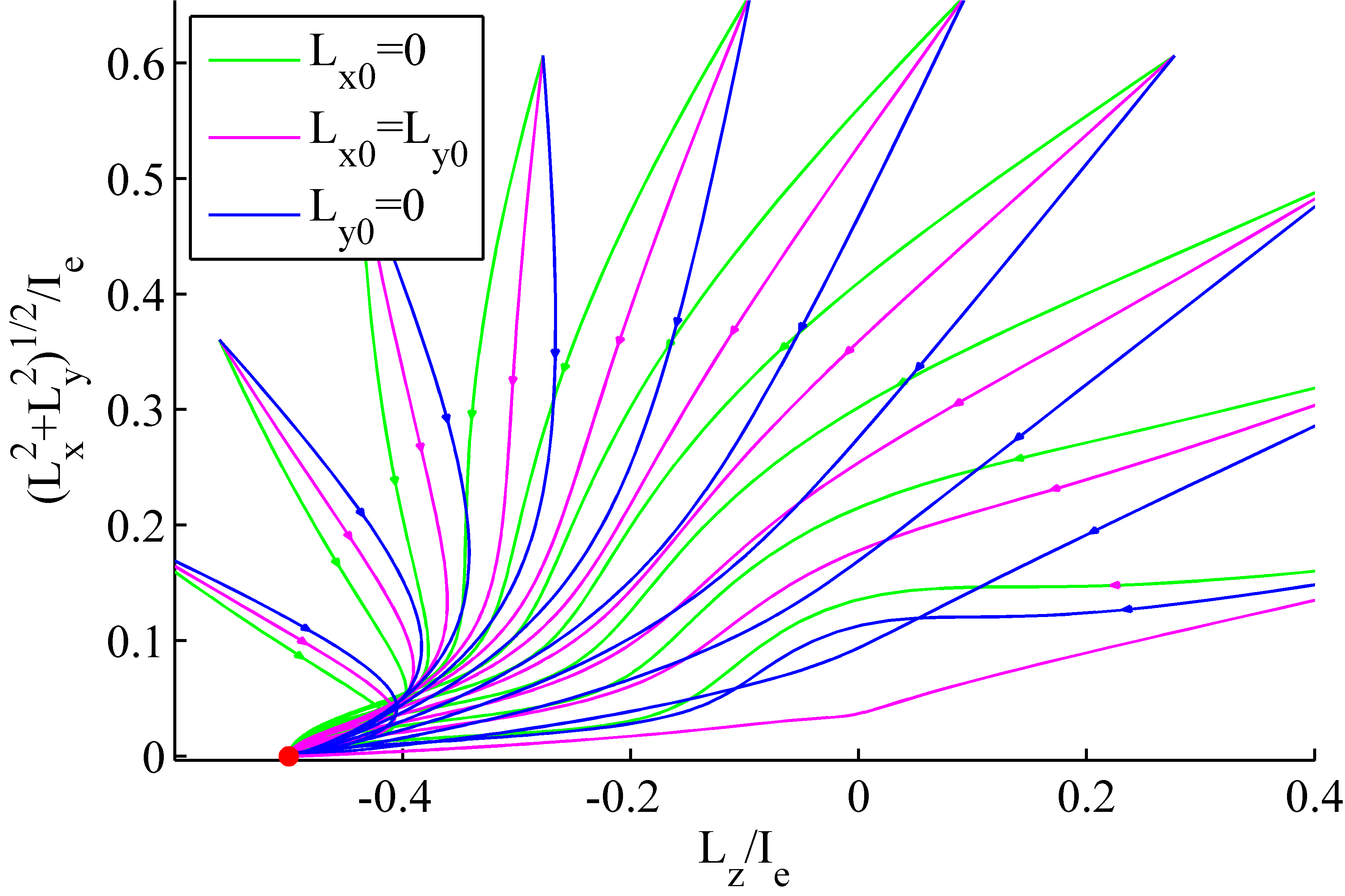}}
	\caption{ \label{finitelxly} The figures show $L_z-E$ (left) and $L_z-(L_x^2+L_y^2)^{1/2}$ (right) projections of slow manifold trajectories covering the entire range of initial conditions for a given initial magnitude of angular momentum ($L_0$). From top to bottom, $\eta_0$ is increased from zero to close to its maximum value ($L_0/I_a$), with each figure showing curves that correspond to three different values of $L_{x0}/L_{y0} \sim (0,1$ and $\infty)$. Regardless of initial condition, the eventual state reached is $((L_x^2+L_y^2)^{1/2},L_z,E)\equiv(0,-I_e/2,I_e/8)$ for a prolate spheroid. Further, the dynamics are qualitatively unaffected by the varying $L_{x0}/L_{y0}$ and $I_a\eta_0/L_0$. }
\end{figure}

\begin{figure}
\subfloat[\label{LEvarly0S}$L-E$ phase plot; ($\kappa=2$) $L_{x0}=0$ ]{\includegraphics[scale=0.55]{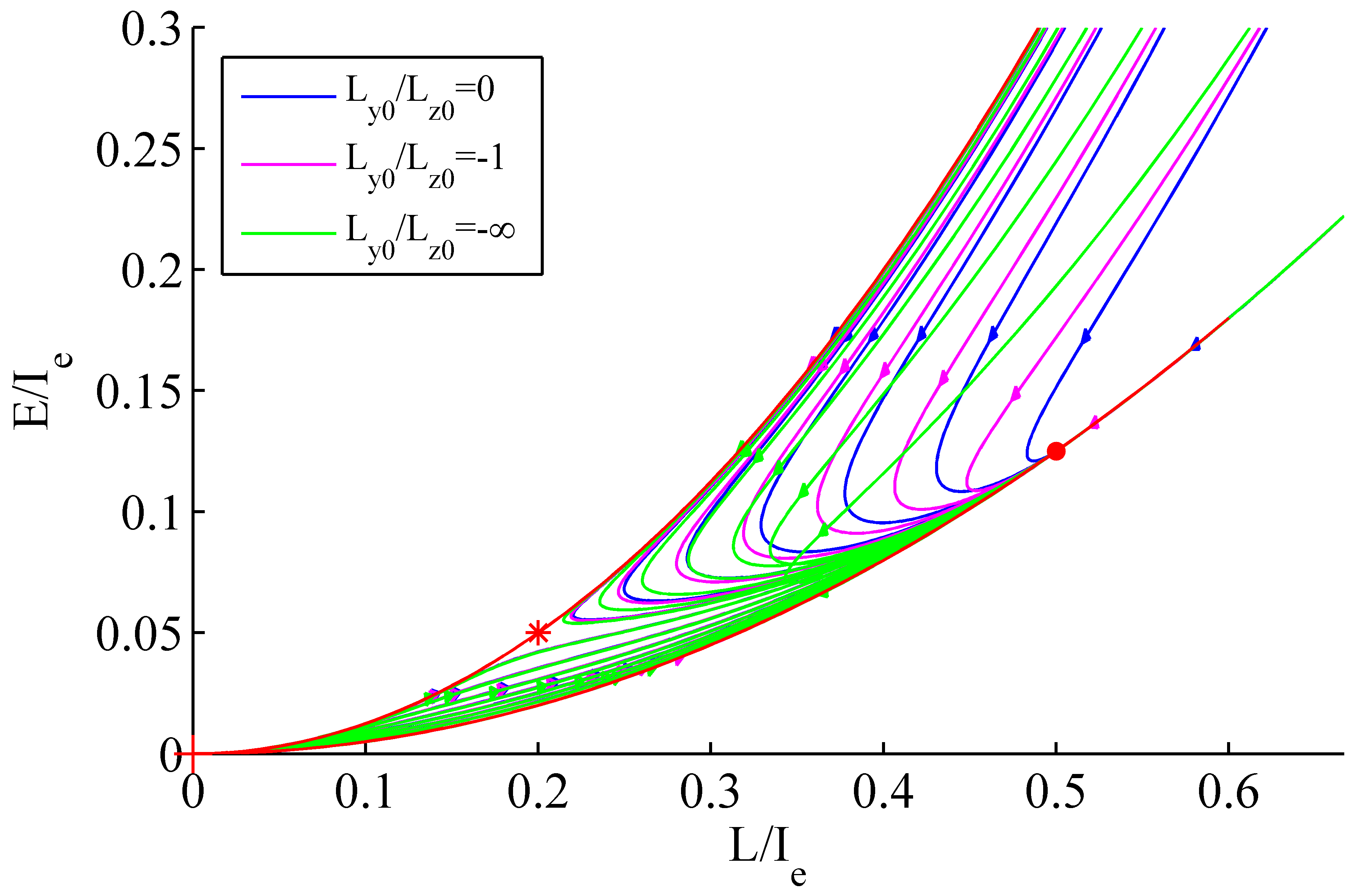}}
\subfloat[\label{LEvarlx0S}$L-E$ phase plot; ($\kappa=2$) $L_{y0}=0$]{\includegraphics[scale=0.55]{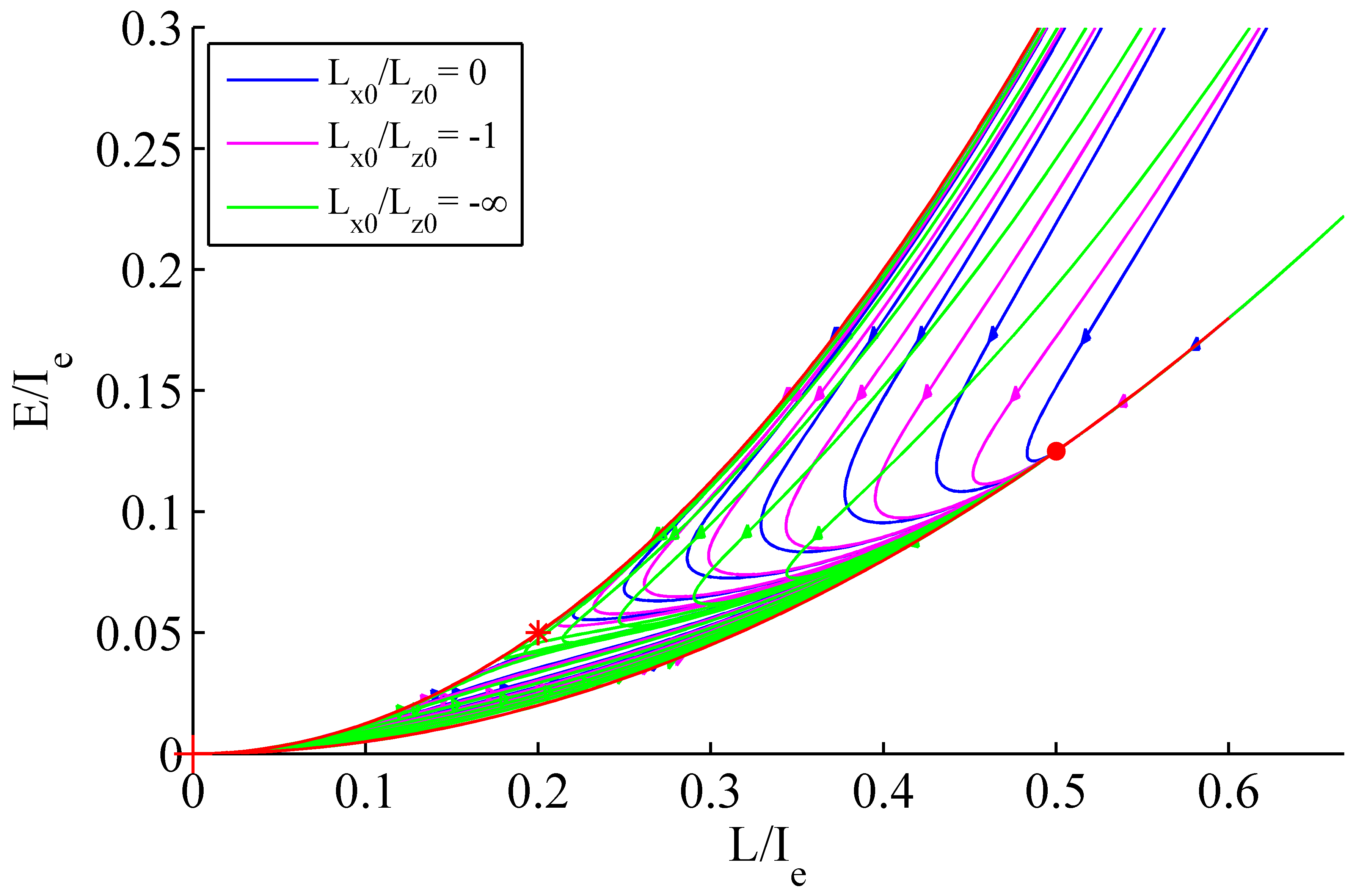}}
\caption{\label{LEplot}The $L-E$ phase plots characterize the dependence of the slow-manifold trajectories on $(L^{2}_{x0}+L^{2}_{y0})^{1/2}/L_{z0}$; the two limiting cases shown in (a) $L_{x0}=0$ and (b) $L_{y0}=0$ point to the weak influence of $L_{x0}/L_{y0}$ on the overall trajectory topology.}
\end{figure}

Figure \ref{finitelxly} studies the effect of non-zero $L_{x0}$ and $L_{y0}$ on the large-$St$ orientation dynamics, by considering the projections of slow-manifold trajectories in both the $L_z-E$\,(subfigures \ref{finitelxly}a, c and e) and $L_z-(L_x^2+L_y^2)^{1/2}$\,(subfigures \ref{finitelxly}b, d and f) planes. Each of these subfigures corresponds to a given $(L_0,E_0)$, with the different sets of trajectories corresponding to different values of $L_{x0}/L_{y0}$. In figures \ref{lzen0}, \ref{lzen0_5} and \ref{lzen0_87b}, all trajectories are seen to eventually converge to $(L_z/I_e,E/I_e)\equiv(-1/2,1/8)$, the stable node identified earlier, regardless of initial conditions and the ratio $L_{x0}/L_{y0}$; each trajectory being bounded on either side by the limiting cases $L_x=0\,(L_{x0}/L_{y0} =0)$ and $L_y=0\,(L_{x0}/L_{y0} = \infty)$. Correspondingly, in figures \ref{lzlxyn0}, \ref{lzlxyn0_5} and \ref{lzlxyn0_87}, all trajectories converge to $((L_x^2+L_y^2)^{1/2}/I_e,L_z/I_e)\equiv(0,-1/2)$. 
In addition to $L_{x0}/L_{y0}$, the effect of the ratio $(L^2_{x0}+L^2_{y0})^{1/2}/L_{z0}$ must also be examined to fully appreciate the effect of $L_x$ and $L_y$ being non-zero, and this is done based via the $L-E$ phase plots in figures \ref{LEvarly0S} and \ref{LEvarlx0S}. Since $L >0$, and since the slow manifold trajectories are bounded between the parabolae $L^2<2E I_e$ and $L^2>2E I_a$, the phase plots in these figures resemble the positive half of the one in figure \ref{LZEfullphase}. 
The trajectory topology evidently remains unchanged regardless of $(L^{2}_{x0}+L^{2}_{y0})^{1/2}/L_{z0}$. 


\section{Large-$St$ Dynamics in simple shear flow: Triaxial Ellipsoids}\label{lsde}
Proceeding along lines similar to that for the spheroid in the earlier section, the space-fixed components of the torque are obtained from (\ref{ultima}), and equated to the rates of change of the corresponding component of the angular momentum; the rate of change of kinetic energy being $\overrightarrow{T}_{viscous}\cdot \overrightarrow{\omega}$, the power dissipated by the viscous torque. The rates of change of the slow variables are therefore given in the following compact form: 
\begin{align}\label{levo}
\dot{L_i}&=-\frac{8\pi}{St}[t_{ji} X_j(\omega_j-\omega_{j,J})],\\\label{vispow}
\dot{E}&=-\frac{8\pi}{St}[X_j(\omega_j-\omega_{j,J})\omega_j].\\
\intertext{The indices $i$ and $j$ run from 1 to 3 in space-fixed coordinates, with the $X_i$'s referring to the principal resistances defined in section \ref{gove}, and $t_{ji}$ defined by (\ref{tiju}). (\ref{levo}) and (\ref{vispow}) 
are the analogs of (\ref{levosx}-\ref{nos}) for the spheroid case examined in the previous section. Next, using $I_i\omega_i = L_j t_{ji}$ to express the RHS's in (\ref{levo}-\ref{vispow}) in terms of the angular momenta, one obtains:}
\label{lgen}\dot{L_l}&=-\frac{8\pi}{St}[\frac{X_i}{I_i}(L_k t_{il}t_{ik})+\epsilon_{ijk}\frac{X_i}{l_j^2+l_k^2}(l_k^2 t_{il}t_{j1}t_{k2})],\\\label{egen}
\dot{E}&=-\frac{8\pi}{St}[\frac{X_i}{I_i^2}(L^2_l t^2_{il})+L_l\epsilon_{ijk}\frac{X_i}{I_i(l_j^2+l_k^2)}(l_k^2 t_{il} t_{j1}t_{k2})],
\end{align}
where $l_i = a$, $b$ and $c$ for $i = 1-3$.
As was the case with (\ref{lxs2}-\ref{ns2}) for the spheroid, the system defined by (\ref{lgen}) and (\ref{egen}) for a triaxial ellipsoid is not closed, since the $t_{ij}$'s continue to be functions of the fast variables\,(the Euler angles). For the spheroid, the method of multiple scales, together with the symmetric Euler-top solution, was used at this stage to eliminate this dependence on the fast variables. However, as mentioned earlier, this formalism is inapplicable for an ellipsoid because the Euler-top dynamics described in section \ref{symtop} has in general a quasi-periodic character, owing to the incommensurability of the spin-nutation components with the precessional one. Thus, there is no single period over which the fast-time dependence may be integrated over, as required in the method of multiple scales. The method of averaging \cite{mitropolsky1967,grebenikov1965}, on the other hand, permits a fairly general variation of the fast variables with time, and we use this in what follows.

Based on (\ref{lgen}) and (\ref{egen}), the evolution of any slow variable $S$\,(say) is of the the general form:
\begin{align}\label{Ssys}
\frac{dS}{dt} = \frac{1}{St}F(\{S\},f(t;T(\{S\}),T'(\{S\}))), 
\end{align}
where $f$ denotes the functional dependence on the fast time variables, and $\{S\}$ the set of (four)\,slow variables. 
As mentioned in the previous paragraph, the fast dynamics is characterized by a pair of incommensurate time periods\,($T$ and $T'$), the second and third arguments of $f$ in (\ref{Ssys}), which are themselves functions of the slow variables for large but finite $St$. From (\ref{lgen}-\ref{egen}), $f$ involves products of $t_{ij}$'s which, on using $t^T_{ij}=\alpha_{ij}t^T_{ij,LO}$, may be expressed in terms of the Euler angles $(\theta_{LO},\phi_{LO},\psi_{LO})$ defining the orientation of the angular-momentum-aligned coordinate system. Closed-form expressions for these angles were given in section \ref{symtop} - see (\ref{eq:nut}), (\ref{eq:rota}) and (\ref{eq:prec}-\ref{eq:prec2}). 
Based on these expressions, all relevant terms in $F(\{S\},f_i(t;T,T'))$ are seen to be of the general form $G(\{S\})g(t;T)h(t;T')$ where $g(t;T)\in\{\cos{(l\theta_{LO})}\cos{(m\psi_{LO})}, \sin{(l\theta_{LO})}\cos{(m\psi_{LO})}, \cos{(l\theta_{LO})}\sin{(m\psi_{LO})}, \sin{(l\theta_{LO})}\sin{(m\psi_{LO})} \}$ and $h(t;T') \in \{\cos{(k \phi_{LO})},\sin{(k \phi_{LO})}\}$ with $k,l,m=0,1,2,3$. Thus, one may write (\ref{Ssys}) in the more specific form:
\begin{align} 
\frac{dS}{dt}&=\frac{1}{St}\sum^M_{j=1}G_j(\{S\})g_j(t;T(\{S\}))h_j(t;T'(\{S\})), \label{sumofsines}
\end{align}
for some positive integer $M$. To obtain the closed system of averaged equations governing $\{S\}$, one now averages both sides of (\ref{Ssys}) over a time much greater than $T$ and $T'$\,(based on the argument for the spheroid, at the beginning of section \ref{lsds}, both periods may be regarded as being $O(\dot{\gamma}^{-1}$)), but much less than $O(St\dot{\gamma}^{-1})$. The latter requirement ensures that {$S$} remains virtually unchanged over the time of averaging; the time integral involved in the averaging therefore applies only to the fast variables. Assuming the time of averaging to be a large multiple of one of the fast time periods, say $nT$, with $1 \ll n \ll St$, and denoting the average by $\langle \langle f_i(t;T(\{S\}),T'(\{S\}))\rangle \rangle$, one has:
\begin{equation}\label{avgdef}
\langle \langle f_i(t;T(\{S\}),T'(\{S\}))\rangle \rangle=\lim_{n\rightarrow\infty}\frac{1}{nT}\int^{nT}_0 {f_i(t;T(\{S\}),T'(\{S\}))\text{d}t},
\end{equation}
 for large $St$. Double angular brackets are used, in contrast to the spheroid, to denote averaging over the two incommensurate components of the fast-time dynamics. Using (\ref{avgdef}) in (\ref{sumofsines}), the averaged equation for $S$ may be written as:
\begin{align}
 \frac{dS}{dt_2}&= \sum^M_{j=1}G_j(\{S\})\langle \langle g_j(t;T(\{S\}))h_j(t;T'(\{S\}))\rangle \rangle, \label{avgS:1} \\
 \intertext{where $t_2=t/St$ as before. Further, since $\langle \langle g_j(t;T(S))h_j(t;T'(S))\rangle \rangle $ is only a function of $S$, say $H_j(S)$, one may write (\ref{avgS:1}) in the form:}
 \frac{dS}{dt_2}&=\sum^M_{j=1}G_j(S)H_j(S), \label{avgS:2}
\end{align}
so that, pending the evaluation of the averages, (\ref{avgS:2}) may be regarded as a formally closed description of the dynamics on the slow manifold. 

The averages involved in the aforementioned general form may be evaluated using (\ref{avgdef}) as follows. Considering one of the terms in the summation in (\ref{avgS:2}),
\begin{align}
\langle \langle g(t;T) h(t;T')\rangle \rangle&=\lim_{n\rightarrow\infty}\frac{1}{nT}\int^{nT}_0 {g(t;T) h(t;T')\text{d}t},\\
 &=\lim_{n\rightarrow\infty}\sum_{i=1}^{n} \frac{1}{nT}\int^{iT}_{(i-1)T} {g(t;T) h(t;T')\text{d}t}.
\intertext{Using the change of variable $\tilde{t}=t-(i-1)T$ for the $i^{th}$ integral in the summation,
}
\langle \langle g(t;T) h(t;T')\rangle \rangle&=\lim_{n\rightarrow\infty}\sum_{i=1}^{n} \frac{1}{nT}\int^{T}_{0} {g(\tilde{t}+(i-1)T;T) h(\tilde{t}+(i-1)T;T')\text{d}\tilde{t}}.
\intertext{On account of the periodicities of $g$ and $h$, $g(\tilde{t}+(i-1)T;T)=g(\tilde{t};T)$ and $h(\tilde{t}+(i-1)T;T')=h([\tilde{t}+(i-1)T]\text{mod}(T') ;T')$, so the average takes the form:}
\langle \langle g(t;T) h(t;T')\rangle \rangle&= \frac{1}{T}\int^{T}_{0} g(\tilde{t};T)\bigg(\lim_{n\rightarrow\infty} \frac{1}{n}\sum_{i=1}^{n} { h([\tilde{t}+(i-1)T]\text{mod}(T');T') }\bigg)\text{d}\tilde{t}.\\
\intertext{Since the incommensurability of $T$ and $T'$ translates to their ratio being irrational, the application of Weyl's equidistribution theorem \cite{korner1988} implies that $[\tilde{t}+(i-1)T]\text{mod}(T')$ samples the interval $[0,T')$ uniformly for sufficiently large $n$, and therefore, $\lim_{n\rightarrow\infty}\frac{1}{n}\sum_{i=1}^{n} { h([\tilde{t}+(i-1)T]\text{mod}(T') ;T')}= \frac{1}{T'}\int^{T'}_0{h(t^*;T')}\text{d}{t^*}$. Thus, the average reduces to the product of independent integrals over $h$ and $g$:}
\langle \langle g(t;T) h(t;T')\rangle \rangle&=\bigg(\frac{1}{T}\int^{T}_{0} g(\tilde{t};T)\text{d}\tilde{t}\bigg)\bigg(\frac{1}{T'}\int^{T'}_{0} h(t^*;T') \text{d}{t^*}\bigg), \\
&=\langle g(t;T)\rangle  \langle h(t;T')\rangle,\label{pdtavg} 
\end{align}
for $T/T'$ being irrational; the exceptions correspond to a set of measure zero on the slow manifold. The above result may now be used to evaluate the fast-variable averages for a triaxial ellipsoid. Thus, defining the slow variables $L^{(0)}_{l}=\langle L_{l}\rangle$, $E^{(0)}=\langle E\rangle$, one obtains:

\begin{align}\label{l0}
\frac{dL^{(0)}_{l}}{dt_2}&=-8\pi[\frac{X_i}{I_i}(L^{(0)}_{k}\langle 
t_{il}t_{ik}\rangle)+\epsilon_{ijk}\frac{X_i}{l_j^2+l_k^2}(l_k^2 \langle t_{il}t_{j1}t_{k2}\rangle)],
\\\frac{dE^{(0)}}{dt_2}&=-8\pi[\frac{X_i}{I_i^2}({L^{(0)}_{l}}^2 \langle t^2_{il}\rangle)+L^{(0)}_{l}\epsilon_{ijk}\frac{X_i}{I_i(l_j^2+l_k^2)}(l_k^2 \langle t_{il}t_{j1}t_{k2}\rangle)]\label{e0}.
\end{align}
The expressions for the averages appearing in (\ref{l0}-\ref{e0}) are provided in Appendix \ref{elavgs}. Therein, all of the averages are shown to reduce to linear combinations of $\langle z^2_i\rangle$ with $z_i=t_{i3,LO}$. Substituting these simplified forms into (\ref{l0}) and (\ref{e0}), one obtains the slow-manifold equations in the following form: 
\begin{align}\label{lxm}
\frac{d{L_x^{(0)}}}{dt_2}&=-8\pi {L_x^{(0)}}[(\frac{X_i}{I_i}\langle z_i^2\rangle)+\frac{{L_z^{(0)}}}{2 {L^{(0)}}^2}X_i(\langle z_i^2\rangle-\frac{\langle z_j^2\rangle( l_k\epsilon_{ijk})^2}{l_j^2+l_k^2})],
\\\label{lym}
\frac{d{L_y^{(0)}}}{dt_2}&=-8\pi {L_y^{(0)}}[(\frac{X_i}{I_i}\langle z_i^2\rangle)+\frac{{L_z^{(0)}}}{2 {L^{(0)}}^2}X_i(\langle z_i^2\rangle-\frac{\langle z_k^2\rangle (l_k\epsilon_{ijk})^2}{l_j^2+l_k^2})],
\\\label{lzm}
\frac{d{L_z^{(0)}}}{dt_2}&=-8\pi [\frac{X_i\epsilon_{ijk}}{2(l_j^2+l_k^2){L^{(0)}}^2}(l_k^2(\langle z_i^2\rangle{L_z^{(0)}}^2+\langle z_j^2\rangle{L_x^{(0)}}^2+\langle z_k^2\rangle{L_y^{(0)}}^2))+{L_z^{(0)}}(\frac{X_i}{I_i}\langle z_i^2\rangle)],
\\\label{em}
\frac{d{E^{(0)}}}{dt_2}&=-8\pi [(\frac{X_i}{I_i^2}\langle z_i^2\rangle{L^{(0)}}^2)+{L_z^{(0)}}\frac{X_i \langle z_i^2\rangle}{I_i})].
\end{align}
When ${L^{(0)}}^2>2E^{(0)}I_2$, the $\langle z_i^2\rangle$ may be obtained using closed-form expressions defining the Euler-top solution in section \ref{symtop}, and are given by:
\begin{align}\label{pz}
\langle z_1^2\rangle =&\frac{I_1}{\delta(I_1\mu+I_3)}\bigg(\frac{E(\mu\delta)}{K(\mu\delta)}+\mu\delta-1 \bigg) ,\\ 
\langle z_2^2\rangle =&\frac{I_2}{\delta(I_1\mu+I_3)}\bigg(1-\frac{E(\mu\delta)}{K(\mu\delta)}\bigg)\bigg(\frac{I_3-I_1}{I_3-I_2}\bigg), \\
\langle z_3^2\rangle=&\frac{I_3}{(I_1\mu+I_3)}\frac{E(\mu\delta)}{K(\mu\delta)},
\end{align}
where $K(\mu\delta)$ and $E(\mu\delta)$ are complete elliptic integrals of the first and second kinds, respectively\cite{gradshteyn2007}; here, $\mu=({2{E^{(0)}}I_3-{L^{(0)}}^2})/({{L^{(0)}}^2-2{E^{(0)
}}I_1})$ and $\delta=({I_2-I_1})/({I_3-I_2})$. When ${L^{(0)}}^2<2EI_2$, the suffixes $1$ and $3$ need to be interchanged in the above expressions. Finally, defining $A={X_i}\langle z_i^2\rangle/I_i$, $B=X_i\langle z_i^2\rangle$, $F={X_i}\langle z_i^2\rangle/{I_i^2}$, $C=[X_i\langle z_j^2\rangle(l_k\epsilon_{ijk})^2]/(l_j^2+l_k^2)$, $D=[X_i\langle z_k^2\rangle(l_k\epsilon_{ijk})^2]/(l_j^2+l_k^2)$, (\ref{lxm}-\ref{em}) may be written in the more succinct form:
\begin{align}\label{lxmain}
\frac{d{L_x^{(0)}}}{dt_2}&=-8\pi {L_x^{(0)}}[A+\frac{{L_z^{(0)}}}{2 {L^{(0)}}^2}(B-C)],\\\label{lymain}
\frac{d{L_y^{(0)}}}{dt_2}&=-8\pi {L_y^{(0)}}[A+\frac{{L_z^{(0)}}}{2 {L^{(0)}}^2}(B-D)],\\\label{lzmain}
\frac{d{L_z^{(0)}}}{dt_2}&=-8\pi [A{L_z^{(0)}}+\frac{1}{2{L^{(0)}}^2}(B{L_z^{(0)}}^2+C{L_x^{(0)}}^2+D{L_y^{(0)}}^2)],\\\label{emain}
\frac{d{E^{(0)}}}{dt_2}&=-8\pi [F{L^{(0)}}^2+A\frac{{L_z^{(0)}}}{2}].
\end{align}
Owing to the dependence of the coefficients on $\sign[{L^{(0)}}^2-2E^{(0)}I_2]$, (\ref{lxmain})-(\ref{emain}) may be interpreted as two separate sets of equations governing the dynamics on disjoint halves of the slow manifold, corresponding to ${L^{(0)}}^2 < 2E^{(0)}I_2$ and ${L^{(0)}}^2 > 2E^{(0)}I_2$. The exceptional case, ${L^{(0)}}^2=2EI_2$, denotes the state of intermediate-axis-aligned rotation, and is analyzed in more detail in the following subsection. Therein, it is shown that 
the limiting forms of (\ref{lxmain}-\ref{emain}), in the vicinity of the fixed point corresponding to the above rotation, remain the same  for $L^{(0)^2}\rightarrow 2EI_2^{\pm}$, provided the suffixes $1$ and $3$ are appropriately interchanged. 

Equations (\ref{lxmain}-\ref{emain}) are the analog of (\ref{lxsmain}-\ref{nsmain}) for a spheroid, and will form the basis for interpreting the large-$St$ orientation dynamics of a triaxial ellipsoid below. Note that the coefficients $A$, $B$, $C$, $D$ and $F$ in (\ref{lxmain}-\ref{emain}) still depend on the slow-manifold variables via $\mu$; this mimics the dependence on $h$ in (\ref{lxsmain}-\ref{nsmain}). This relation between $\mu$ and $h$ is expected, since the former describes the shape of the short-time orbit for a triaxial ellipsoid, in the same manner as $h$\,(in being the cosine of Euler cone angle) does for a spheroid; indeed, $\mu$ reduces to $I_a\bar{h}^2/I_eh^2$ for the limiting cases of prolate and oblate spheroids.


\subsection{Fixed points and Stability analysis}\label{elsa}
Equating the right hand sides of (\ref{lxmain}-\ref{emain}) to zero, and analyzing the resulting set of algebraic relations, the following criteria for an equilibrium (fixed point) are obtained :
$$({L_x^{(0)}},{L_y^{(0)}},{L_z^{(0)}})\equiv{\bigg(0,0,\frac{-B}{2A}=\frac{-A}{2F}\bigg)},$$ 
which implies $A^2-BF=0$. On expansion, this relation gives:
\begin{equation}
\begin{split}
\frac{(I_1-I_2)^2\langle z_1^2\rangle\langle z_2^2\rangle}{(I_1I_2)^2}+\frac{(I_3-I_1)^2\langle z_1^2\rangle\langle z_3^2\rangle}{(I_3I_1)^2}+\frac{(I_2-I_3)^2\langle z_2^2\rangle\langle z_3^2\rangle}{(I_2I_3)^2}=0.
\end{split}
\end{equation}
The positive-definite coefficients imply an equilibrium is possible if and only if any two of $\langle z_1^2\rangle,\langle z_2^2\rangle,\langle z_3^2\rangle$ are zero. The three ways in which this can happen correspond to the principal-axis-aligned rotations, which we consider in turn below.    
\subsubsection*{\underline{${L^{(0)}}^2=2{E^{(0)}}I_1$}}
This case arises when $\langle z_1^2,z_2^2,z_3^2\rangle\equiv(1,0,0)$, 
and corresponds to rotation with the longest axis aligned with the ambient vorticity. The fixed point coordinates are $({L_x^{(0)}},{L_y^{(0)}},{L_z^{(0)}},{E^{(0)}})\equiv(0,0,-I_1/2,I_1/8)$. For $I_1=I_2$, this reduces to the tumbling mode of an oblate spheroid, while for $I_2=I_3$, one obtains the spinning mode of a prolate one. Consideration of the linearized equations, in the vicinity of the above fixed point, gives the following eigenvalues:
 $$(\lambda_1,\lambda_2,\lambda_3,\lambda_4)\equiv-\frac{8\pi}{I_1}\bigg(\frac{X_2 c^2}{a^2+c^2}+\frac{X_3 b^2}{a^2+b^2},\frac{X_2 a^2}{a^2+c^2}+\frac{X_3 a^2}{a^2+b^2}, X_1,\frac{X_3(I_1-I_3)}{ I_3}+\frac{X_2(I_1-I_2)}{I_2 }\bigg).$$
The first three are negative irrespective of the aspect ratios $(b/a,c/a)$ while the fourth is always positive, implying that the longest-axis-aligned rotation is always a saddle point, this being consistent with the results in the earlier section; in that, the prolate spinning and oblate tumbling modes appear as saddle points on the slow manifold. 
 

\subsubsection*{\underline{${L^{(0)}}^2=2{E^{(0)}}I_3$}
}
This case arises when $\langle z_1^2,z_2^2,z_3^2\rangle\equiv(0,0,1)$, 
and corresponds to rotation with the shortest axis aligned  with ambient vorticity. The fixed point is given by $({L_x^{(0)}},{L_y^{(0)}},{L_z^{(0)}},{E^{(0)}})\equiv(0,0,-I_3/2,I_3/8)$. For $I_1=I_2$, one obtains the spinning mode of an oblate spheroid, while for $I_2=I_3$, one obtains the tumbling mode of a prolate one. A linear stability analysis gives the following eigenvalues:
 $$(\lambda_1,\lambda_2,\lambda_3,\lambda_4)\equiv-\frac{8\pi}{I_3}\bigg(\frac{X_2a^2}{a^2+c^2}+\frac{X_1 b^2}{c^2+b^2},\frac{X_1c^2}{b^2+c^2}+\frac{X_2 c^2}{a^2+c^2}, X_3,\frac{X_1(I_3-I_1)}{I_1 }+\frac{X_2(I_3-I_2)}{I_2}\bigg).$$
All four eigenvalues are negative, so rotation about the shortest axis is always a stable node; this again being consistent with our earlier results for a spheroid, and with the findings of earlier numerical investigations\cite{lundell2011}.

\subsubsection*{$\underline{{L^{(0)}}^2=2{E^{(0)}}I_2}$}

This case arises when $\langle z_1^2,z_2^2,z_3^2\rangle\equiv(0,1,0)$, and the associated fixed point, $({L_x^{(0)}},{L_y^{(0)}},{L_z^{(0)}},{E^{(0)}})\equiv(0,0,-I_2/2,I_2/8)$, corresponds to intermediate-axis-aligned rotation about the ambient vorticity. Now, ${L^{(0)}}^2=2{E^{(0)}}I_2$ implies $\mu\delta=1$, and the non-analytic\,(logarithmic) behavior of $K(\mu\delta)$ for $\mu\delta \rightarrow 1$ ensures that this is a singular fixed point, not admitting a local linear approximation. The neighborhood of the point is best analyzed in $(L_x^{(0)},L_y^{(0)},L_z^{(0)},\mu)$ space, since this simplifies the terms involving the elliptic integrals; the fixed point now being given by $(0,0,-I_2/2,1/\delta)$. Linearizing (\ref{lxmain}) and (\ref{lymain}) directly yields two of the four eigenvalues as:
$$(\lambda_1,\lambda_2)\equiv\bigg(-\frac{8\pi}{I_2}\bigg(\frac{X_1c^2}{b^2+c^2}+\frac{X_3a^2}{a^2+b^2}\bigg),-\frac{8\pi}{I_2}\bigg(\frac{X_1b^2}{b^2+c^2}+\frac{X_3 b^2}{a^2+b^2}\bigg)\bigg),$$
which are negative for all aspect ratio pairs, implying that any trajectory rapidly converges to the $(L_z^{(0)},\mu)$ plane. It is therefore sufficient to consider the limiting forms of (\ref{lzmain}) and (\ref{emain}) in the vicinity of the fixed point projected onto this plane. Note, however, that unlike the other two equilibrium points above, the definitions of $\mu$ and ($\delta,A,B,C,D,F$) differ for approach towards $(-I_2/2,1/\delta)$ from either side\,(corresponding to $L^{(0)^2}\rightarrow 2E^{(0)}I_2^{\pm}$). Defining $L^{(0)'}_z=L^{(0)}_z+I_2/2$ and $\mu' = 1/\delta- \mu$, and using ${\lim_{\mu\delta\rightarrow1}{E(\mu\delta)}/{K(\mu\delta)}={1}/{\log{(4/\sqrt{\mu'\delta})}}}$ at leading logarithmic order, (\ref{lzmain}) and (\ref{emain}) reduce to the following approximate forms:
 \begin{align}\label{lzasymp}
\frac{d{L_z^{(0)'}}}{dt_2}&=-\frac{8\pi}{I_2}[X_2 {L_z^{(0)'}}+\frac{(I_3-I_2)(I_2-I_1)(X_3-X_1)}{2(I_3-I_1)}\frac{1}{\log(4/\sqrt{\mu'\delta})}],\\
\label{muasymp}
\frac{d\mu'}{dt_2}&=\frac{16\pi}{I_2\delta}[\bigg(\frac{X_1(I_2-I_1)}{I_1}+\frac{X_3(I_3-I_2)}{I_3}\bigg)\frac{1}{\log{(4/\sqrt{\mu'\delta})}}],
\end{align}
for ${L_z^{(0)}}', \mu' \ll 1$ and ${L^{(0)^2}}>2{E^{(0)}}I_2$; the forms for $L^{(0)^2}<2{E^{(0)}}I_2$ result from the interchange $(I_1,X_1) \leftrightarrow (I_3,X_3)$ in the above equations. 
The $\mu'$-dynamics, as governed by (\ref{muasymp}), is independent of ${L_z^{(0)}}'$, and may be solved in closed form to obtain:
\begin{align}\label{muassol}
	\mu'\delta\log{\frac{16}{\mu'\delta}}&=\mu_0'\delta\log{\frac{16}{\mu_0'\delta}}+2Wt_2.
\end{align}
Here, 
$W=\pm16\pi[\frac{X_1(I_2-I_1)}{I_1 I_2}+\frac{X_3(I_3-I_2)}{I_3 I_2}]$, so that $\mu'$ in (\ref{muassol}) decreases to\,(increases from) zero from\,(to) a value $\mu'_0\,(>0)$ in a finite time $t_{2_0}= \pm \frac{\mu_0'\delta}{2 W}\log{\frac{\mu_0'\delta}{16}}$; the plus and minus signs pertaining to $L^{(0)^2}>2{E^{(0)}}I_2$ and $L^{(0)^2}<2{E^{(0)}}I_2$, respectively. (\ref{lzassol}) may now be solved to obtain:
\begin{align}
	\label{lzassol}
	L^{(0)'}_z&=L^{(0)'}_{z0}\textrm{e}^{-\frac{8\pi X_2t_2}{I_2}}- \hat{W} \textrm{e}^{-\frac{8\pi X_2t_2}{I_2}}\int^{t_2}_{0}\frac{\textrm{e}^{\frac{8\pi X_2t_2'}{I_2}} \mu'(t_2')\delta }{\mu_0'\delta\log{\frac{16}{\mu_0'\delta}}+2Wt_2-\mu'(t_2')\delta}\text{d}t_2',
\end{align}
where $\hat{W}=\pm \frac{8\pi(X_1-X_3)(I_2-I_3)(I_1-I_2)}{I_2(I_1-I_3)}$ with $\mu'(t_2)$ in the integral being defined by (\ref{muassol}). The expression for $L^{(0)'}_{z0}$ obtained from setting $L^{(0)'}_z$ to zero in (\ref{lzassol}), together with (\ref{muassol}), implicitly define the the leading order approximation, $(L^{(0)'}_{z0},\mu_0)$, for the two halves  of the critical trajectory that start\,($\mu_0 >0$) and end\,($\mu_0 <0$) at the intermediate fixed point.

\begin{figure}
	\subfloat[]{\includegraphics[scale=0.45]{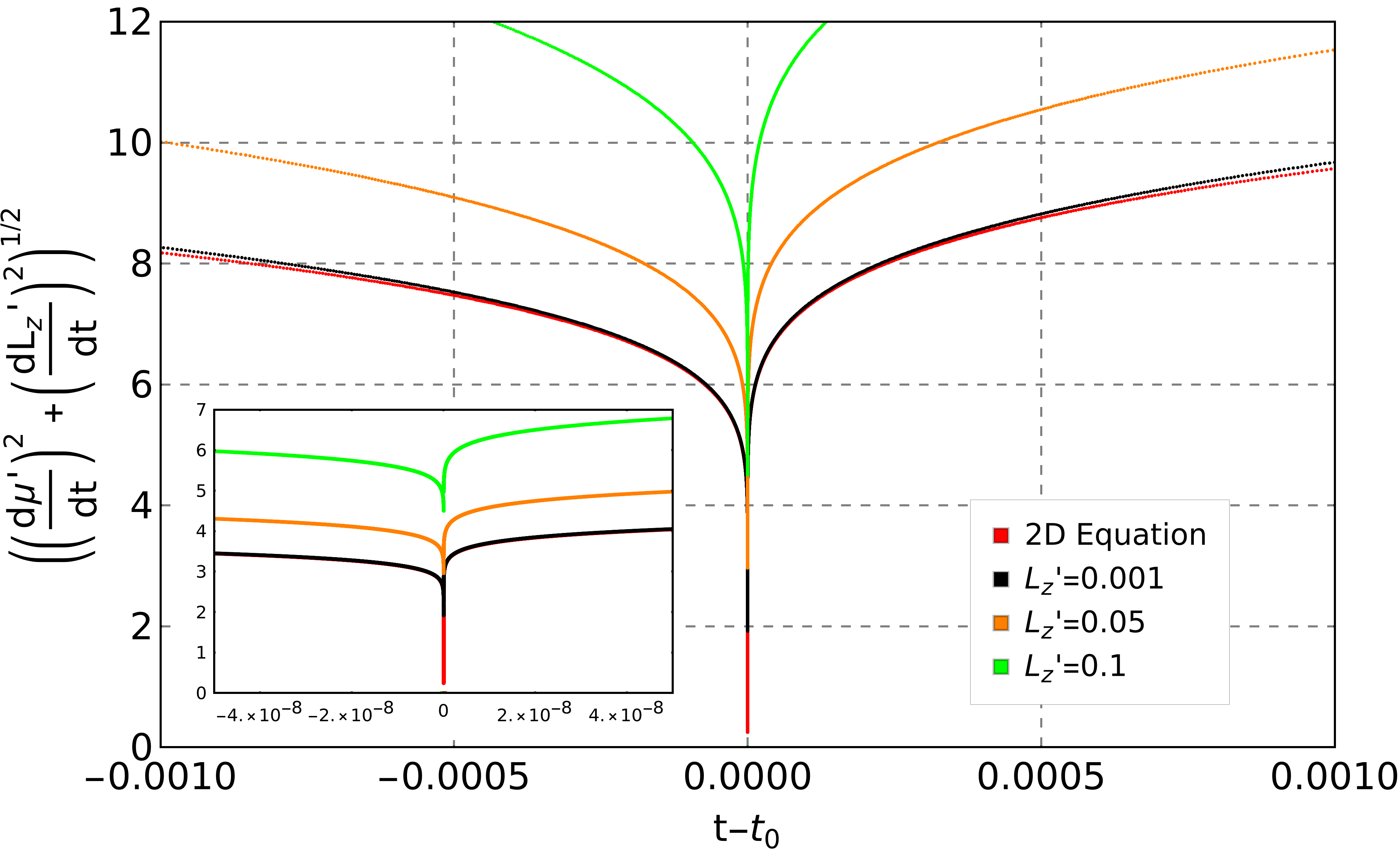}}
    \subfloat[]{\includegraphics[scale=0.41]{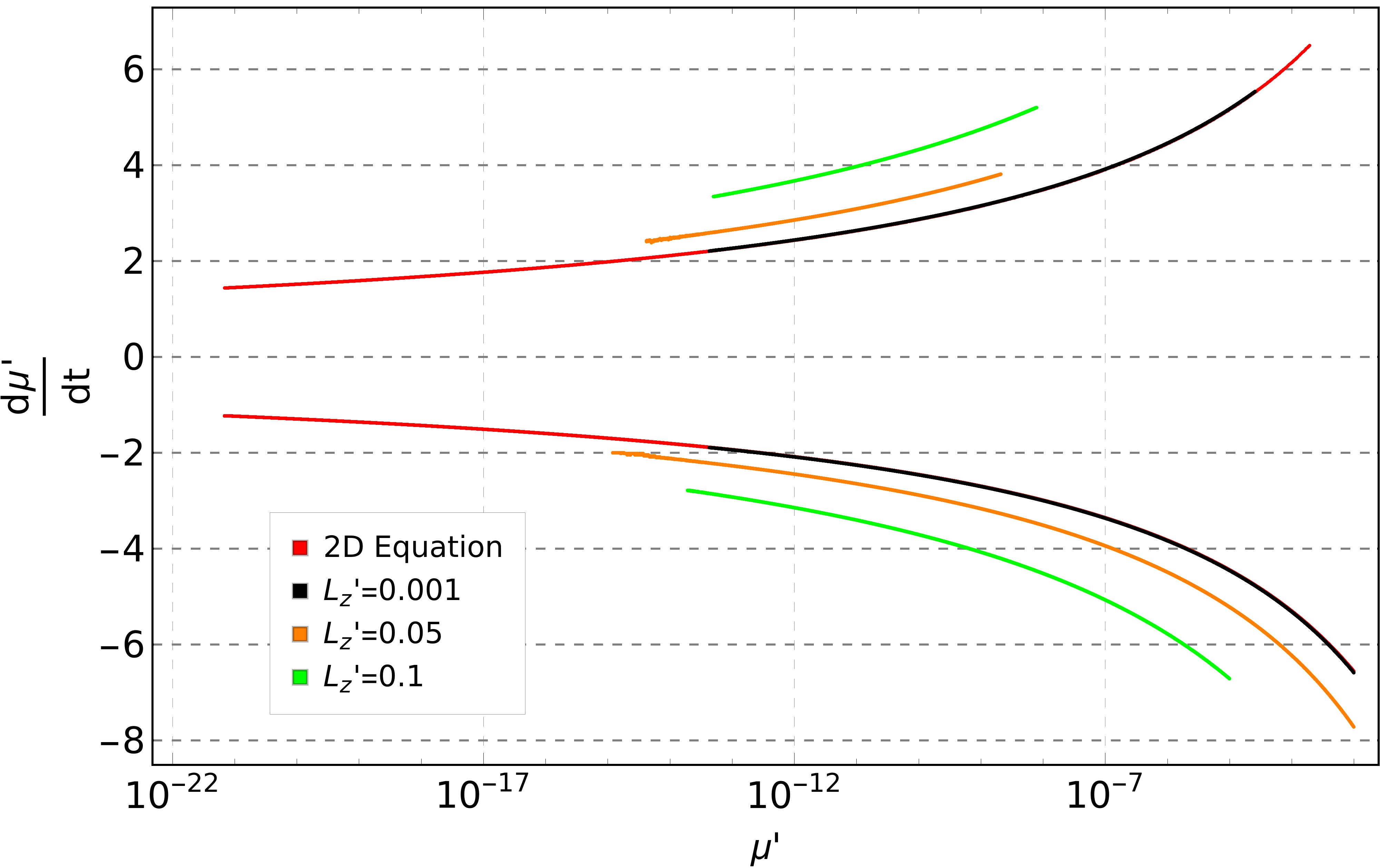}}
	\caption{Modulus of rate of change of $\mu'$ and $L'_z$ as a function of $L'_z$. The red curve is obtained from (\ref{lzasymp}) and (\ref{muasymp}), the green, orange, and black curves are obtained from (\ref{lxmain}-\ref{emain}). }
	\label{Velocity_near_intermediate_point}
\end{figure}

Figures \ref{Velocity_near_intermediate_point}a and b explore the nature of the trajectories in the immediate neighborhood of the intermediate fixed point. The former shows the trajectories converging to, and diverging from, the fixed point in a finite time, with the latter showing the velocity decreasing to zero, in an inverse logarithmic fashion, during approach. The solid and dashed curves in these figures correspond to the full 4D slow-manifold solution, and a 2D approximation based on (\ref{muassol}) and (\ref{lzassol}).

In summary, 
the intermediate-axis-aligned rotation is a singular fixed point, accessible\,(in a finite time) only along a single critical trajectory. This trajectory divides the slow manifold, for $L_z < 0$, into two parts: trajectories in one part converge directly to the stable node, while those in the other pass by the saddle on the way to the node. The aforesaid finite-time feature is important since it implies that neighboring trajectories pass by the singular point without perceptibly slowing down, in turn implying that the point, although singular, is largely passive as far its influence on the ellipsoidal orientation dynamics is concerned. For spheroids, this fixed point coalesces with either of the other two fixed points. In the prolate case, the coalescence is with the stable node, while in the oblate case, it is with the saddle point.


An issue not addressed in our derivation of equations (\ref{lxmain}-\ref{emain}) is the role of transient resonances arising from the variation of $T$ and $T'$ with time owing to their dependence on the slow variables. In the calculation of the Euler-top averages, we had assumed $T$ and $T'$ to always be incommensurate (implicit in the application of the Weyl's equidistribution theorem). However, with time, the system will invariably pass through an (countable) infinity of transient resonances when the ratio $T/T'$ is a rational number, possibly leading to contributions apart from those accounted for. This is true, especially since the periodic functions in this case are elliptic functions, expressible as infinite Fourier sums, leading to a resonant contribution for any rational ratio of $T$ and $T'$. The error in this case, will be $O(St^{1+\upsilon})$ with $0 <\upsilon< 1$, implying that a larger $St$ is required to produce the same level of agreement between the fast and slow manifold for a triaxial ellipsoid, as compared to a spheroid. The precise value of $\upsilon$ must be obtained by asymptotic matching of the solution for pre-resonant, resonant and post-resonant dynamics. If the relatively simple examples given in textbooks \cite{kevorkian1996} are any indication, an analysis of even one of these resonances is of a formidable complexity. Thus, the prudent way of assessing their importance is by a comparison of the leading order theory above to the exact numerics. As shown in section \ref{numres} below, the agreement of the leading order solution, for $St \geq 1000$, does seem to indicate that the aforementioned resonances likely play a relatively minor role.

\subsection{The $L_z-E$ Phase Plane : Ellipsoid}\label{numres}

\begin{figure}
\subfloat[\label{LxvstvarSt0b} $L_{x}/I_3$ vs $t_2$ for different $St$; $(a,b,c)\equiv(1,0.8,0.5)$ 
$(\theta_0,\phi_0,\psi_0,\dot{\theta_0},\dot{\phi_0},\dot{\psi_0}) \equiv(\pi/3,\pi/6,0,1,1,-1)$]{\includegraphics[scale=0.45]{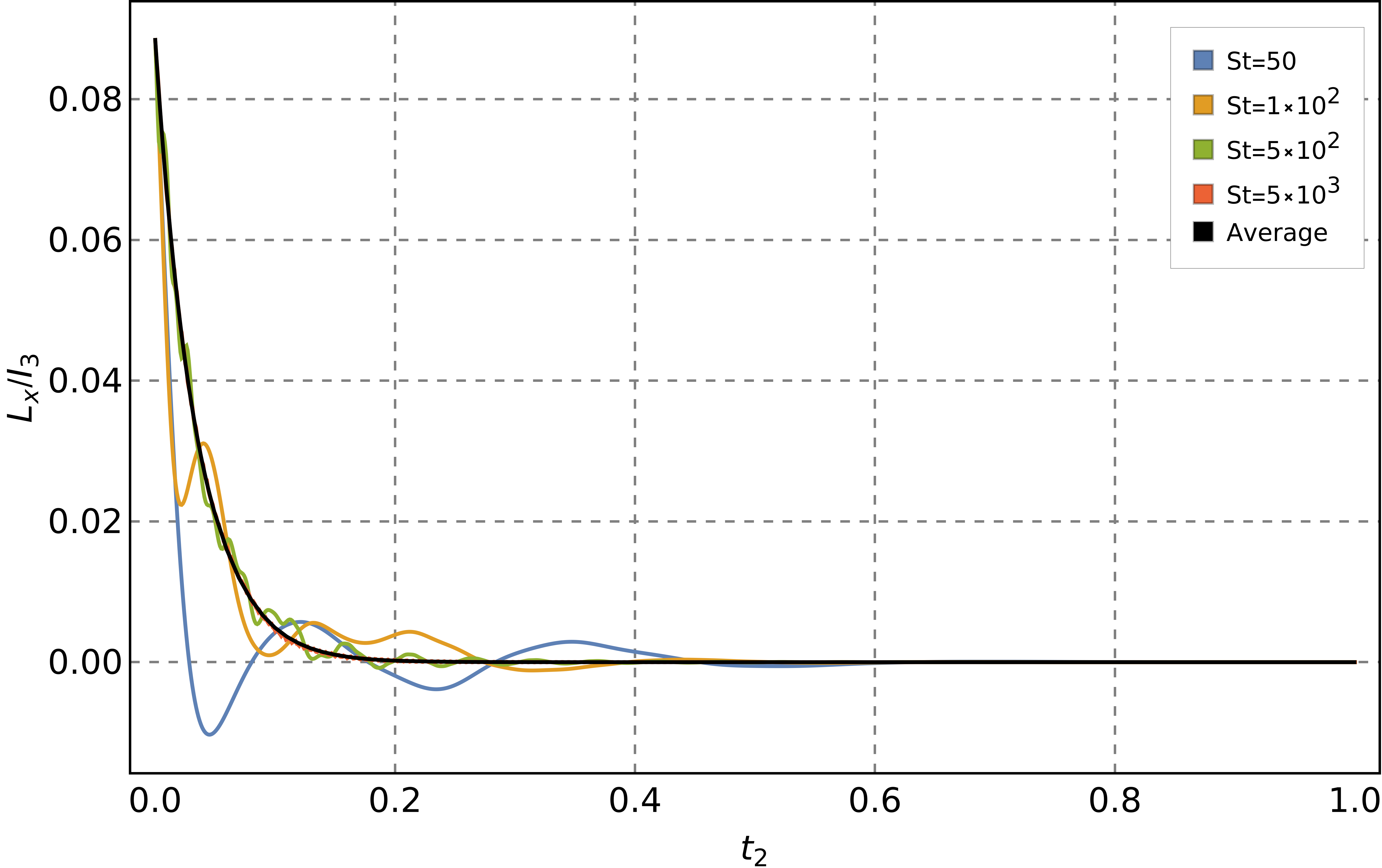}}
\subfloat[\label{LyvstvarSt0b} $L_{y}/I_3$ vs $t_2$ for different $St$; $(a,b,c)\equiv(1,0.8,0.5)$ 
$(\theta_0,\phi_0,\psi_0,\dot{\theta_0},\dot{\phi_0},\dot{\psi_0}) \equiv(\pi/3,\pi/6,0,1,1,-1)$]{\includegraphics[scale=0.45]{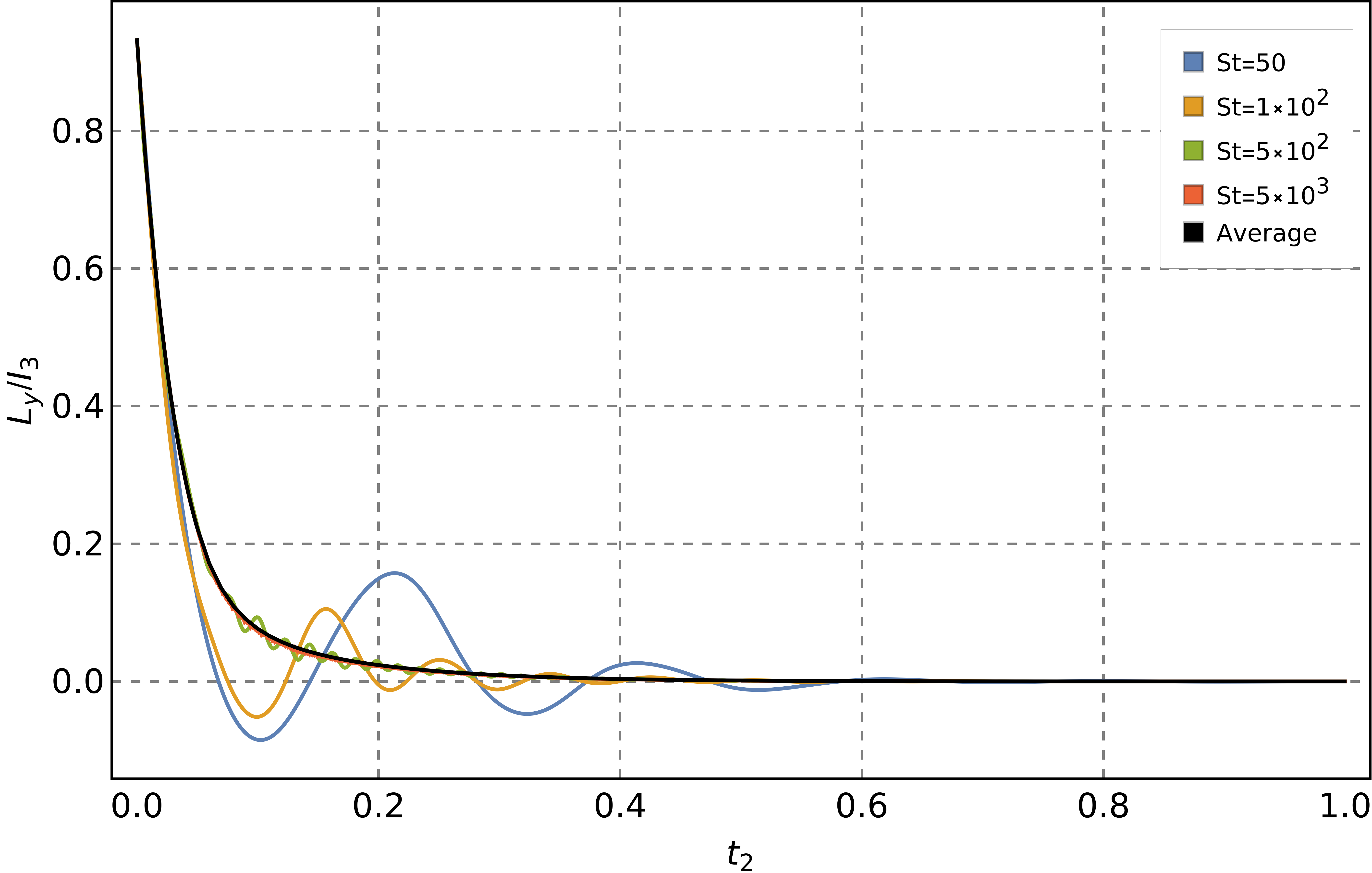}}\\
\subfloat[\label{LzvstvarSt0b} $L_{z}/I_3$ vs $t_2$ for different $St$; $(a,b,c)\equiv(1,0.8,0.5)$ 
$(\theta_0,\phi_0,\psi_0,\dot{\theta_0},\dot{\phi_0},\dot{\psi_0}) \equiv(\pi/3,\pi/6,0,1,1,-1)$]{\includegraphics[scale=0.45]{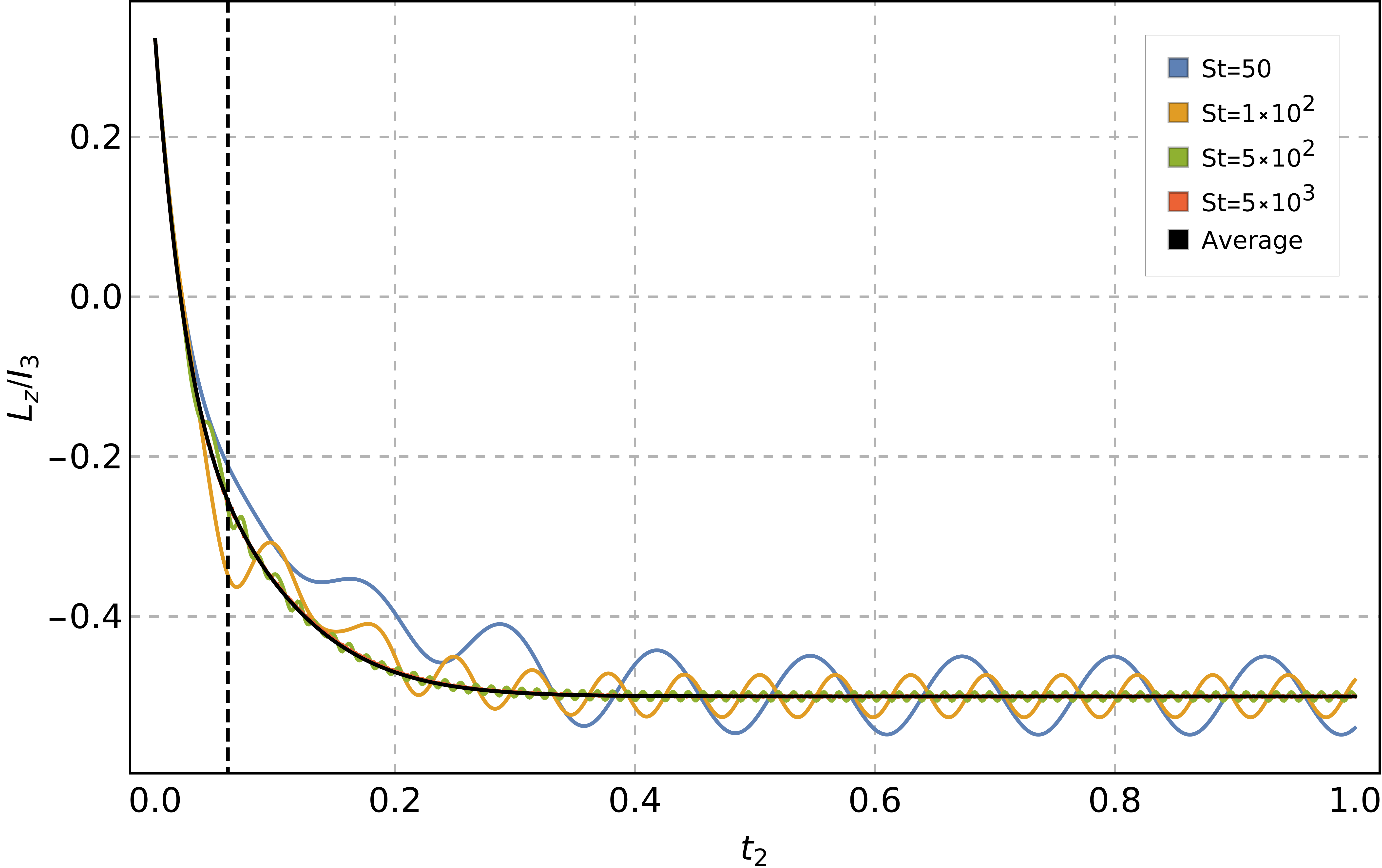}}
\subfloat[\label{EvstvarSt0b} $E/I_3$ vs $t_2$ for different $St$; $(a,b,c)\equiv(1,0.8,0.5)$ 
$(\theta_0,\phi_0,\psi_0,\dot{\theta_0},\dot{\phi_0},\dot{\psi_0}) \equiv(\pi/3,\pi/6,0,1,1,-1)$]{\includegraphics[scale=0.45]{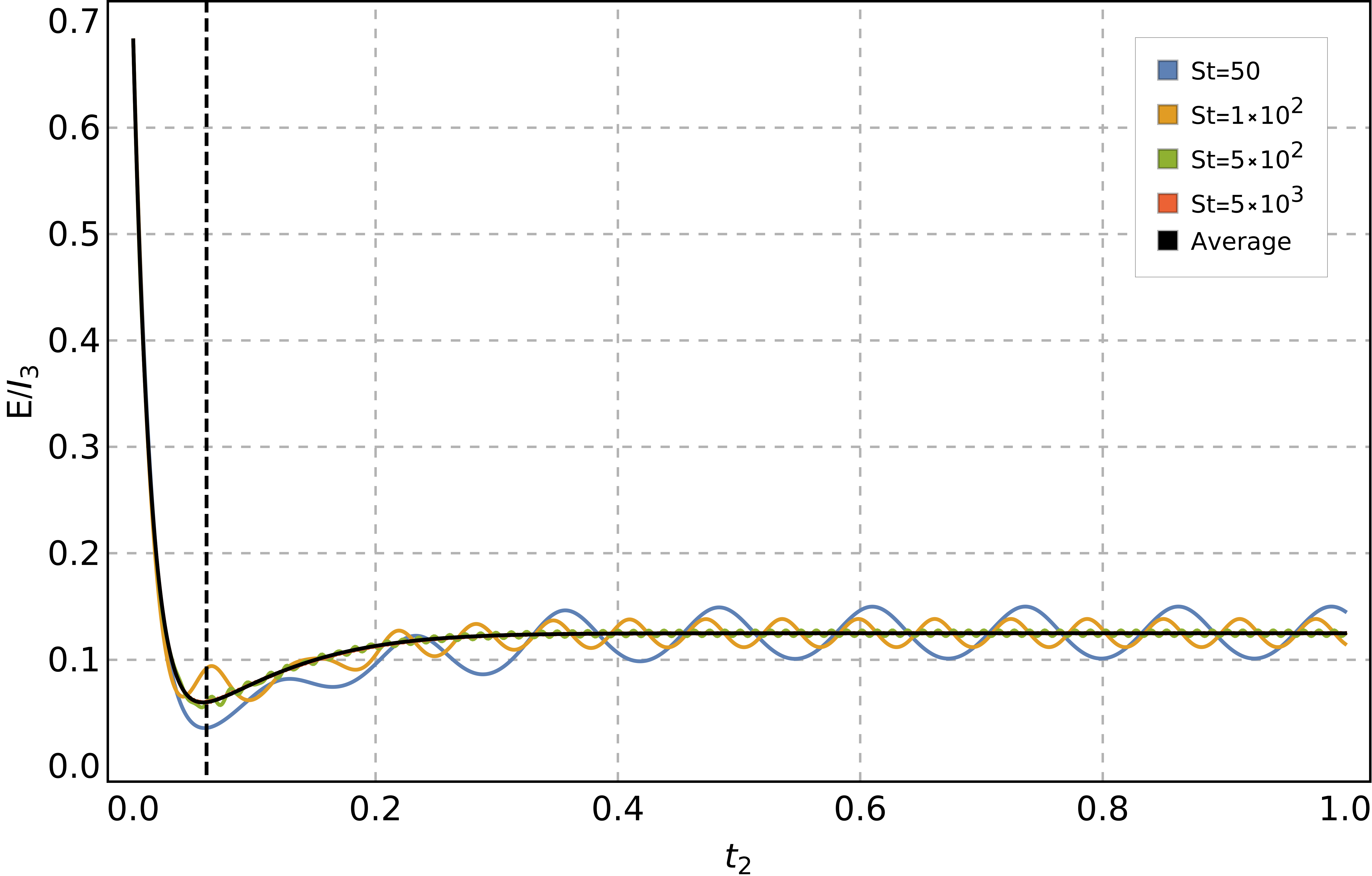}}
\caption{Evolution of normalized angular momenta and energy with time for a triaxial ellipsoid obtained through full solution for $St=50, 100, 500,5000$ and its comparison with slow manifold. The eventual state is given by $(L_x/I_3,L_y/I_3,L_z/I_3,E/I_3)\equiv (0,0,-1/2,1/8)$ as predicted by the analysis.} 
\label{elev}
\end{figure}

\begin{figure}
\subfloat[\label{Lxt2_near_Saddle_ellip} $L_{x}/I_3$ vs $t_2$ for different $St$; $(a,b,c)\equiv(1,0.8,0.5)$ 
$(\theta_0,\phi_0,\psi_0,\dot{\theta_0},\dot{\phi_0},\dot{\psi_0}) \equiv(1.57,0,\pi/2,0,-0.61,0.0003)$]{\includegraphics[scale=0.45]{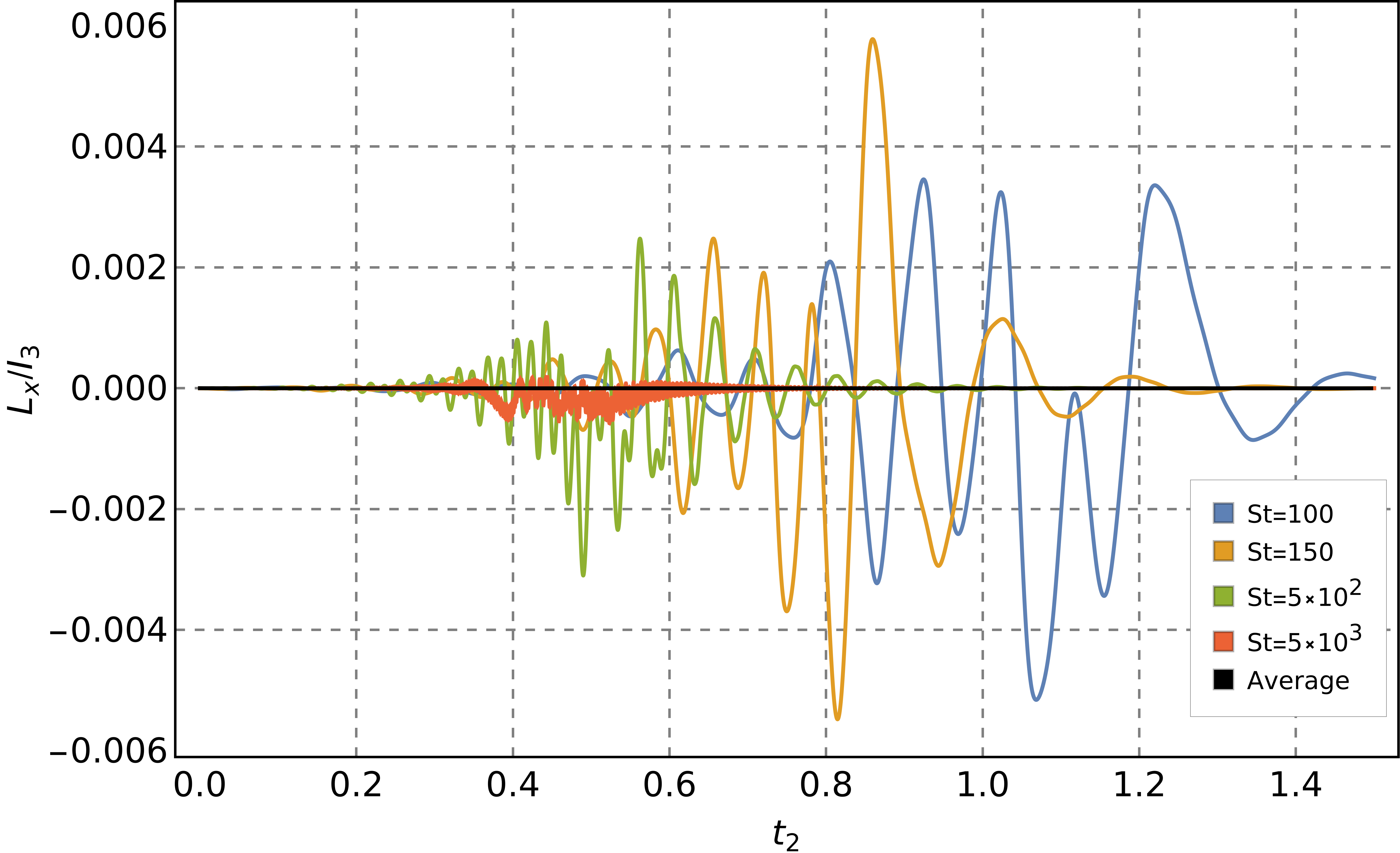}}
\subfloat[\label{Lyt2_near_Saddle_ellip} $L_{y}/I_3$ vs $t_2$ for different $St$; $(a,b,c)\equiv(1,0.8,0.5)$ 
$(\theta_0,\phi_0,\psi_0,\dot{\theta_0},\dot{\phi_0},\dot{\psi_0}) \equiv(1.57,0,\pi/2,0,-0.61,0.0003)$]{\includegraphics[scale=0.45]{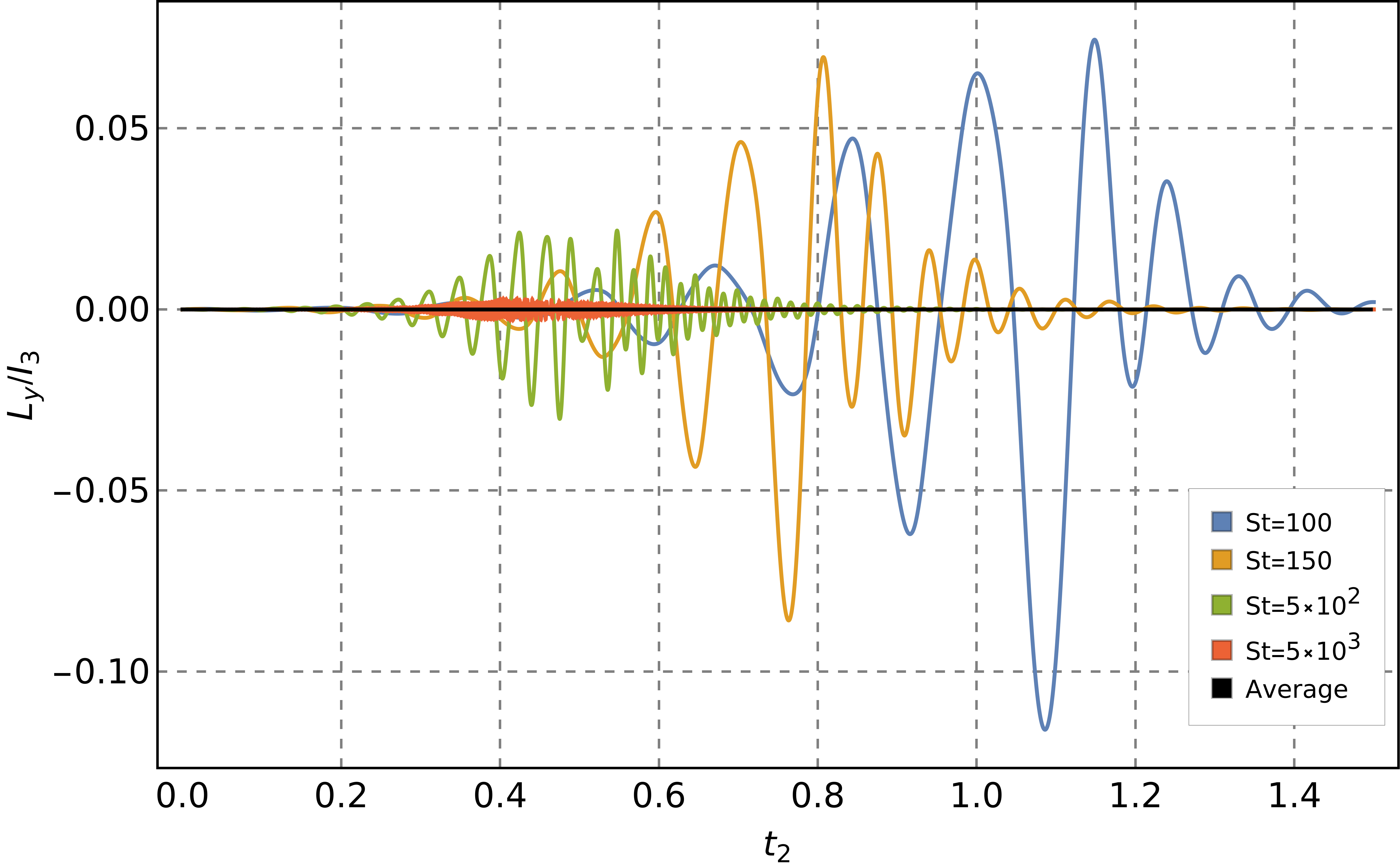}}\\
\subfloat[\label{Lzt2_near_Saddle_ellip} $L_{z}/I_3$ vs $t_2$ for different $St$; $(a,b,c)\equiv(1,0.8,0.5)$ 
$(\theta_0,\phi_0,\psi_0,\dot{\theta_0},\dot{\phi_0},\dot{\psi_0}) \equiv(1.57,0,\pi/2,0,-0.61,0.0003)$]{\includegraphics[scale=0.45]{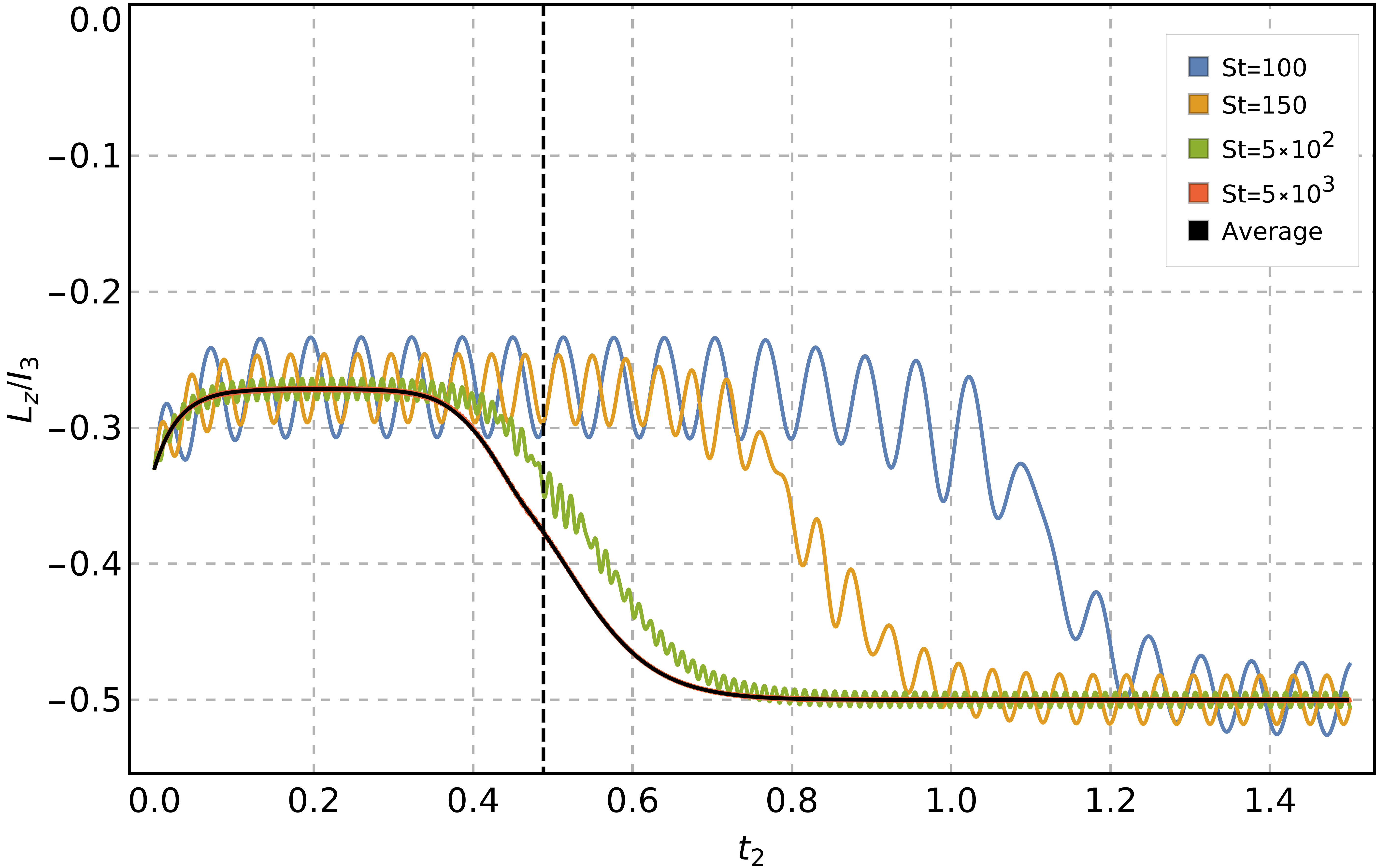}}
\subfloat[\label{Et2_near_Saddle_ellip} $E/I_3$ vs $t_2$ for different $St$; $(a,b,c)\equiv(1,0.8,0.5)$ 
$(\theta_0,\phi_0,\psi_0,\dot{\theta_0},\dot{\phi_0},\dot{\psi_0}) \equiv(1.57,0,\pi/2,0,-0.61,0.0003)$]{\includegraphics[scale=0.45]{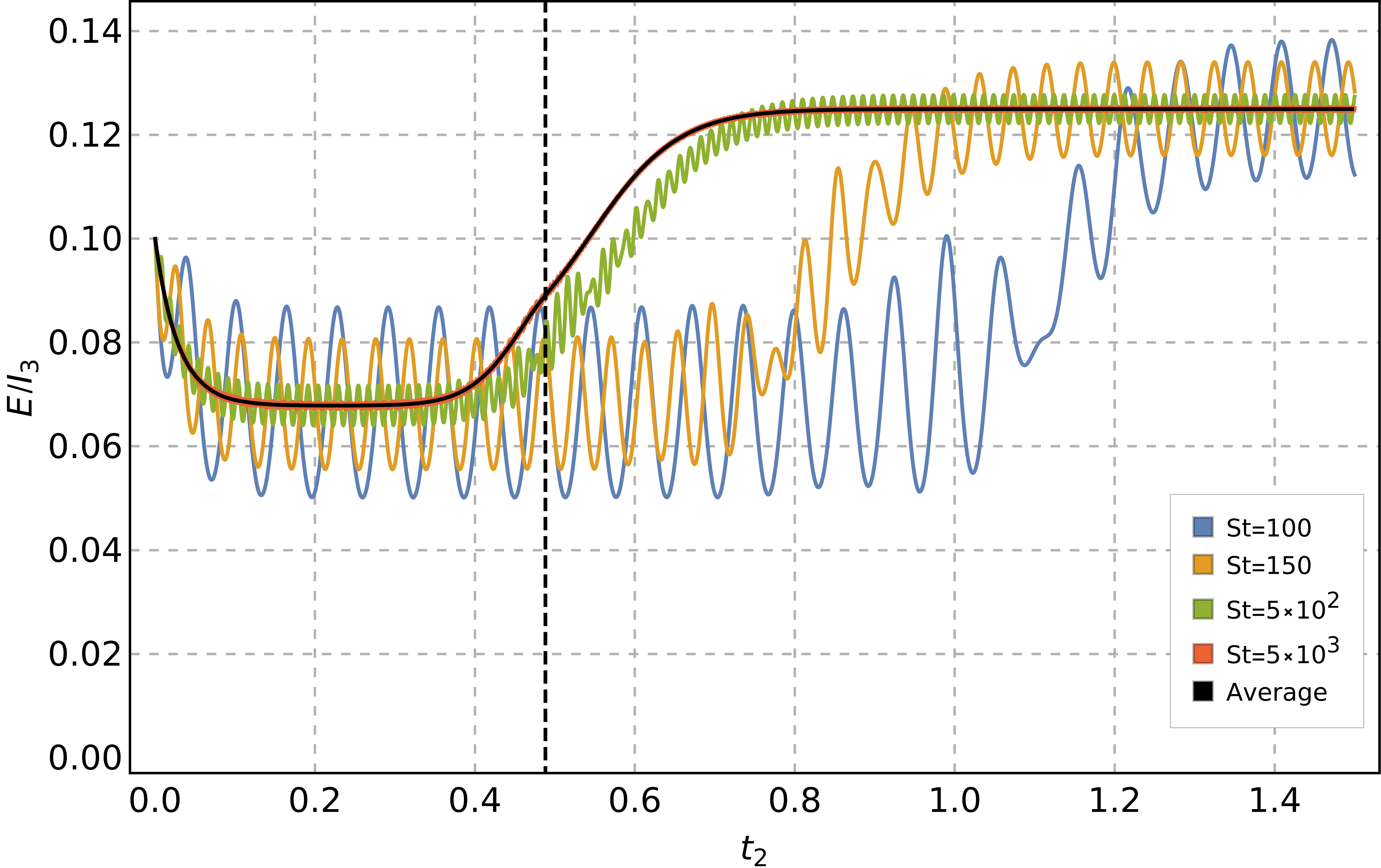}}
\caption{Evolution of normalized angular momenta and energy with time for a triaxial ellipsoid obtained through full solution for $St=100,150,500,5000$ and its comparison with slow manifold. The eventual state is given by $(L_x/I_3,L_y/I_3,L_z/I_3,E/I_3)\equiv (0,0,-1/2,1/8)$ as predicted by the analysis.} 
\label{elev_saddle}
\end{figure}

We follow along the lines of the spheroid in section \ref{numress}, and in Figures \ref{elev} and \ref{elev_saddle}, first present a comparison between slow-manifold and full-solution trajectories, plotted as a function of time, for a pair of initial conditions. As for a spheroid, $L_x^{(0)}$ and $L_y^{(0)}$ decay to zero in both cases, with $L_z^{(0)}$ and $E^{(0)}$ converging to $-I_3/2$ and $I_3/8$, respectively, in agreement with the fixed-point analysis above. The comparison is good for the range of $St$ chosen, and as expected, improves with increasing $St$, owing to the decrease in the amplitude of the fast oscillations of the full-solution trajectories; note that the oscillations are more irregular compared to those in Figures \ref{elevS} and \ref{elevS2}, reflecting the quasi-periodic short-time dynamics of a triaxial ellipsoid. The initial condition chosen for Figure \ref{elev_saddle} above leads to trajectories passing in the close vicinity of the saddle point\,(${L^{(0)}}^2=2{E^{(0)}}I_1$), which manifests as intermediate plateaus in Figures \ref{elev_saddle}c and d, similar to those seen in Figures \ref{elevS2}c and d for a spheroid. Further, the slow-manifold trajectories for both initial conditions also pass fairly close to the intermediate-axis fixed point(${L^{(0)}}^2=2{E^{(0)}}I_2$), and the vertical lines in each of the subfigures correspond to the times at which each of these trajectories is closest to this singular fixed point; evidently, the singular point has very little effect on the temporal evolution of either the slow-manifold or full-solution trajectories, consistent with the results of the local analysis above in the neighborhood of this fixed point. 

\begin{figure}


{\includegraphics[scale=0.8]{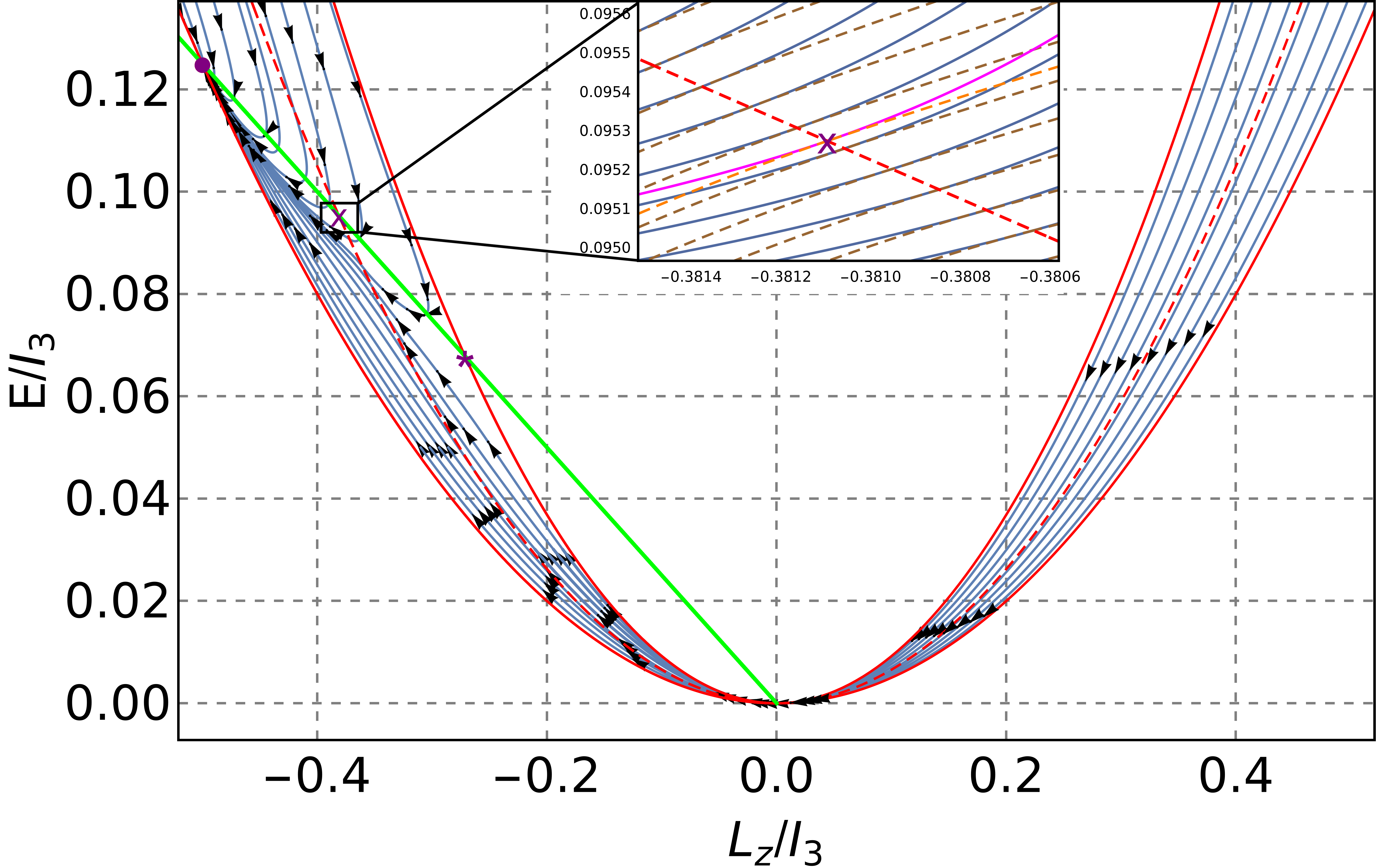}}\\
\caption{\label{LZEfullphaseel}Slow-manifold trajectories on the $L_z-E$ plane for a triaxial ellispoid\,$(b/a=0.8,c/a=0.5)$. The pair of parabolae\,(solid red curves), defined by $L_z^2=2EI_1$ and $L_z^2=2EI_3$, bound all phase-plane trajectories. The three fixed points correspond to the stable node $(L_z/I_3,E/I_3)\equiv(-1/2,1/8)$\,(purple dot), the saddle point $(L_z/I_3,E/I_3)\equiv(I_1/I_3)(-1/2,1/8)$\,(purplr asterisk) and the intermediate fixed point $(L_z/I_3,E/I_3)\equiv(I_2/I_3)(-1/2,1/8)$\,(purple cross); all three points lie on the (green)\,line given by $L_z+4E=0$. The parabola $L_z^2=2EI_2$\,(dashed red curve) demarcates the plane into two distinct halves. The magnified view in the inset shows both the numerical and semi-analytical trajectories in the neighborhood of $(L_z/I_3,E/I_3)\equiv(I_1/I_3)(-1/2,1/8)$; the semi-analytical forms being defined by the solution of (\ref{lzasymp}) and (\ref{muasymp}).}
\end{figure}

For reasons already mentioned in section \ref{numress}, the large-$St$ orientation dynamics is well characterized by trajectories in the $L_z-E$ plane. Slow-manifold trajectories on this plane are shown in Figure \ref{LZEfullphaseel}, for an ellipsoid with $(b/a,c/a)\equiv(0.8,0.5)$, assuming $L_{x0}=L_{y0}=0$. These trajectories are again bounded by a pair of parabolae, defined by $L_z^2 = 2EI_1$\,(upper) and $L_z^2 = 2EI_3$\,(lower). The overall phase-plane topology is similar to a spheroid, with disjoint halves connected at a common pinch point, $(L_z,E) \equiv (0,0)$. While the right half consists solely of trajectories heading towards the pinch point, reaching it in the limit of infinite time, the left half now has three fixed points - the stable node $(L_z/I_3,E/I_3)\equiv(-1/2,1/8)$, the saddle point $(L_z/I_3,E/I_3)\equiv(I_1/I_3)(-1/2,1/8)$, and the singular fixed point $(L_z/I_3,E/I_3)\equiv(I_2/I_3)(-1/2,1/8)$. A third parabola, $L_z^2 = 2EI_2$, passing through the latter point, is the boundary across which the governing system of equations, (\ref{lxmain}-\ref{emain}), changes form owing to indicial\,($1 \leftrightarrow 3$) interchange; it is shown as a dashed red curve in Fig.\ref{LZEfullphaseel}. As already mentioned in the context of  the approximate $({L_z^{(0)}}',\mu')$-system, in the neighborhood of the intermediate-axis-aligned rotation mode, the singular point is seen to only play a passive role in  organizing trajectories in its immediate vicinity.
 The magnified view in the inset highlights the exceptional set of initial conditions that end up converging to, or diverging from, the singular fixed point. These comprise the critical trajectory whose local representation, in a semi-analytical form, was given by solving (\ref{lzasymp}) and (\ref{muasymp}), and is shown as an dashed orange curve; this is seen to agree well with the numerics\,(pink curve). 
 Apart from this curve, and the points on the upper parabola\,(which constitute the invariant manifolds of the saddle point), all other trajectories in the negative-$L_z$ half converge to the stable node. 

\begin{figure}
\subfloat[\label{fig:SI1005550} $L_{z}/I_3$ vs $E/I_3$ for different $St$; $(a,b,c)\equiv(1,0.55,0.5)$ ]{\includegraphics[trim=0 0 0 25, clip, scale=0.6]{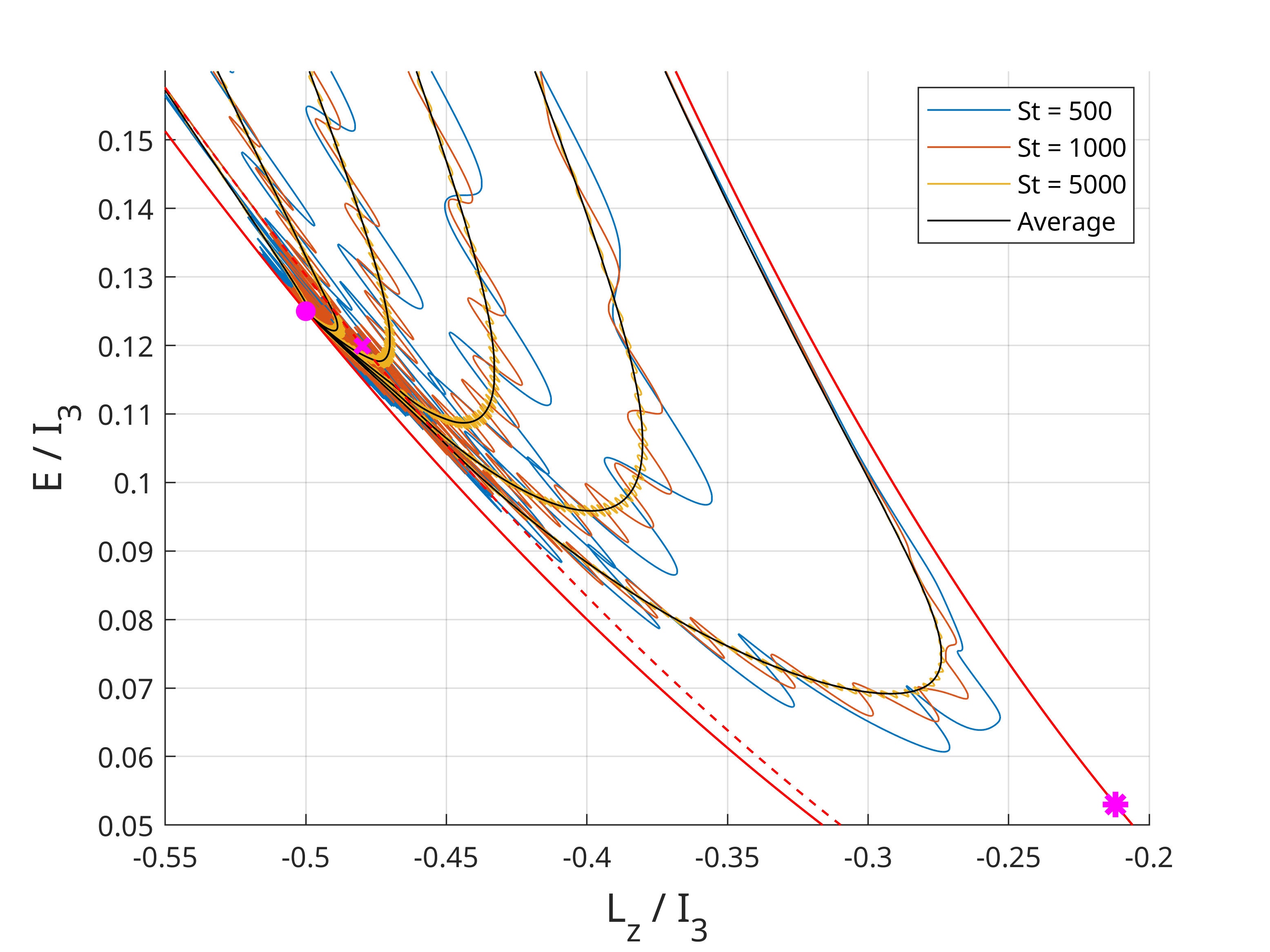}}
\subfloat[\label{fig:SI1005530} $L_{z}/I_3$ vs $E/I_3$ for different $St$; $(a,b,c)\equiv(1,0.55,0.3)$ ]{\includegraphics[trim=0 0 0 25, clip, scale=0.6]{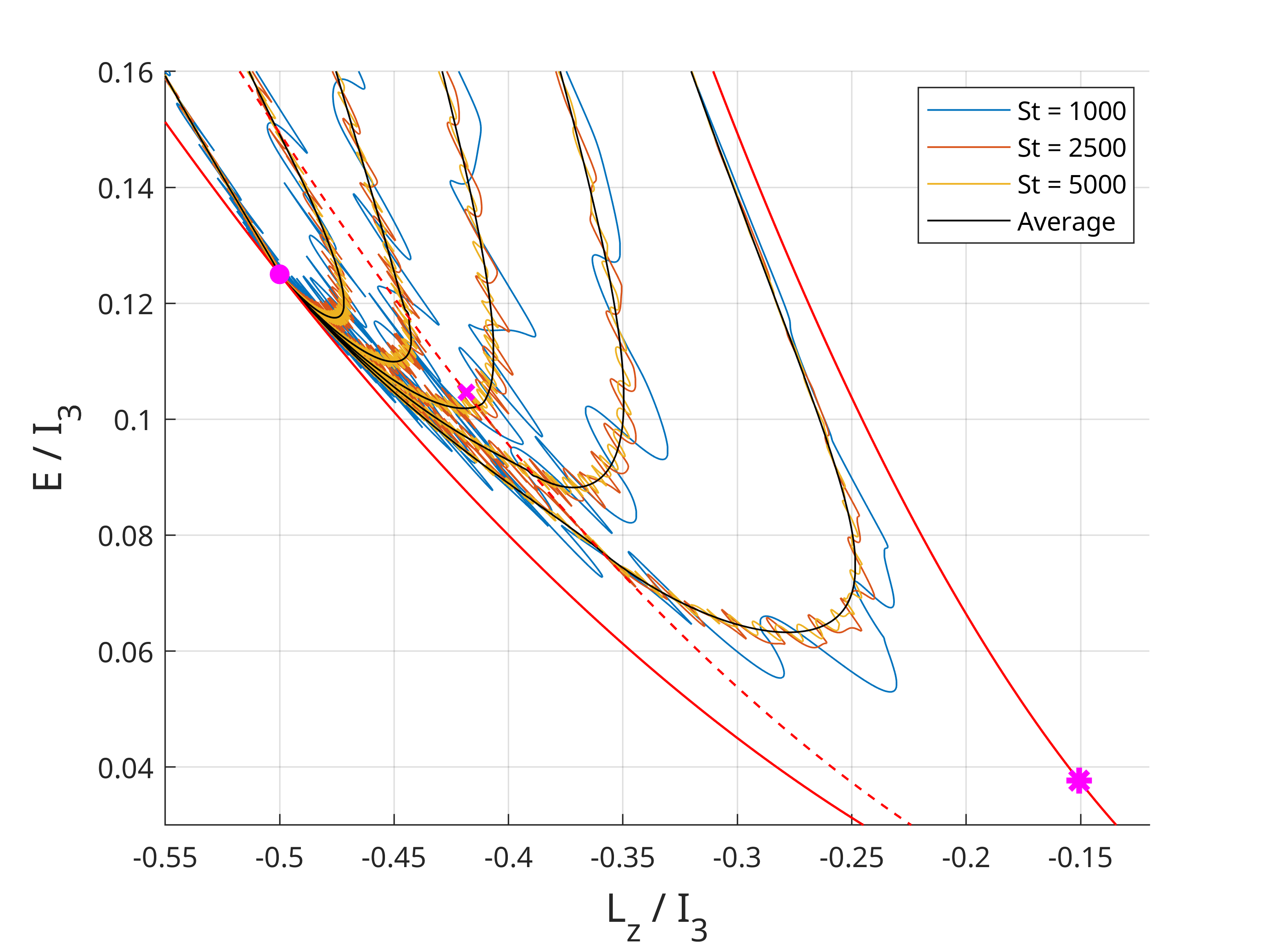}}\\
\subfloat[\label{fig:SI1007550} $L_{z}/I_3$ vs $E/I_3$ for different $St$; $(a,b,c)\equiv(1,0.75,0.5)$ ]{\includegraphics[trim=0 0 0 25, clip, scale=0.6]{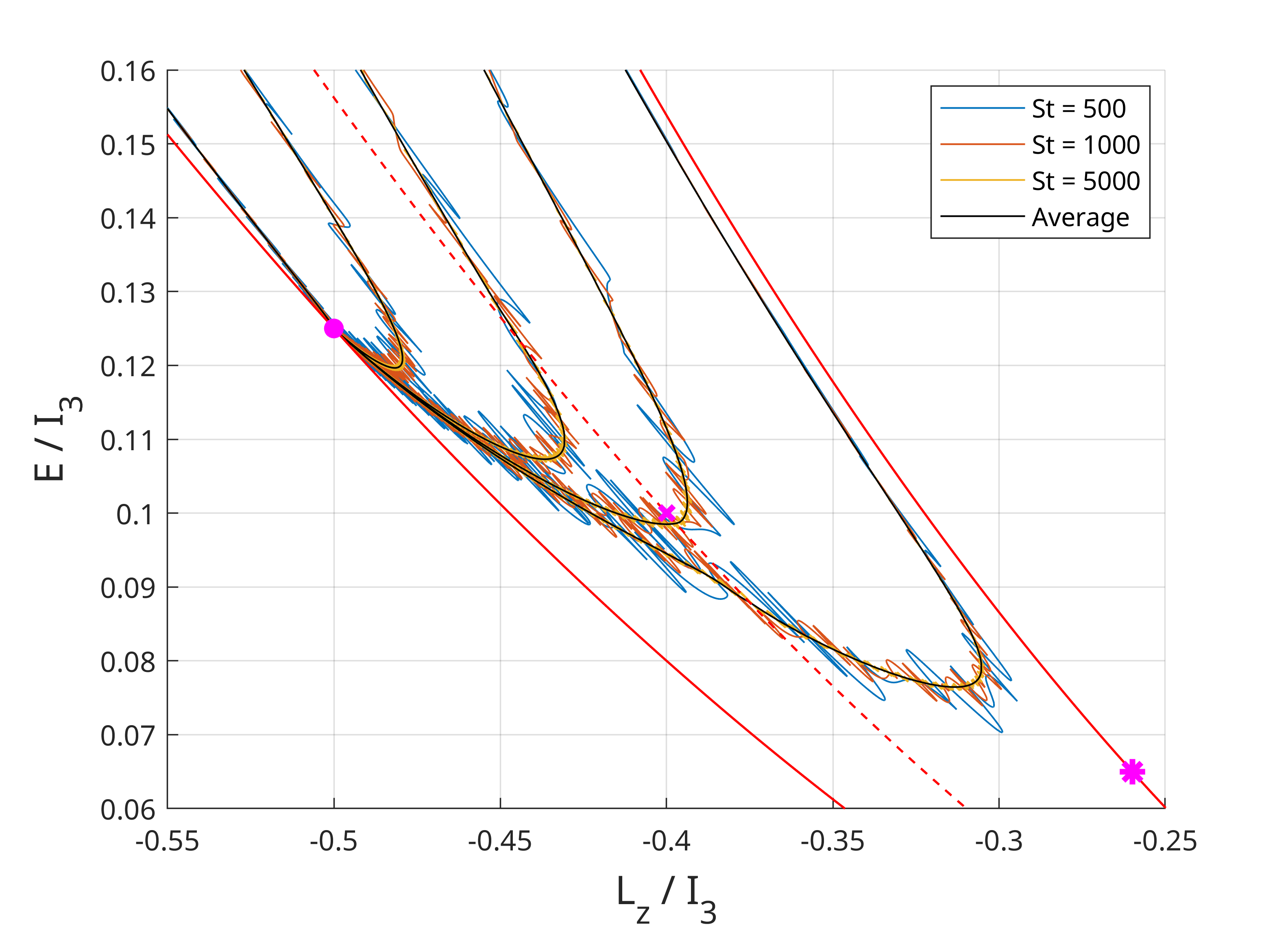}}
\subfloat[\label{fig:SI1007530} $L_{z}/I_3$ vs $E/I_3$ for different $St$; $(a,b,c)\equiv(1,0.75,0.3)$ ]{\includegraphics[trim=0 0 0 25, clip, scale=0.6]{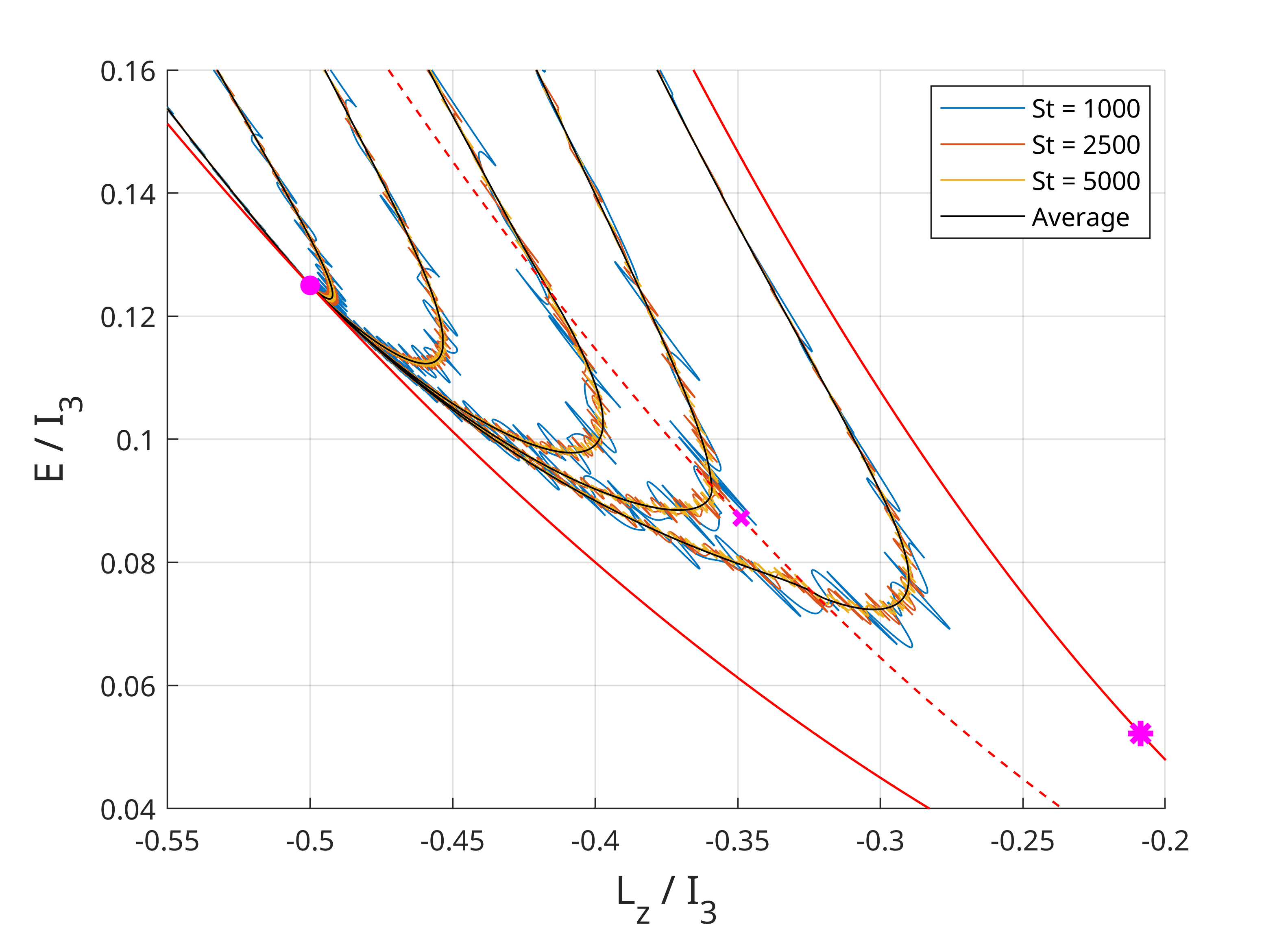}}\\
\subfloat[\label{fig:SI1009550} $L_{z}/I_3$ vs $E/I_3$ for different $St$; $(a,b,c)\equiv(1,0.95,0.5)$]{\includegraphics[scale=0.6]{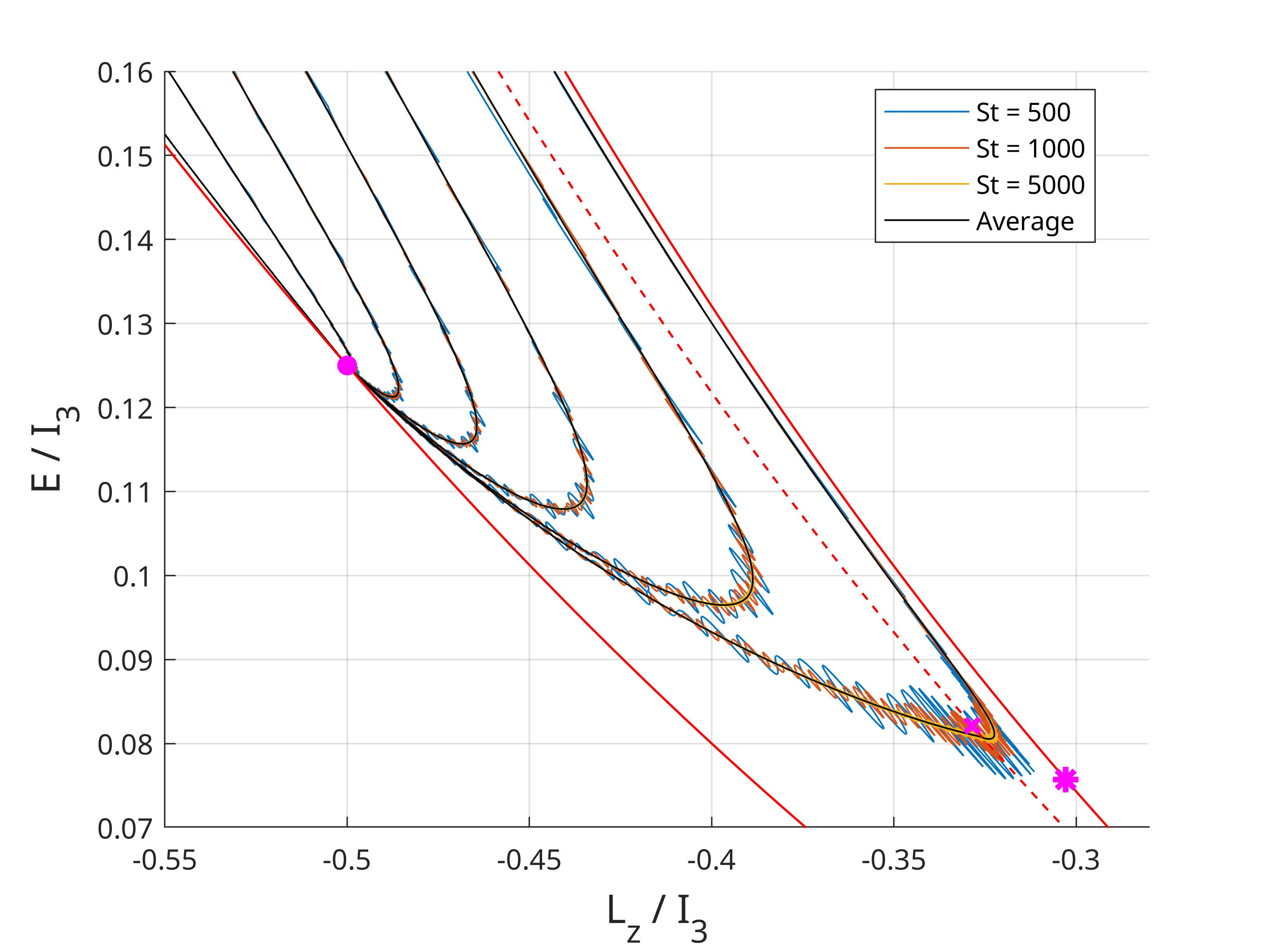}}
\subfloat[\label{fig:SI1009530} $L_{z}/I_3$ vs $E/I_3$ for different $St$; $(a,b,c)\equiv(1,0.95,0.3)$]{\includegraphics[scale=0.6]{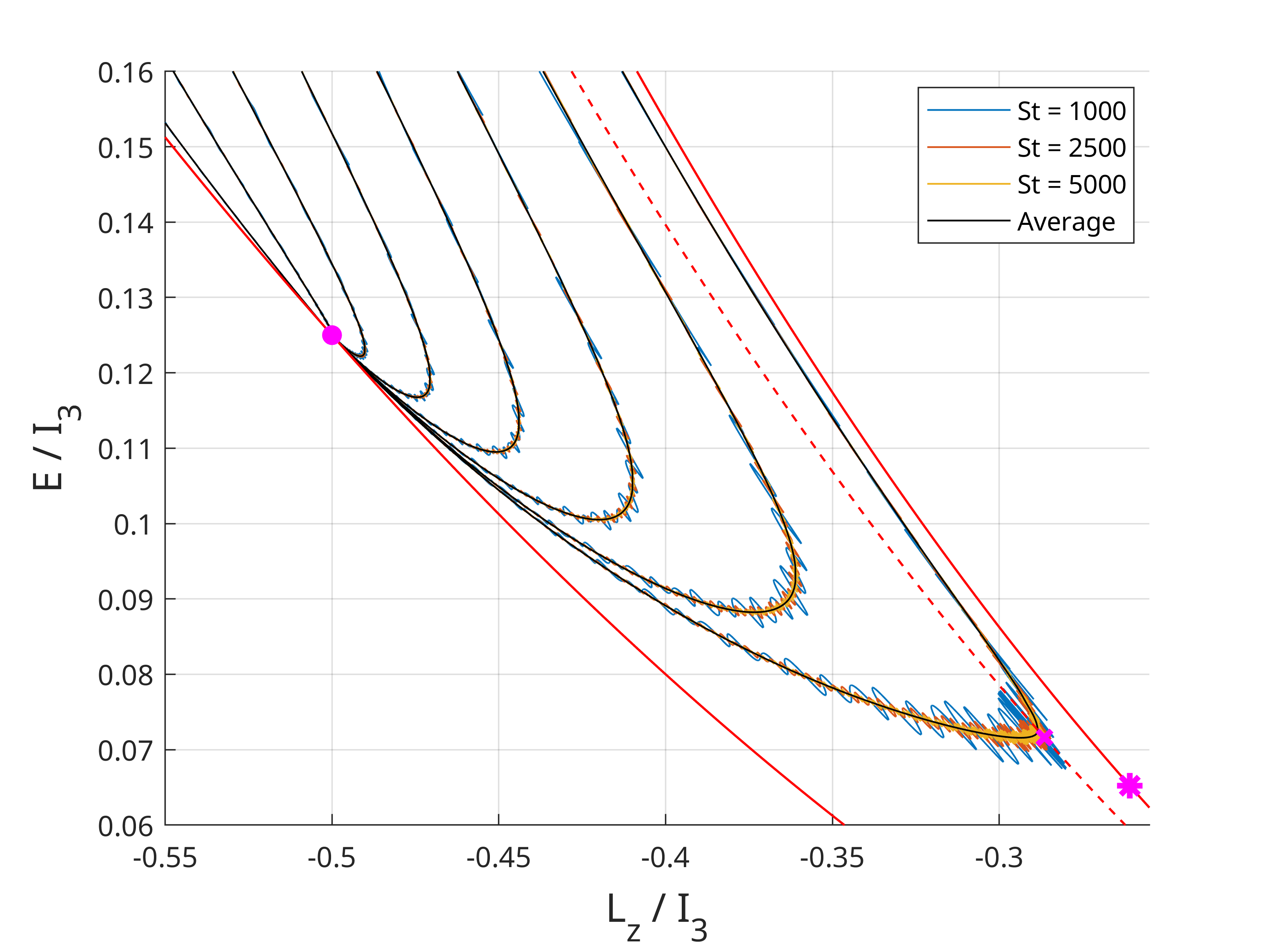}}\\
\caption{Comparison of slow-manifold and full-solution trajectories on the $L_z-E$ phase plane, for a range of ellipsoid aspect ratios, and $St$. The first\,(a,c and e) and second\,(b,d and f) columns correspond to ellipsoids of small and moderate eccentricities, respectively. Along each column, the particle changes from near-prolate at the top, to a near-oblate spheroid at the bottom. The initial conditions for the full-solution trajectories are constrained by $(\theta_0,\phi_0,\psi_0,\dot{\theta_0},\dot{\phi}_0,\dot{\psi_0})\equiv(\theta_0,0,0,0,L^{(0)}_{z0}/I_1,L^{(0)}_{z0} \cos{\theta_0} (1/I_3-1/I_1))$.}
\label{ellze}
\end{figure}

 
Figures \ref{fig:SI1005550}-\ref{fig:SI1009530} show the comparison between slow-manifold and full-solution trajectories on the $L_z-E$ plane, for a wide range of aspect ratio pairs $(b/a,c/a)$, and for different large $St$. The subfigures are organized as follows: in each column, the aspect ratios go from those of a near-prolate spheroid at the top, to the complementary near-oblate spheroid at the bottom; the two columns differ in the eccentricities of the limiting near-spheroids. On the whole, there is good agreement between the averaged and the full orientation dynamics for $St\gtrsim 1000$. Deviations from the averaged approximation reduce in amplitude\,(and increase in frequency), either with increasing $St$ or as the particle grows `fatter'. The latter happens as one goes down each column, owing to the oblate spheroid being more massive than the complementary prolate one; for a given $St$, the oscillations also have a higher amplitude for the second-column figures owing to the greater eccentricity. As for the spheroid, these variations can be partly accounted by defining an effective Stokes number that, unlike $St$, accounts for the changing moment of inertia and resistance coefficient, for a fixed magnitude of the largest axis length. 
Finally, as already pointed out in the context of Figure \ref{elev_saddle}, the deviations of the full-solution trajectories have a well defined periodicity for the near-spheroidal cases, while being more irregular for the triaxial ellipsoid - for instance compare the $St=1000$ full-solution trajectories in figures \ref{fig:SI1005550}\,(near-prolate) and \ref{fig:SI1009530}\,(near-oblate) with those in figure \ref{fig:SI1009550}. 

\begin{figure}
\subfloat[\label{LxLzE_phaseplot} $L_{x}$-$L_{z}$-$E$ phase plot]{\includegraphics[scale=0.15]{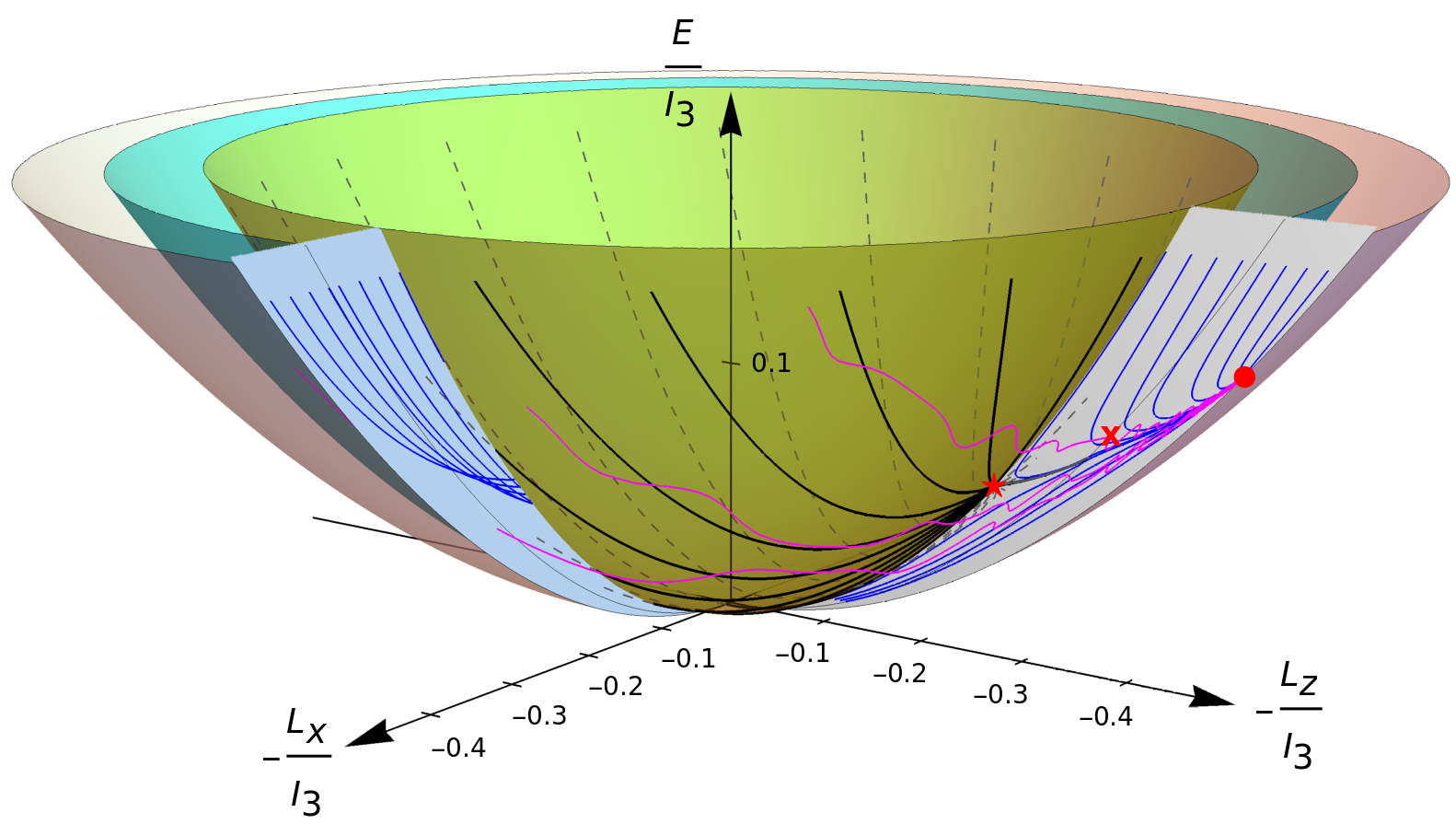}}
\subfloat[\label{LyLzE_phaseplot} $L_{y}$-$L_{z}$-$E$ phase plot]{\includegraphics[scale=0.15]{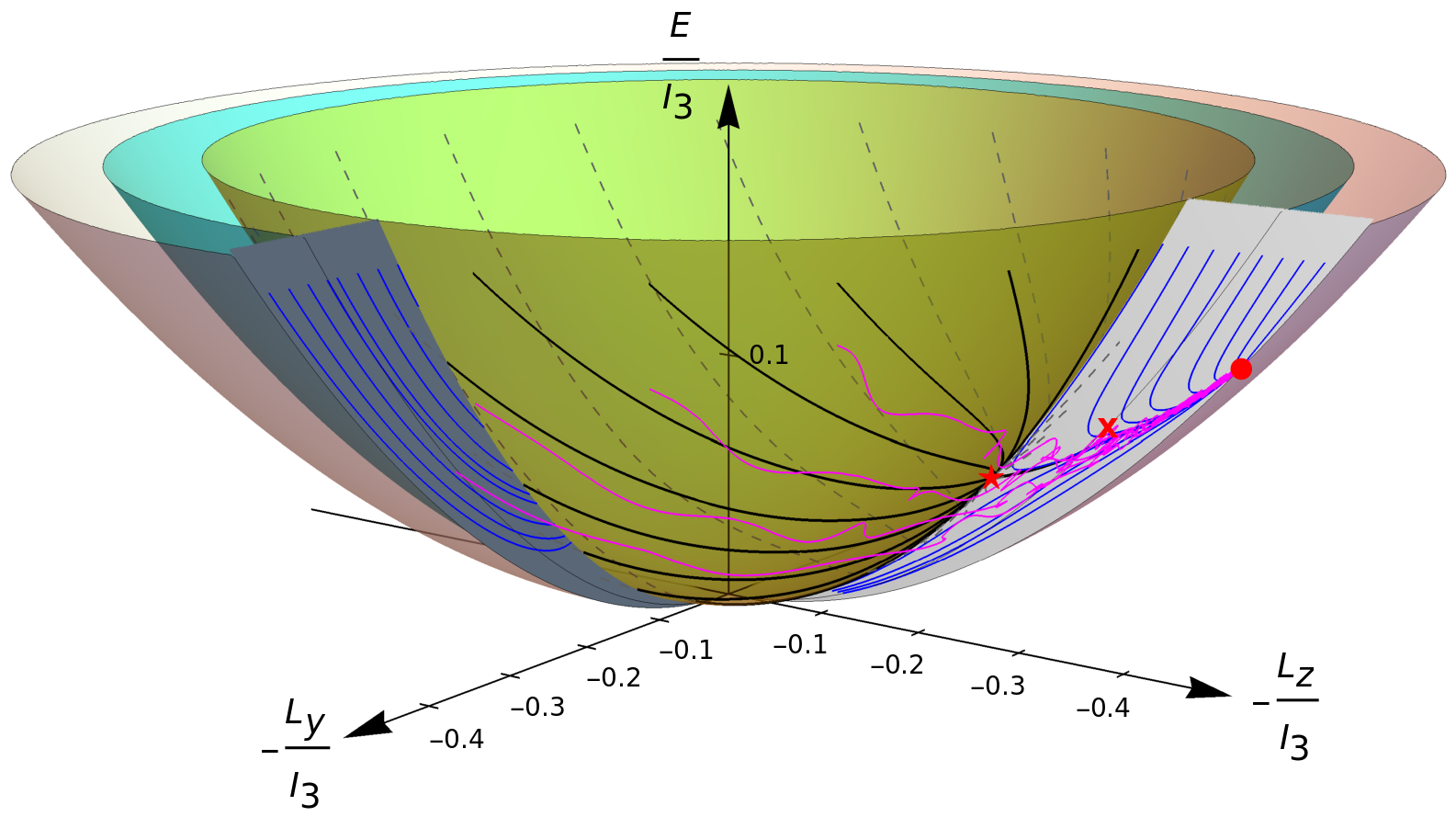}}\\
\caption{Evolution of slow-manifold trajectories in the $L_x-L_z-E$ and $L_y-L_z-E$ subspaces. A single 3D trajectory is depicted in blue, with its projections on the three bounding surfaces shown in black; the three fixed points in the $L_z-E$ plane are shown in red. The three bounding surfaces are the $L_x-E$ plane, the $L_z-E$ plane and the paraboloid $L_x^2 +L_z^2 = 2EI_3$ in (a); and the $L_y-E$ plane, the $L_z-E$ plane and the paraboloid $L_y^2 +L_z^2 = 2EI_3$ in (b).}
\label{3D_phaseplot}
\end{figure}

Since the slow-manifold dynamics occurs in a 4D hyper-space, that is not readily visualizable, Figure \ref{3D_phaseplot} shows projections of this dynamics onto a pair of 3D subspaces - the $(L_x,L_z,E)$ and $(L_y,L_z,E)$ spaces. The choice of subspaces exploits the property of the slow-manifold equations that, if $L_x$ or $L_y$ equals zero at the initial time, it is zero for all times. The trajectories in Figures \ref{3D_phaseplot}a and b arise from choosing $L_x =0$ and $L_y = 0$ at the initial time, and the above property implies that they are always confined to the respective subspaces. These figures show 3D dynamical trajectories being "guided" by their projections on the different bounding surfaces.

\section{Discussions and Conclusions}\label{DAC}

In this work we have examined in detail the inertial orientation dynamics of both spheroids and ellipsoids in a simple shear flow, with the emphasis being on large Stokes numbers. For the first time, it is shown that the orientation dynamics is best characterized in the space of infinite-$St$ invariants - these are the constants of motion in the classical mechanics problem that arises from neglecting the effect of the ambient shear flow. In the present case, this problem corresponds to the Euler top, and the infinite-$St$ invariants are the three components of the angular momentum and the rotational kinetic energy. The resulting 4D slow manifold allows for an economical depiction of the orientation dynamics. The governing autonomous system of equations on the slow manifold is obtained via the method of multiple scales for a spheroid, and using the method of averaging for an ellipsoid. Note that the usual physical-space representation of large-$St$ trajectories would in general lead to both self-intersections, as well as to different trajectories crossing each other. In contrast, uniqueness considerations imply that averaged trajectories on the slow manifold can only `intersect' at fixed points; even the crossings of full-solution trajectories are progressively minimized with increasing $St$, when plotted on the slow manifold. This idea of an economical slow-manifold representation may be generalizable to other scenarios. For instance, there have been several recent computational efforts that concern the orientation dynamics of a spheroid, in an ambient simple shear flow, in present of gravity \citep{cui_effect_2024}. It is likely that the best representation of the orientation dynamics, particularly for large $St$, is in the space of invariants corresponding to the Lagrange top\citep{landau1976}. The slow-manifold equations obtained here - (\ref{lxsmain}-\ref{nsmain}) for a spheroid and (\ref{lxmain}-\ref{emain}) for an ellipsoid - are far more amenable to numerical integration on account of eliminating the fast time scale, which renders the full system of equations increasingly stiff in the limit of large $St$. Detailed comparisons of the full-solution and averaged trajectories, carried out in earlier sections, show that the latter serve as a good approximation of the orientation dynamics for $St$ as small as $1000$.

The slow manifold for an ellipsoid has three fixed points, corresponding to vorticity-aligned rotation about each of its principal axes. Rotation about the shortest axis is a stable node, and corresponds to the long-time limit of trajectories starting from almost any point on the slow manifold; this being consistent with earlier numerical investigations\,\citep{lundell2011}. The remaining two fixed points correspond to vorticity-aligned rotation about the longest and intermediate axes, with the latter in particular being a singular point that does not allow for a local linear approximation. The singular behavior is higlighted by the possibility of this fixed point being approached in a finite-time, when moving along a critical trajectory. We expect the intermediate-axis-aligned rotation to be regularized with the addition of other physical effects, for instance, a gravitational torque induced by fluid inertial effects\,\citep{dabade2015,anand_motion_2022}. In the limiting case of a spheroid, this aforesaid singular point merges with either of the other fixed points, with the result that vorticity-aligned rotation about the shorter spheroid axis continues to be a stable node, while that about the longer one is a saddle point. 

It is worth emphasizing that we have examined the orientation dynamics of an ellipsoid for large but finite $St$, {\it and} over asymptotically long time scales of $O(St\dot{\gamma}^{-1})$. In the limit $St \rightarrow \infty$, but with $t$ fixed, one would approach the classical Euler top, in which case intermediate-axis-aligned rotation is a saddle point, with the longest and shortest-axis-aligned rotations being centers. Thus the aforementioned long-time constraint renders the large-$St$ limit a singular one. The singular nature of this limit is evident from what has already been mentioned in the paragraph above - that, in contrast to the Euler top, the intermediate-axis-aligned rotation is a singular albeit passive fixed point, while the shortest-axis-aligned rotation is a stable node. It is worth adding that the evolution of the spheroid towards its final equilibrium orientation is slow for both small and large $St$, as was already evident from the numerical investigations of Lundell and co-workers\citep{lundell2010}; the associated time scales being $O(St^{-1}\dot{\gamma}^{-1})$ and $O(St\dot{\gamma}^{-1})$. The slowdown at large $St$ may be interpreted as arising due to the infinite-$St$ limit\,(the Euler top) being an integrable one, and thereby only allowing for closed trajectories in the relevant phase space.


\begin{figure}
\subfloat[\label{spheroid_tra_topo} Trajectory topology of a spheroid $(\kappa=10)$$(St=1000)$  $(\theta_0,\phi_0,\psi_0,\dot{\theta_0},\dot{\phi_0},\dot{\psi_0}) \equiv(0.10472,5.4978,0,0,0,0.0)$]{\includegraphics[scale=0.29]{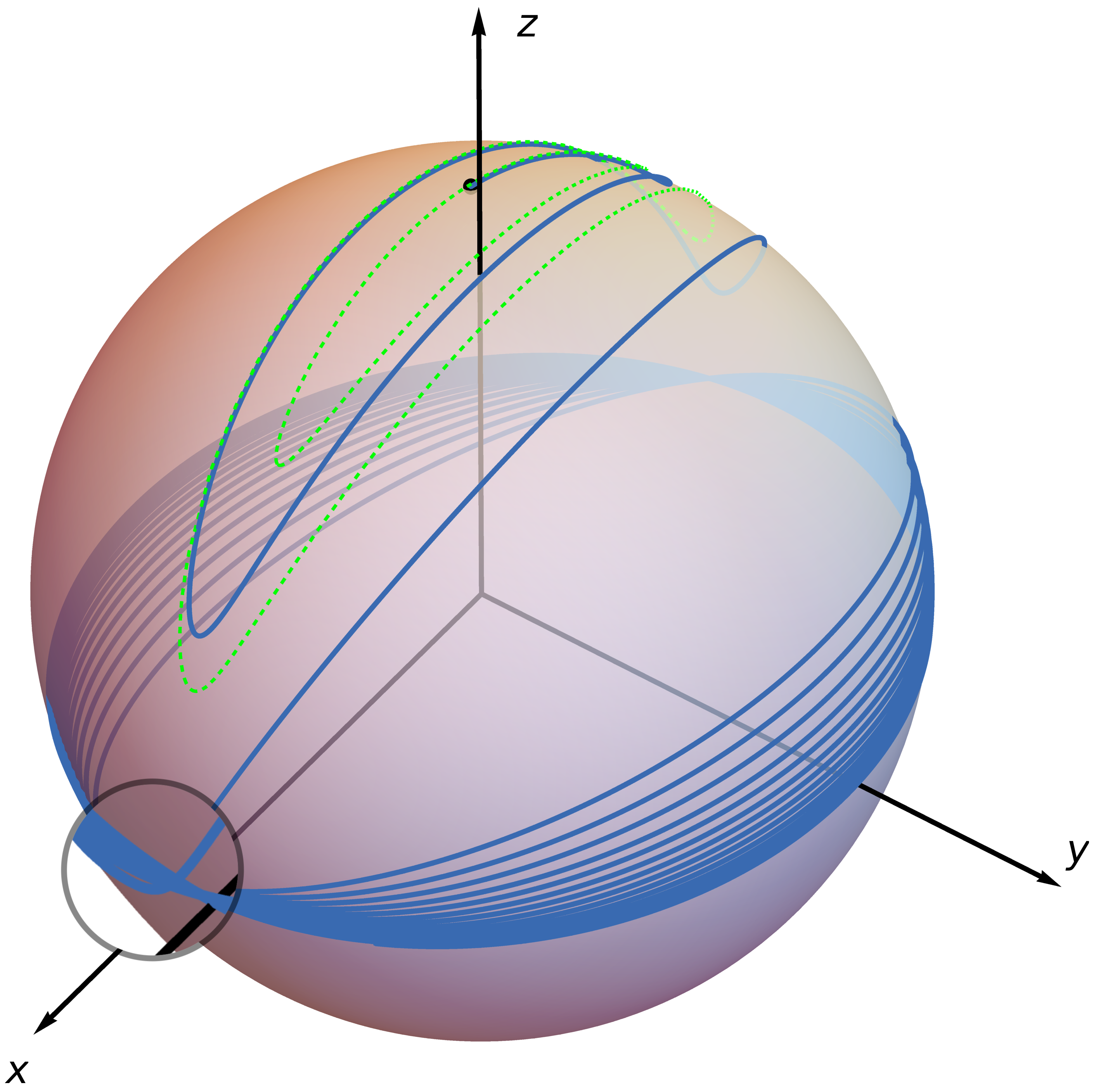}}
\hspace{0.08\linewidth}
\subfloat[\label{ellipsoid_tra_topo}Trajectory topology of an ellipsoid $(a,b,c)\equiv(0.1,0.15,1)$$(St=1000)$  $(\theta_0,\phi_0,\psi_0,\dot{\theta_0},\dot{\phi_0},\dot{\psi_0}) \equiv(0.10472,2.35619,0,0,0,0)$]{\includegraphics[scale=0.29]{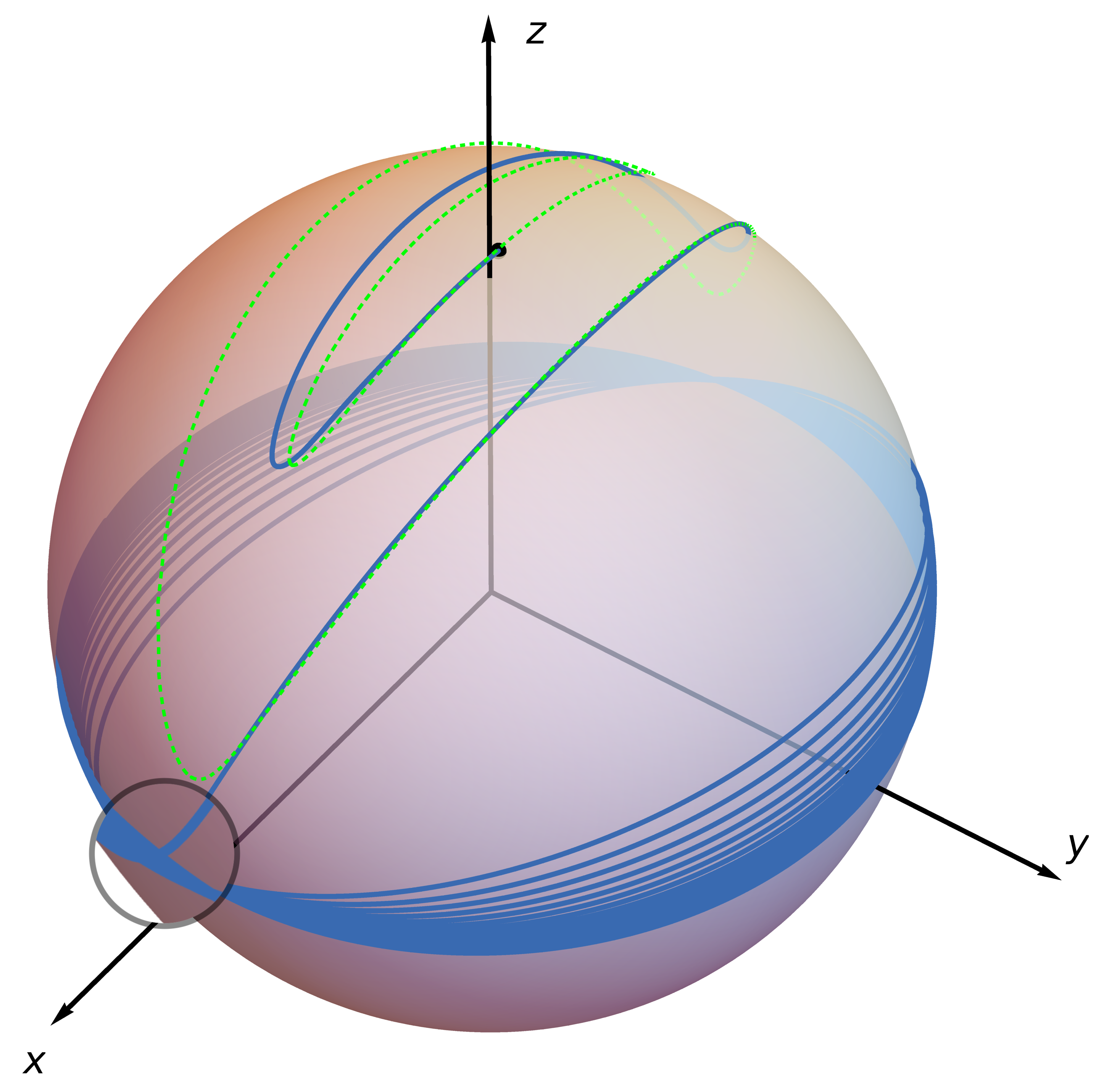}}\\
\caption{Unit sphere trajectories of a spheroid and an ellipsoid starting from rest, exhibiting a transition from an initial portion with Jeffery-like turns, to an eventual Euler-spiral. Black dots corresponds to the initial orientations chosen, and green dashed curves are Jeffery orbits drawn for comparison with the initial Jeffery-spiral; the circles depict a magnified view at the instant of the Jeffery-Euler transition.}
\label{unit_sphere_tra_top}
\end{figure}

Recall that the analysis in sections \ref{lsds} and \ref{lsde} was predicated on the fast time scale being $O(\dot{\gamma}^{-1})$, rather than $O(\Omega^{-1})$ - all of the particle trajectories shown thus far assumed an initial angular velocity of $O(\dot{\gamma})$, so as to conform to this scenario. For $\Omega \ll \dot{\gamma}$, there may no longer be a separation between the fast\,(of $O(\Omega^{-1}$)) and slow\,(of $O(St\dot{\gamma}^{-1})$) time scales, potentially rendering the methods of multiple scales and averaging inapplicable. An important example of this limit is a spheroid or ellipsoid starting from rest. Figures \ref{spheroid_tra_topo} and b show the unit-sphere trajectories in these cases, for $St =1000 $. In contrast to the large-$St$ trajectory topologies seen earlier, both the ellipsoid and spheroid start off moving along rather tightly spiralling trajectories, with individual turns resembling Jeffery orbits. The direction  of spiralling is towards the flow-gradient plane, and the behavior as a whole is consistent with what one expects for weak particle inertia\citep{dabade2016}. The latter makes sense since $\Omega$ is evidently small for short times, and therefore, the actual Stokes number based on $\Omega$ is small, despite the largeness of $St$. In both the aforementioned figures, one sees an abrupt transition from `Jeffery-spirals' with elliptical turns, to `Euler-spirals' with approximately circular turns, which marks the onset of strong particle inertial effects. It is of interest to estimate this onset time, because the averaged dynamics described in earlier sections would potentially be rendered irrelevant if the onset time turns out to be of order, or longer than, the slow time scale - the drift of an anisotropic particle starting from rest, towards the eventual stable orientation, would then be dominated by the initial transient, as opposed to the asymptotic averaged dynamics.

\begin{figure}
	\subfloat[$L_x-t/St^{1/2}$ plot; ($\kappa=2$) $(\dot{\theta}_0, \dot{\phi}_0,\dot{\psi}_0,\theta_0,\phi_0) \equiv(0,0,0,0.314,0.52)$.]{\includegraphics[scale=0.45]{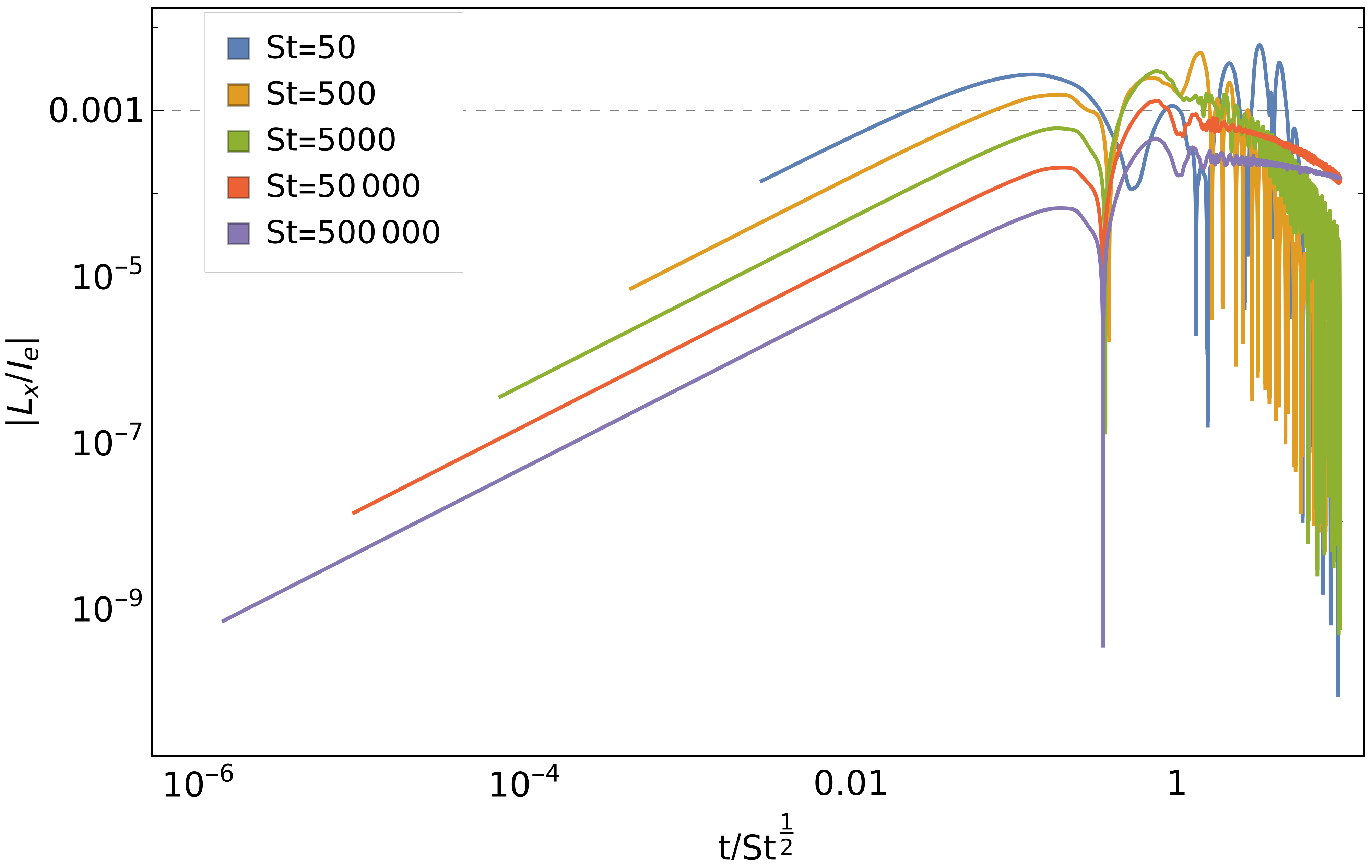}}
        \subfloat[$L_y-t/St^{1/2}$ plot; ($\kappa=2$) $(\dot{\theta}_0, \dot{\phi}_0,\dot{\psi}_0,\theta_0,\phi_0) \equiv(0,0,0,0.314,0.52)$.]{\includegraphics[scale=0.45]{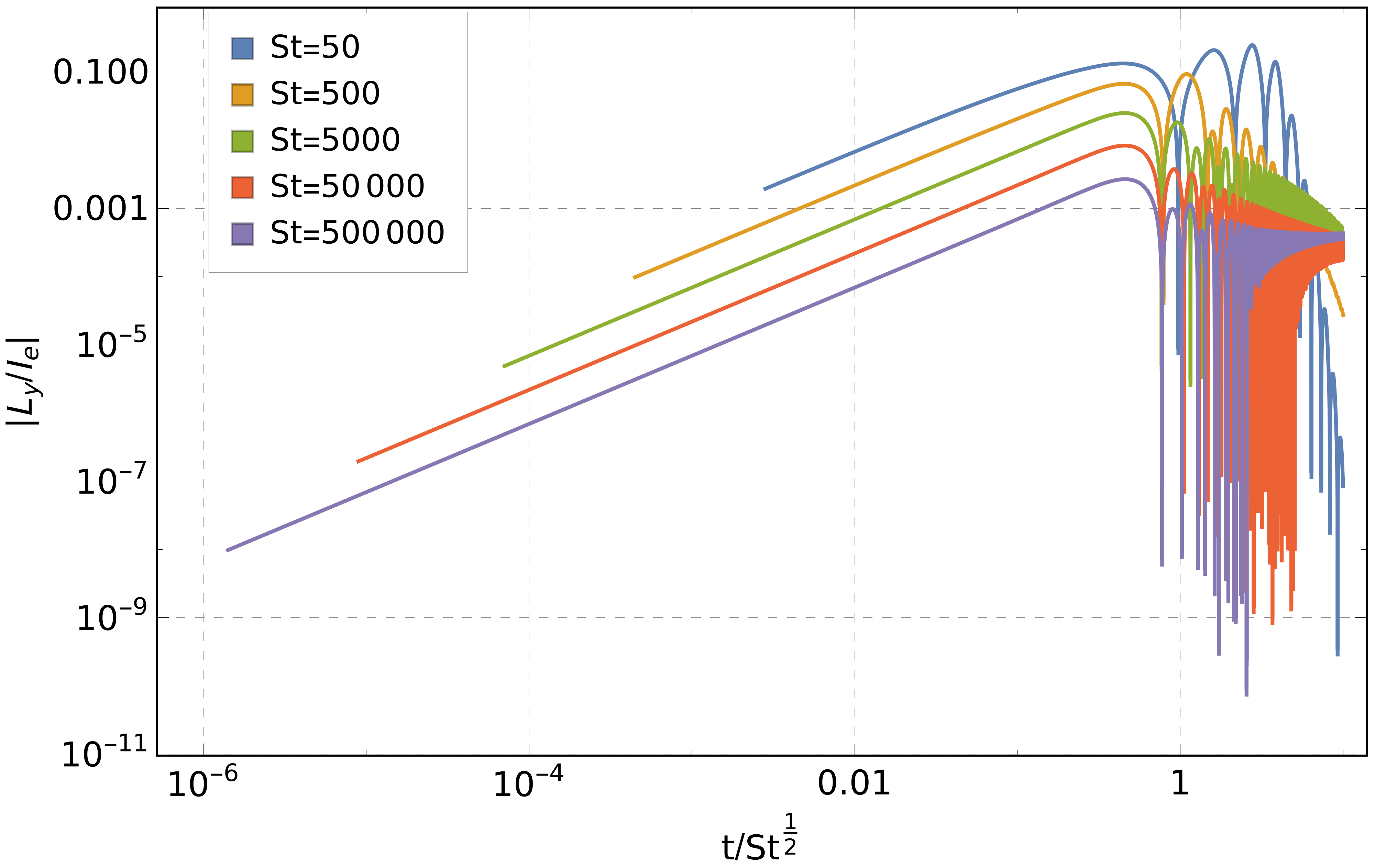}}\\
	\subfloat[$L_z-t/St^{1/2}$ plot; ($\kappa=2$) $(\dot{\theta}_0, \dot{\phi}_0,\dot{\psi}_0,\theta_0,\phi_0) \equiv(0,0,0,0.314,0.52)$.]{\includegraphics[scale=0.45]{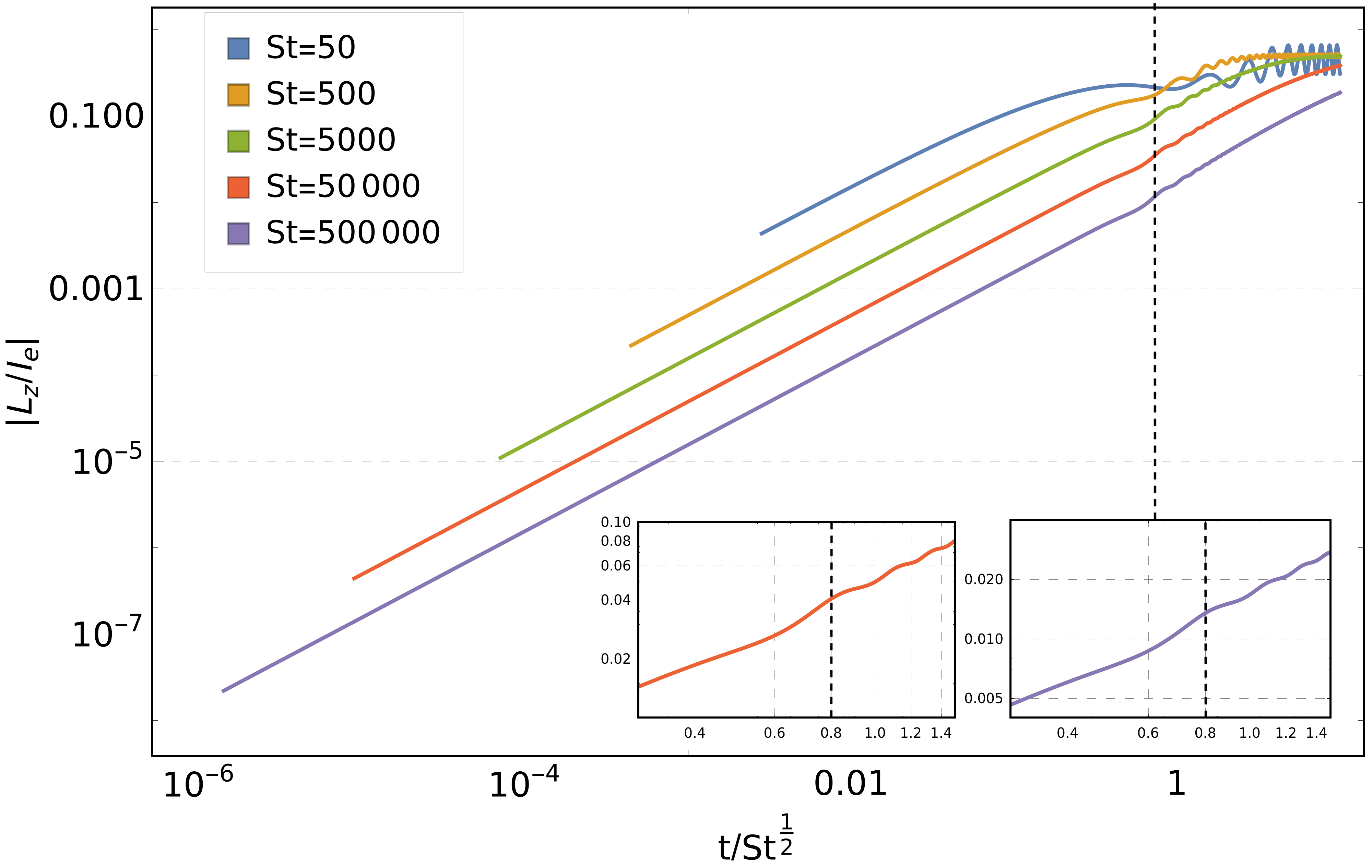}}
	\subfloat[$E-t/St^{1/2}$ plot; ($\kappa=2$)  $(\dot{\theta}_0, 
   \dot{\phi}_0,\dot{\psi}_0,\theta_0,\phi_0) \equiv(0,0,0,0.314,0.52)$.]{\includegraphics[scale=0.45]{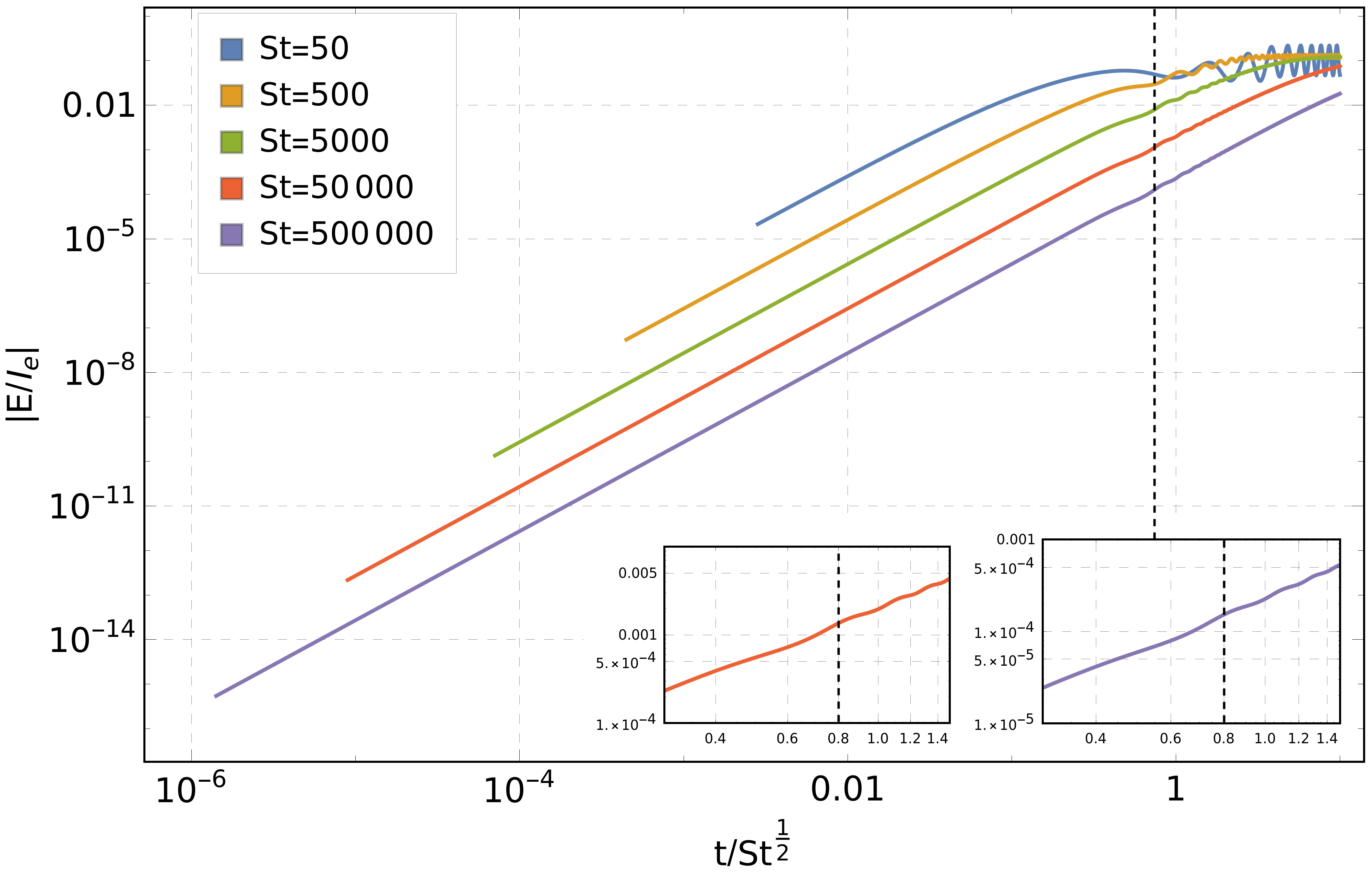}}
	\caption{Angular momentum and energy plotted as a function of $t/St^{\frac{1}{2}}$, for a spheroid with $\kappa=2$ starting from rest; $St = 50, 500, 5000, 50000, 500000$. The onset of strong particle inertial effects is marked vertical dashed lines; magnified views in the insets show the transition to averaged-dynamics, characterized by small amplitude oscillations, for the highest $St$.}
	\label{spheroid_transient_plot}
\end{figure}

\begin{figure}
	\subfloat[$L_x-$time Plot; ($a=3, b=2, c=1$) $(\dot{\theta}_0, \dot{\phi}_0,\dot{\psi}_0,\theta_0,\phi_0, \psi_0) \equiv(0,0,0,0.314,0.52,1.05)$.]{\includegraphics[scale=0.45]{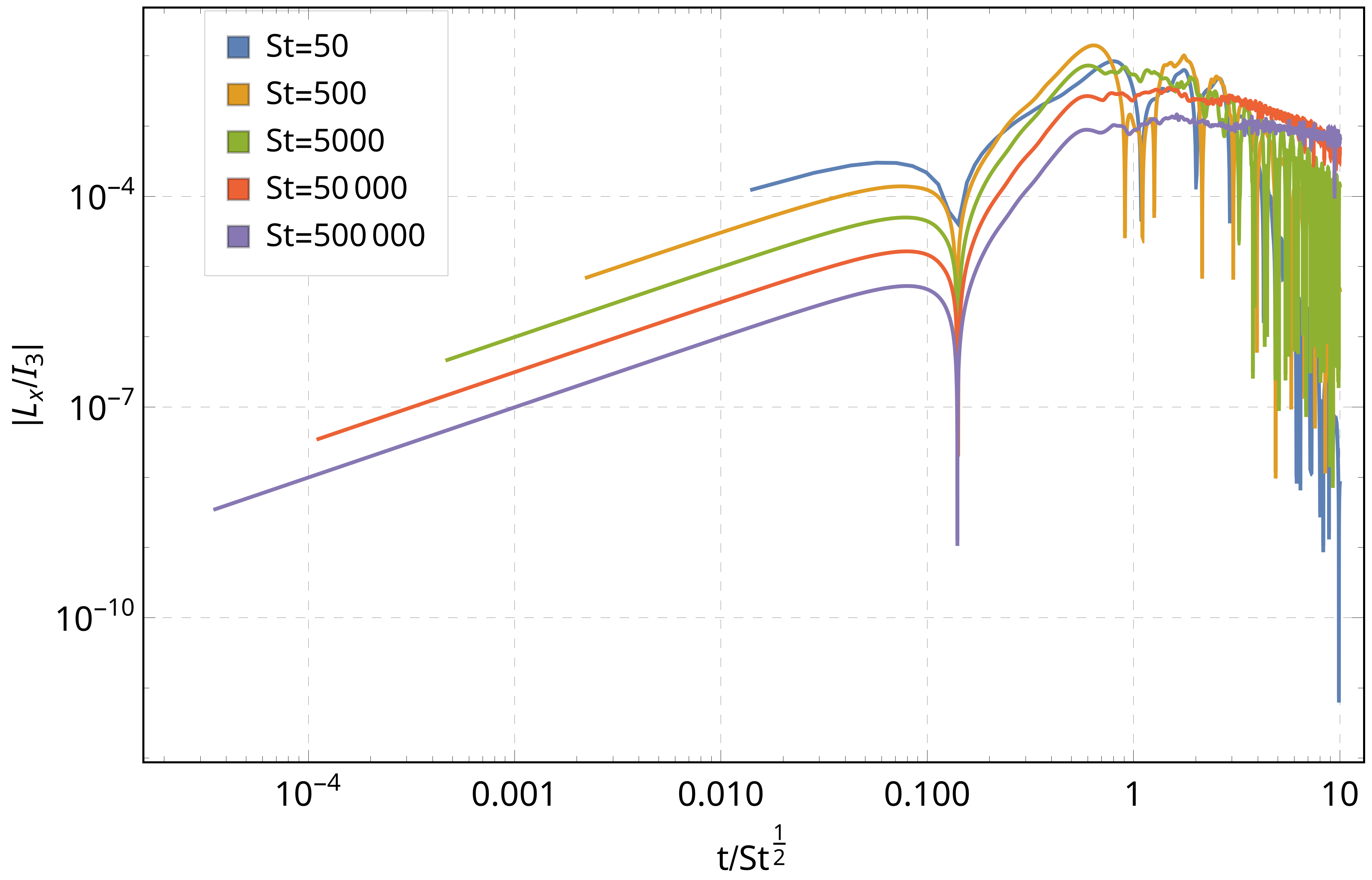}}
        \subfloat[$L_y-$time Plot; ($a=3, b=2, c=1$) $(\dot{\theta}_0, \dot{\phi}_0,\dot{\psi}_0,\theta_0,\phi_0, \psi_0) \equiv(0,0,0,0.314,0.52,1.05)$.]{\includegraphics[scale=0.45]{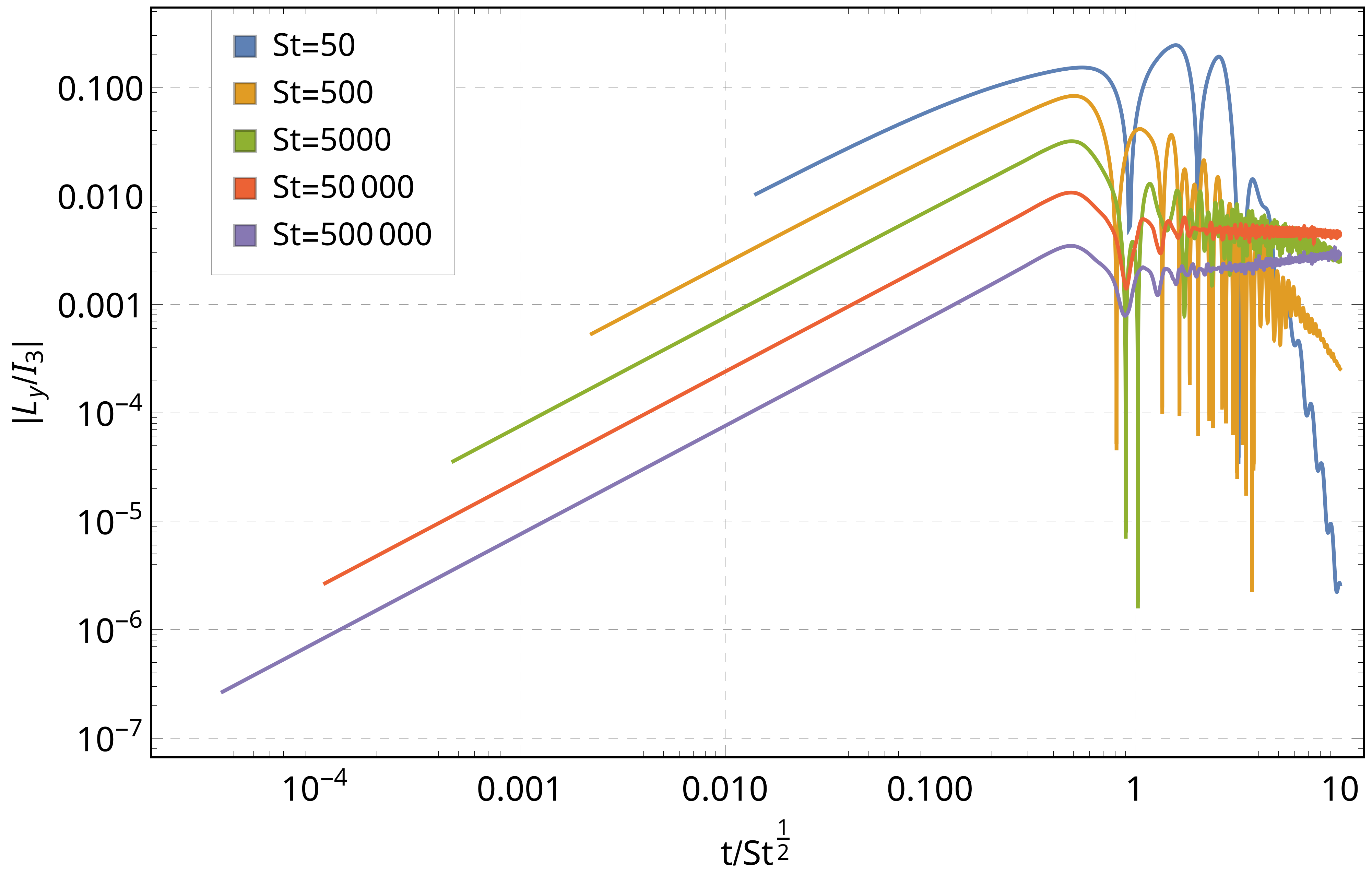}}\\
	\subfloat[$L_z-$time Plot; ($a=3, b=2, c=1$) $(\dot{\theta}_0, \dot{\phi}_0,\dot{\psi}_0,\theta_0,\phi_0, \psi_0) \equiv(0,0,0,0.314,0.52,1.05)$.]{\includegraphics[scale=0.45]{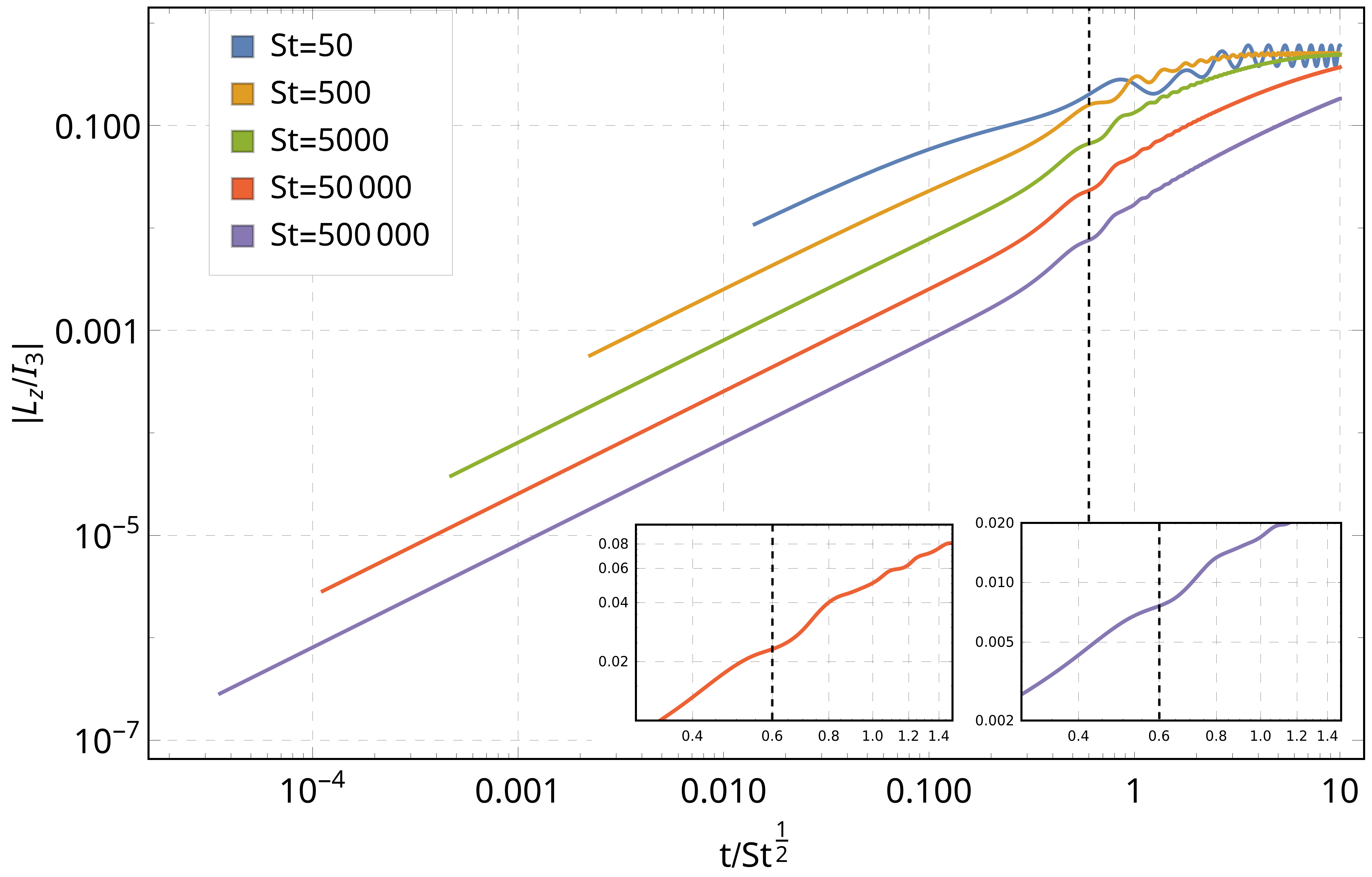}}
	\subfloat[$E-$time Plot; ($a=3, b=2, c=1$) $(\dot{\theta}_0, \dot{\phi}_0,\dot{\psi}_0,\theta_0,\phi_0, \psi_0) \equiv(0,0,0,0.314,0.52,1.05)$.]{\includegraphics[scale=0.45]{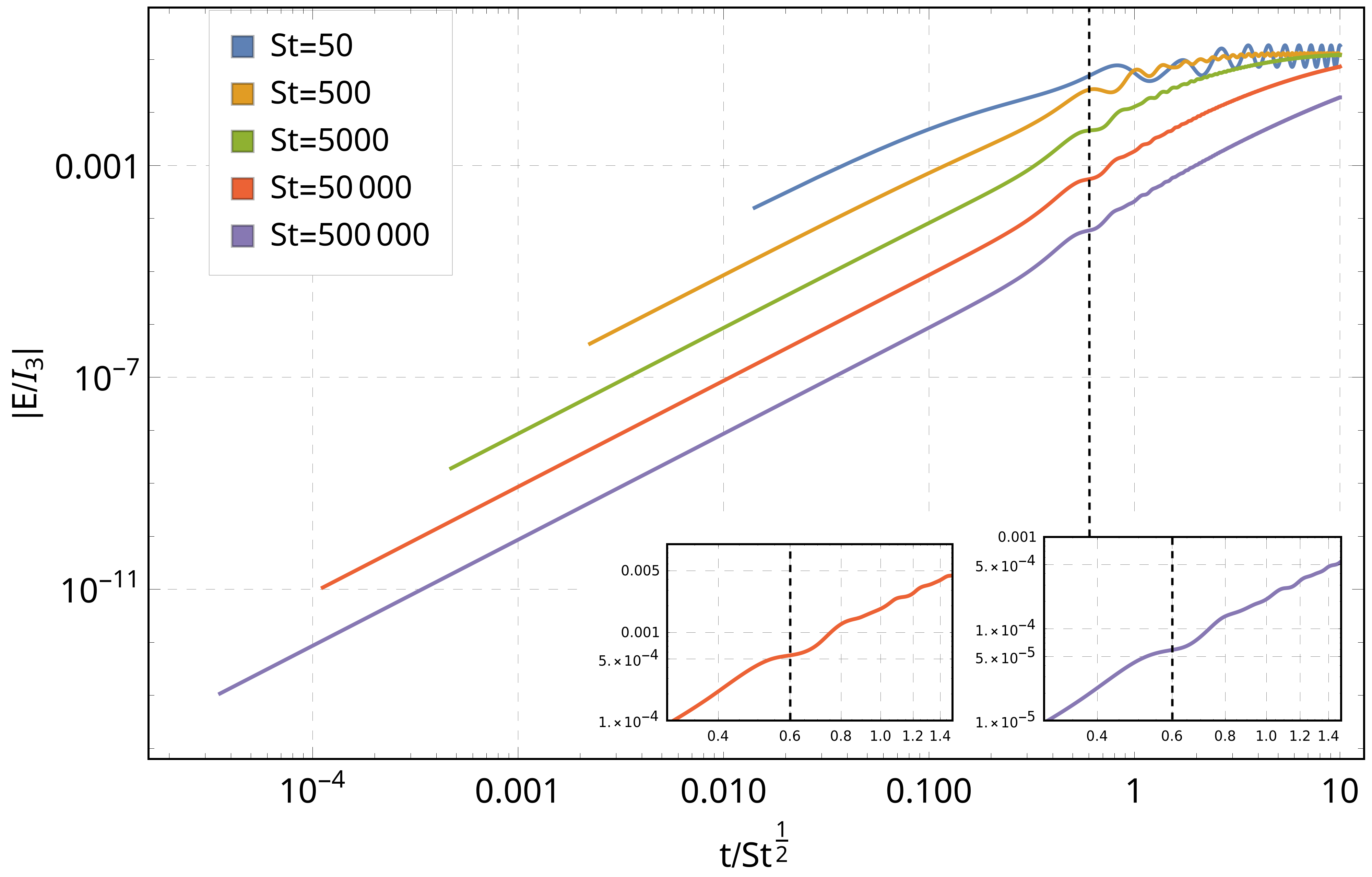}}
	\caption{Angular momenta and energy plotted as a function of $t/St^{\frac{1}{2}}$, for an ellipsoid with $b/c=2, a/c=3$ starting from rest; $St = 50, 500, 5000, 50000, 500000$. The onset of strong particle inertial effects is marked vertical dashed lines; magnified views in the insets show the transition to averaged-dynamics, characterized by small amplitude oscillations, for the highest $St$.}
	\label{Ellipsoid_transient_plot}
\end{figure}

\begin{figure}
\subfloat[\label{spheroid_tra_St} Trajectory topology of a spheroid $(\kappa=2)$  $(\theta,\phi,\psi,\dot{\theta},\dot{\phi},\dot{\psi}) \equiv(0.78539,2.35619,0,0,0,0.0)$]{\includegraphics[scale=0.45]{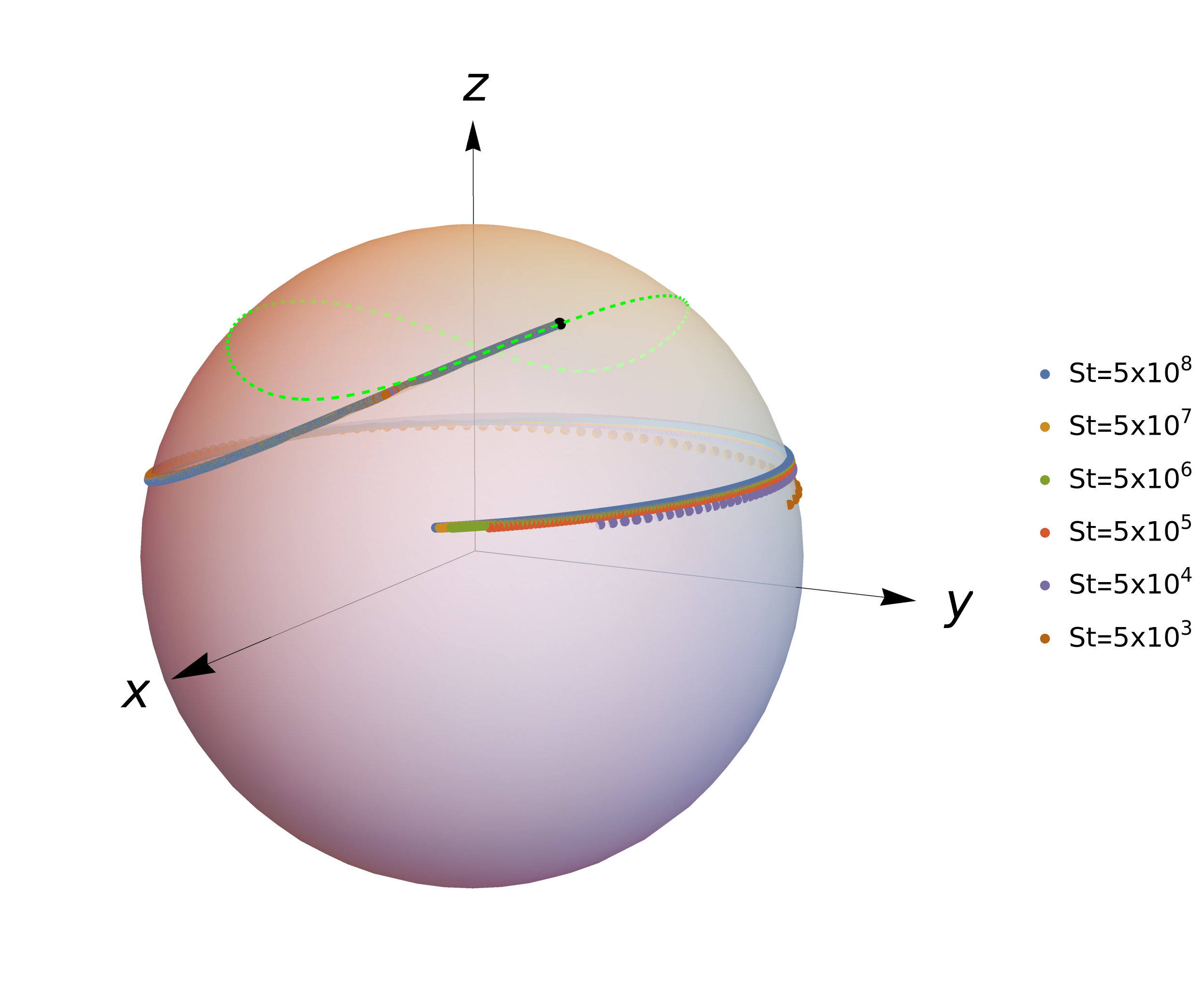}}
\subfloat[\label{ellipsoid_tra_topo}Trajectory topology of an ellipsoid $(a,b,c)\equiv(0.5,0.55,1)$ $(\theta,\phi,\psi,\dot{\theta},\dot{\phi},\dot{\psi}) \equiv(0.78539,2.35619,0,0,0,0)$]{\includegraphics[scale=0.45]{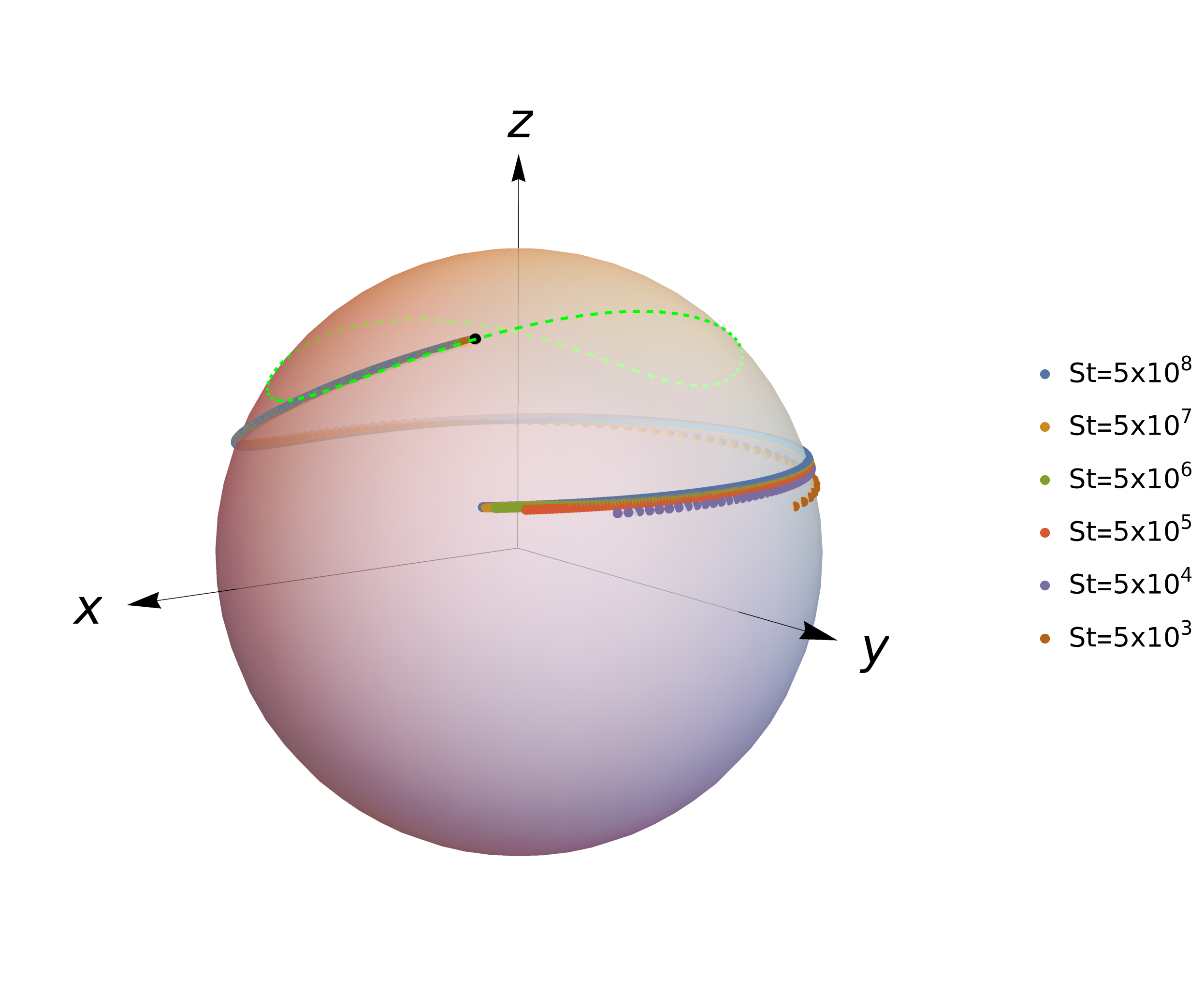}}\\
\caption{Unit sphere trajectories of a spheroid and an ellipsoid, starting from rest, until a time $t = St^{1/2} \dot{\gamma}^{-1}$ for different large $St$; the trajectory length asymptotes to an order unity value for $St \rightarrow \infty$. The different trajectories depart from a Jeffery orbit\,(green dashed curve), starting from the same initial orientation\,(black dot), at times much shorter than a Jeffery period.}
\label{unit_tra_St}
\end{figure}

To obtain an estimate for the onset time, we return to the governing equation, (\ref{ultima}), written in expanded form as:
\begin{equation}\label{govg}
I_i\dot{\omega_i}+\epsilon_{ijk}I_k\omega_j\omega_k+\frac{8\pi X_i}{St}(\omega_i-\omega_{i,J})=0.
\end{equation} 
For a particle starting from rest, the initial balance is between $I_i\dot{\omega}_i$ and the driving viscous torque of $O(X_i St^{-1}\omega_{i,J})$; the spheroid orientation remains virtually unchanged in this initial period, with the angular velocity increasing linearly with time as $O(St^{-1}\dot{\gamma}^2t)$. This linear increase continues until a time of $O(St^{\frac{1}{2}}\dot{\gamma}^{-1})$, when the angular velocity becomes $O(St^{-\frac{1}{2}}\dot{\gamma})$, and the term $\epsilon_{ijk}I_k\omega_j\omega_k$ enters the leading order balance. For larger times, $\omega \gg O(St^{-\frac{1}{2}}\dot{\gamma})$, with the first two terms in (\ref{govg}) now being dominant and of the same order, and with the viscous torque corresponding to a slow damping - this, of course, is the large-$St$ averaged dynamics already analyzed. The initial transient is thus expected to persist for a time of $O(St^{\frac{1}{2}}\dot{\gamma}^{-1})$ which is asymptotically small in relation to the slow time scale for $St \gg 1$. Figures \ref{spheroid_transient_plot} and \ref{Ellipsoid_transient_plot} show the evolutions of $L_x, L_y, L_z$ and $E$ for a prolate spheroid\,($\kappa=2$) and an ellipsoid\,($b/c=2, a/c=3$), as a function of time scaled by $St^{\frac{1}{2}}\dot{\gamma}^{-1}$, over a range of large $St$. The onset of small-amplitude wiggles \,(representative of fast oscillations superposed on a slow drift) is seen to begin at virtually the same time for all of these curves, corroborating the estimate of the onset time above.
It is worth noting that the aforementioned scaling estimates of an $O(St^{\frac{1}{2}}\dot{\gamma}^{-1})$ transient, and the associated $O(St^{-\frac{1}{2}}\dot{\gamma})$ transitional angular velocity implies that, even as $St \rightarrow \infty$, the particle undergoes an angular displacement of order unity from its initial rest state until the onset of strong inertial effects. The order unity angular displacement implies that, in contrast to the trajectories in figures \ref{unit_sphere_tra_top}a and b, the transient for $St \rightarrow \infty$ cannot involve multiple Jeffery-like turns, and that the idea of a transition from Jeffery-like to Euler-like dynamics may therefore be misleading. Figures \ref{unit_tra_St}a and b, which show trajectories integrated for one transition time, confirm this aspect; although, one needs to go to a $St$ of $O(10^8)$ for this purpose. The large-$St$ trajectory in these figures is seen to deviate from the Jeffery orbit, corresponding to the same initial orientation, well before executing a single turn.

On a final note, it is worth mentioning that a complete solution of the large-$St$ rotation problem, from an initial state of rest, requires a matched asymptotic expansions approach, where the `initial' conditions for the averaged dynamics arise from the long-time limit of the transitional dynamics on times of $O(St^{\frac{1}{2}}\dot{\gamma}^{-1})$. During the transitional period, all of the terms in (\ref{govg}) are comparable in magnitude, with the only simplification being that $\omega_i - \omega_{i,J}$ in the viscous torque may be approximated as $\omega_{i,J}$. The latter is of little consequence since, as pointed out above, the order unity change in the spheroid orientation implies that $\omega_{i,J}$ cannot be regarded as a constant even in the infinite-$St$ limit, as a result of which the transitional dynamics is not solvable in closed form.

\section{Acknowledgments}
The authors would like to acknowledge the valuable support and guidance of Dr. Shyama Prasad Das, Indian Institute of Technology Madras, without whom this work would not be possible. 
\bibliography{Latest_Draft.bib}{}
\bibliographystyle{apsrev4-1}
\appendix
\section{The Euler-top averages}\label{LOA}
\subsection{Spheroid}\label{spheravg}
Here, we list all relevant Euler angle averages necessary to derive the slow manifold description given by (\ref{lxsmain}-\ref{nsmain}):\\
\begin{center}
\begin{tabular}{ |c|c| }
	\hline
	\multicolumn{2}{ |c| }{}\\
		\multicolumn{2}{ |c| }{Table of Averages: Spheroid}\\
	\multicolumn{2}{ |c| }{} \\
	\hline
	  & \\
	$\langle \cos{\theta}\rangle =\frac{{L_z^{(0)}} h}{{L^{(0)}}}$ & $\langle \cos^2{\theta}\rangle =\frac{{L_z^{(0)}}^2 h^2}{{L^{(0)}}^2}+\frac{{L_{xy}^{(0)}}^2 \bar{h}^2}{2 {L^{(0)}}^2}$ \\
	& \\
	$\langle \sin{\theta}\sin{\phi}\rangle =\frac{{L_x^{(0)}} h}{{L^{(0)}}}$& $\langle \sin^2{\theta}\sin^2{\phi}\rangle =\frac{{L_x^{(0)}}^2 h^2}{{L^{(0)}}^2}+\frac{{L_y^{(0)}}^2 \bar{h}^2}{2{L_{xy}^{(0)}}^2}+\frac{{L_x^{(0)}}^2 {L_z^{(0)}}^2 \bar{h}^2}{2{L_{xy}^{(0)}}^2 {L^{(0)}}^2}$ \\
	&  \\
	$\langle \sin{\theta}\cos{\phi}\rangle =-\frac{{L_y^{(0)}} h}{{L^{(0)}}}$&	$\langle \sin^2{\theta}\cos^2{\phi}\rangle =\frac{{L_y^{(0)}}^2 h^2}{{L^{(0)}}^2}+\frac{{L_x^{(0)}}^2 \bar{h}^2}{2{L_{xy}^{(0)}}^2}+\frac{{L_y^{(0)}}^2 {L_z^{(0)}}^2 \bar{h}^2}{2{L_{xy}^{(0)}}^2 {L^{(0)}}^2}$\\
	& \\
	 $\langle \sin{\theta}\cos{\phi}\cos{\theta}\rangle =-\frac{{L_z^{(0)}}{L_y^{(0)}} h^2}{{L^{(0)}}^2}+\frac{{L_z^{(0)}}{L_y^{(0)}} \bar{h}^2}{2 {L^{(0)}}^2}$  	& $\langle \sin{\theta}\sin{\phi}\cos{\theta}\rangle =\frac{{L_z^{(0)}}{L_x^{(0)}} h^2}{{L^{(0)}}^2}-\frac{{L_z^{(0)}}{L_x^{(0)}} \bar{h}^2}{2 {L^{(0)}}^2}$\\
  	&   \\
	\hline
\end{tabular}
\end{center}

\subsection{Ellipsoid}\label{elavgs}
This section includes a complete list of all averages that need to be obtained to derive the slow manifold description given by (\ref{lxmain}-\ref{emain}). All the required averages reduce to linear combinations of $\langle t^2_{i3,LO}\rangle $ or $\langle z^2_i\rangle $, with the coefficients being functions of the slow variables in the form $\alpha_{j3}\alpha_{k3}$. Simplification to the aforementioned final form, makes extensive use of $t_{ij}^T=\alpha_{ij}t_{ij,LO}^T$, (\ref{pdtavg}) and the properties of $\alpha_{ij}$ as a rotation matrix. For the ease of reading and intrepretation we define $t_{i1,LO}=x_i$, $t_{i2,LO}=y_i$ and $t_{i3,LO}=z_i$ and $t_{i1}=x'_i$, $t_{i2}=y'_i$, $t_{i3}=z'_i$. 

\begin{center}
\begin{small}
\begin{tabular}{ |c| }
\hline\\
	\multicolumn{1}{ |c| }{Table of Averages :Ellipsoid} \\\\\hline\\
	\multicolumn{1}{ |c| }{$\langle t_{ki,LO}t_{kj,LO}\rangle =\delta_{ij}(\frac{1}{2}(1-\delta_{i3})(1-\langle z_k^2\rangle )+\delta_{i3}\langle z_k^2\rangle  $} \\\\\hline\\	$\langle x_1y_1,x_2y_2,x_3y_3,x_1z_1,x_2z_2,x_3z_3,y_1z_1,y_2z_2,y_3z_3\rangle =0$ \\
	$\langle x_1^2,y_1^2\rangle =(1-\langle z_1^2\rangle )/2$ \\
	$\langle x_2^2,y_2^2\rangle =(1-\langle z_2^2\rangle )/2$ \\
	$\langle x_3^2,y_3^2\rangle =(1-\langle z_3^2\rangle )/2$ \\\hline\\
	\multicolumn{1}{ |c| }{$\langle t_{il,LO}t_{jm,LO}t_{kn,LO}\rangle =\frac{1}{2}\epsilon_{ijk}\epsilon_{lmn}(\delta_{i3}\langle z_i^2\rangle +\delta_{j3}\langle z_j^2\rangle +\delta_{k3}\langle z_k^2\rangle )$} \\\\\hline\\
	$\langle x_1x_2z_3,x_1z_2x_3,z_1x_2x_3,y_1y_2z_3,y_1z_2y_3,z_1y_2y_3\rangle =0$ \\
	$\langle x_1x_2y_3,x_1y_2x_3,y_1x_2x_3,y_1y_2x_3,y_1x_2y_3,x_1y_2y_3\rangle =0$ \\ 				            	 $\langle z_1z_2x_3,z_1x_2z_3,x_1z_2z_3,z_1z_2y_3,z_1y_2z_3,y_1z_2z_3\rangle =0$ \\
	$\langle x_1x_2x_3\rangle =\langle y_1y_2y_3\rangle =\langle z_1z_2z_3\rangle =0$\\
	 $\langle z_1x_2y_3,-z_1y_2x_3\rangle =\langle z_1^2\rangle /2$ \\
	 $\langle y_1z_2x_3,-x_1z_2y_3\rangle =\langle z_2^2\rangle /2$\\
	$\langle x_1y_2z_3,-y_1x_2z_3\rangle =\langle z_3^2\rangle /2$ \\\hline	\\
	\multicolumn{1}{ |c| }{$\epsilon_{ijk}\langle t_{i3,LO}t_{j1}t_{k2}\rangle =\epsilon_{ijk}^2\langle t_{i3,LO}^2\rangle =\epsilon_{ijk}^2\langle z_i^2\rangle $} \\\\\hline\\
$\langle z_1x_2'y_3'\rangle =-\langle z_1x_3'y_2'\rangle =\langle z_1^2\rangle $ \\
$\langle z_2x_3'y_1'\rangle =-\langle z_2x_1'y_3'\rangle =\langle z_2^2\rangle $ \\ 	
$\langle z_3x_1'y_2'\rangle =-\langle z_3x_2'y_1'\rangle =\langle z_3^2\rangle $\\\hline\\	
$\epsilon_{ijk}\langle t_{il}t_{j1}t_{k2}\rangle =\delta_{l3}\epsilon_{ijk}^2(\alpha_{13}\langle z_j^2\rangle $\\
$+\alpha_{23}\langle z_k^2\rangle +\alpha_{33}\langle z_i^2\rangle )/2+\epsilon_{ijk}^2(1-\delta_{l3})\alpha_{l3}\alpha_{33}(\langle z_i^2\rangle -\delta_{l1}\langle z_j^2\rangle -\delta_{l2}\langle z_k^2\rangle )/2$\\\\\hline
\begin{tabular}{ c|c }
\\
$\langle x_1'x_2'y_3'\rangle =\alpha_{13}\alpha_{33}(\langle z_1^2\rangle -\langle z_2^2\rangle )/2$ & $\langle z_1'x_2'y_3'\rangle =(\alpha_{13}^2\langle z_2^2\rangle +\alpha_{23}^2\langle z_3^2\rangle +\alpha_{33}^2\langle z_1^2\rangle )/2$\\
$\langle x_2'x_3'y_1'\rangle =\alpha_{13}\alpha_{33}(\langle z_2^2\rangle -\langle z_3^2\rangle )/2$ & $\langle z_2'x_3'y_1'\rangle =(\alpha_{13}^2\langle z_3^2\rangle +\alpha_{23}^2\langle z_1^2\rangle +\alpha_{33}^2\langle z_2^2\rangle )/2$ \\
$\langle x_3'x_1'y_2'\rangle =\alpha_{13}\alpha_{33}(\langle z_3^2\rangle -\langle z_1^2\rangle )/2$ & $\langle z_3'x_1'y_2'\rangle =(\alpha_{13}^2\langle z_1^2\rangle +\alpha_{23}^2\langle z_2^2\rangle +\alpha_{33}^2\langle z_3^2\rangle )/2$\\
$\langle y_1'x_3'y_2'\rangle =\alpha_{23}\alpha_{33}(\langle z_2^2\rangle -\langle z_1^2\rangle )/2$ &
$\langle z_1'x_3'y_2'\rangle =-(\alpha_{13}^2\langle z_3^2\rangle +\alpha_{23}^2\langle z_2^2\rangle +\alpha_{33}^2\langle z_1^2\rangle )/2$ \\
$\langle y_2'x_1'y_3'\rangle =\alpha_{23}\alpha_{33}(\langle z_3^2\rangle -\langle z_2^2\rangle )/2$ &
$\langle z_2'x_1'y_3'\rangle =-(\alpha_{13}^2\langle z_1^2\rangle +\alpha_{23}^2\langle z_3^2\rangle +\alpha_{33}^2\langle z_2^2\rangle )/2$ \\
$\langle y_3'x_2'y_1'\rangle =\alpha_{23}\alpha_{33}(\langle z_1^2\rangle -\langle z_3^2\rangle )/2$ &
$\langle z_3'x_2'y_1'\rangle =-(\alpha_{13}^2\langle z_2^2\rangle +\alpha_{23}^2\langle z_1^2\rangle +\alpha_{33}^2\langle z_3^2\rangle )/2$ \\\hline
	\multicolumn{2}{ |c| }{}\\
	\multicolumn{2}{ |c| }{$\langle t_{il}t_{im}\rangle =\alpha_{l3}\alpha_{m3}(3\langle z_i^2\rangle -1)/2+\delta_{lm}(1-\langle z_i^2\rangle )/2$}\\
		\multicolumn{2}{ |c| }{}\\\hline\\
  $\langle y_i'z_i'\rangle =\alpha_{23}\alpha_{33}(3\langle z_i^2\rangle -1)/2$ & $\langle x_i'^2\rangle =\alpha_{13}^2(3\langle z_i^2\rangle -1)/2+(1-\langle z_i^2\rangle )/2$\\
 $\langle x_i'z_i'\rangle =\alpha_{33}\alpha_{13}(3\langle z_i^2\rangle -1)/2$ & $\langle y_i'^2\rangle =\alpha_{23}^2(3\langle z_i^2\rangle -1)/2+(1-\langle z_i^2\rangle )/2$  \\
    $\langle x_i'y_i'\rangle =\alpha_{13}\alpha_{23}(3\langle z_i^2\rangle -1)/2$ & $\langle z_i'^2\rangle =\alpha_{33}^2(3\langle z_i^2\rangle -1)/2+(1-\langle z_i^2\rangle )/2$ \\\hline
\end{tabular}
\end{tabular}
\end{small}
\end{center}

\section{The First Order Dynamics of a Spheroid in Shear flow}\label{FOD}

In addition to the leading order behavior of a spheroid over long times, the method of multiple scales can be used to obtain the $O(1/St)$ corrections to the averaged dynamics. To demonstrate this, we consider the simple case of a vorticity aligned Euler cone ($L^{(0)}_x=0,L^{(0)}_y=0$). 
The above condition on the leading order angular momenta necessitates from (\ref{omg1def}-\ref{omg3def}) that the initial conditions for the Euler angles of the form are $\dot{\theta_0}=0$ and $\dot{\psi_0}=\frac{I_e-I_a}{I_a}\dot{\phi_0}\cos\theta_0$. Note that there are therefore only three initial conditions to the problem in the full solution: $(\theta_0,\dot{\phi_0},\phi_0)$, of which $\phi_0$, although irrelevant for the leading order dynamics, becomes important in determining the first order behavior. Similarly, the choice of the terms $\alpha_{ij}$ with $j\neq3$ are no longer arbitrary and must be specified for a given $\phi_0$. For calculational simplicity we choose the following with $L_{xy}=\sqrt{L_x^2+L_y^2}$ and $\sqrt{L_x^2+L_y^2+L_z^2}$:
\begin{align}\label{fodij}
	\alpha_{ij}=
	\begin{pmatrix}
		\frac{L_{y}}{L_{xy}}&\frac{L_{x} L_{z}}{L L_{xy}}&\frac{L_{x}}{L}\\
		-\frac{L_{x}}{L_{xy}}&\frac{L_{y} L_{z}}{L L_{xy}}&\frac{L_{y}}{L}\\
		0&-\frac{L_{xy}}{L}&\frac{L_{z}}{L}\\
	\end{pmatrix}.
\end{align}
For a spheroid with $L^{(0)}_x=0,L^{(0)}_y=0$ we may impose without loss of generality that $\lim_{L_x,L_y\rightarrow0}L_y/L_{xy}=1$ provided a suitable value of $\phi_0$ is chosen. Using the above, (\ref{alphatranss}) simplifies to:
\begin{equation}\label{fodtraj}
\begin{pmatrix} 
\sin{\theta}\sin{\phi} & -\sin{\theta}\cos{\phi} & \cos\theta
\end{pmatrix}
=
\begin{pmatrix}
\bar{h}\sin{(\int^t_0 \frac{L}{I_e} \text{d}t+\phi_0)} & -\bar{h}\cos{(\int^t_0 \frac{L}{I_e} \text{d}t+\phi_0)}& h
\end{pmatrix},
\end{equation}

The rate of change of $L_z$ and $\eta$ correct to the first order in the multiple scales formulation, are governed by (\ref{lzms}) and (\ref{nms}). On combining these with the expressions for leading order dynamics (equations (\ref{lzsmain}) and (\ref{nsmain})) give the following expressions for the rate of change of ${L_z^{(1)}}$ and ${\eta^{(1)}}$:\\
\begin{align}\label{lzsfod}
\frac{\partial{L_z^{(1)}}}{\partial t_1}&=-4\pi Y_c \big( \frac{\kappa^2-1}{1+\kappa^2}\big) \bar{h}^2 \cos{2(\int^t_0 \frac{L}{I_e} \text{d}t+\phi_0)}, \\\label{nsfod}
\frac{\partial{\eta^{(1)}}}{\partial t_1}&=0.\\
\intertext{For our assumption that $<{L_z^{(1)}}>={\eta^{(1)}}=0$ made in obtaining the leading order solution to be consistent with these equations, no constants or functions of the slow time scale can arise on integration. This implies that ${\eta^{(1)}}$ is zero for all time. Integrating (\ref{lzsfod}) is made more complex by the slowly varying time period. The integral can be evaluated using the WKB approximation for a periodic function with a slowly varying frequency which approximates $(\int^{t_1}_0 \frac{L}{I_e} \text{d}t+\phi_0)$  to $(t_1 / t_2 \int^{t_2}_0 \frac{L^{(0)}(\xi)}{I_e} \text{d}\xi +\phi_0)$ . Using this result, we obtain the evolution of $L_z^{(1)}$ as: }
\label{lzsfod2}
{L_z^{(1)}}&=-2\pi Y_c \big( \frac{\kappa^2-1}{1+\kappa^2}\big) \frac{\bar{h}^2 t_2}{\int^{t_2}_0 \frac{L^{(0)}(\xi)}{I_e} d\xi} \sin{2 (St \int^{t_2}_0 \frac{L^{(0)}(\xi)}{I_e} \text{d}\xi + \phi_0)}.
\end{align}

\begin{figure}
 \centering
 \includegraphics[scale=0.06]{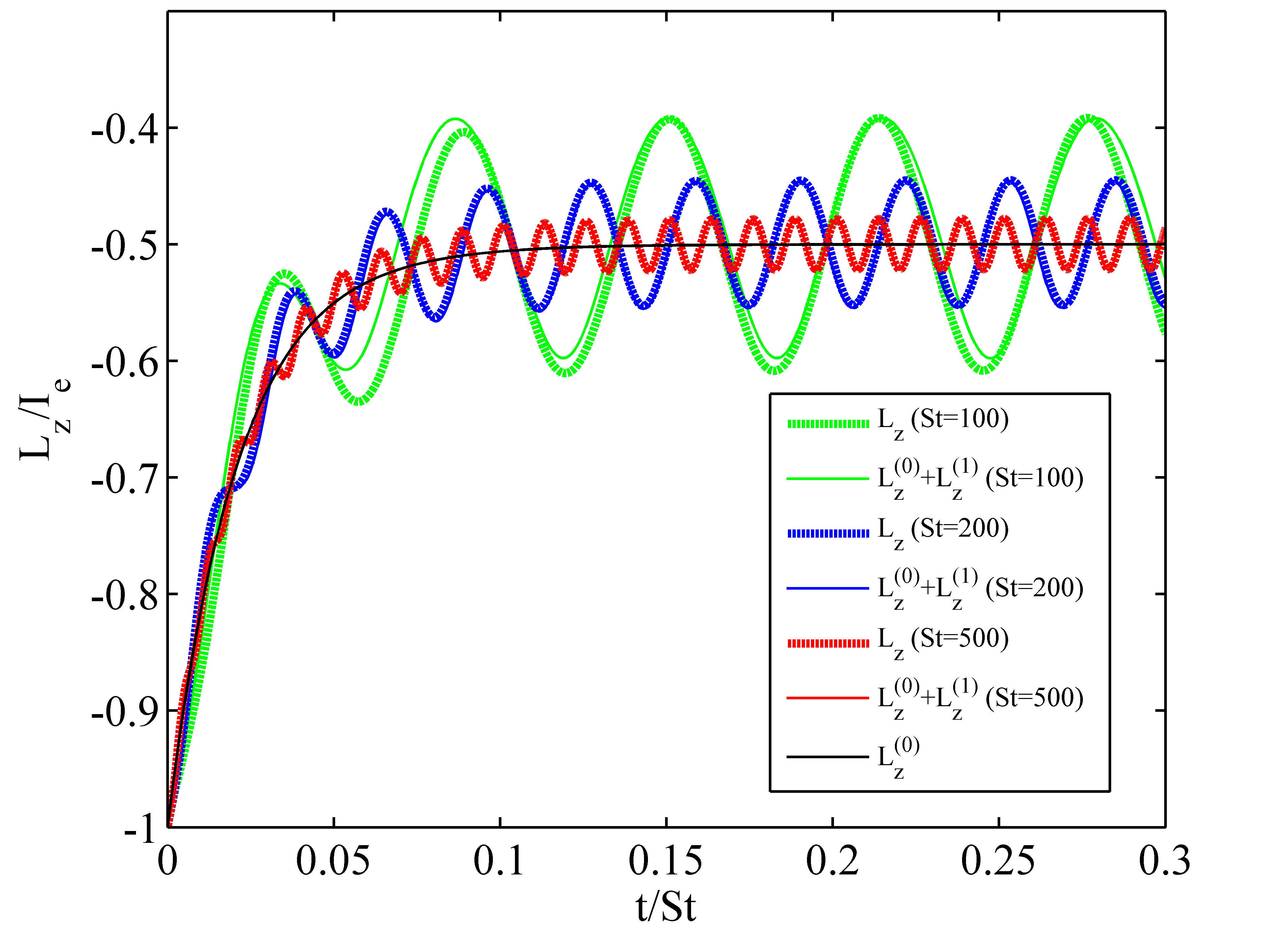}
\caption{\label{fodplot} Comparison between the first order approximation to $L_z$ using the method of multiple scales and the full solution for a $\kappa=2$  prolate spheroid with angular momentum aligned with vorticity, illustrating the $O(1/St)$ scaling of the deviation from the mean.}
 \end{figure}

Figure \ref{fodplot} shows the comparison between the full solution, the slow manifold and the first order correction using the method of multiple scales. We can similarly obtain the first order correction for the generalized spheroid system.

\end{document}